\documentclass[%
 reprint,
superscriptaddress,
preprintnumbers,
nofootinbib,
 amsmath,amssymb,
 aps,
prd,
]{revtex4-2}

\usepackage{graphicx}
\usepackage{dcolumn}
\usepackage{bm}
\usepackage{aas_macros}
\usepackage{hyperref}

\newcommand{\redmapper}{redMaPPer}

\newcommand{\Msun}{\rm{M}_\odot}

\newcommand{\de}{\text{d}}

\newcommand{\Mpc}{\rm{Mpc}}

\newcommand{\LCDM}{$\Lambda$CDM }

\newcommand{\ob}{^{\rm ob}}
\newcommand{\true}{^{\rm true}}

\newcommand{\zmin}{z_{\rm min}}
\newcommand{\zmax}{z_{\rm max}}
\newcommand{\lmin}{\lambda_{\rm min}}
\newcommand{\lmax}{\lambda_{\rm max}}

\newcommand{\lob}{\lambda^{\rm ob}}
\newcommand{\ltrue}{\lambda^{\rm true}}
\newcommand{\zob}{z^{\rm ob}}
\newcommand{\ztrue}{z^{\rm true}}
\newcommand{\lsat}{\lambda^{\rm sat}}
\newcommand{\lcen}{\lambda^{\rm cen}}
\newcommand{\Mmin}{M_{\rm min}}

\newcommand{\sint}{\sigma_{\rm intr}}

\usepackage[usenames]{color}
\definecolor{purple}{RGB}{150,0,200}

\newcommand{\planck}{{\it Planck}}

\newcommand{\avg}[1]{\langle #1 \rangle}
\newcommand{\Var}{\mbox{Var}}
\newcommand{\hMpc}{h^{-1}\ \Mpc}
\newcommand{\hMsun}{h^{-1}\ M_{\odot}}
\newcommand{\hatn}{\boldsymbol{\hat{n}}}
\newcommand{\hinv}{h^{-1}}
\newcommand{\Lmin}{L_{\rm min}}
\newcommand{\fcen}{f_{\rm cen}}

\newcommand{\Omegam}{\Omega_{{\rm m}}}

\newcommand{\Planck}{{\it Planck}}
\newcommand{\metacal}{\textsc{metacalibration}\ }
\newcommand{\IMthree}{\textsc{IM3SHAPE}\ }

\newcommand{\msun}{M_{\odot}}
\defcitealias{desy1wl}{MV19}

\begin{document}

\preprint{DES-2017-0320}
\preprint{FERMILAB-PUB-20-084}

\title[DES Y1 Cluster Cosmology]{Dark Energy Survey Year 1 Results: Cosmological Constraints from Cluster Abundances and Weak Lensing}


\author{T.~M.~C.~Abbott}
\affiliation{Cerro Tololo Inter-American Observatory, National Optical Astronomy Observatory, Casilla 603, La Serena, Chile}
\author{M.~Aguena}
\affiliation{Departamento de F\'isica Matem\'atica, Instituto de F\'isica, Universidade de S\~ao Paulo, CP 66318, S\~ao Paulo, SP, 05314-970, Brazil}
\affiliation{Laborat\'orio Interinstitucional de e-Astronomia - LIneA, Rua Gal. Jos\'e Cristino 77, Rio de Janeiro, RJ - 20921-400, Brazil}
\author{A.~Alarcon}
\affiliation{Institut d'Estudis Espacials de Catalunya (IEEC), 08034 Barcelona, Spain}
\affiliation{Institute of Space Sciences (ICE, CSIC),  Campus UAB, Carrer de Can Magrans, s/n,  08193 Barcelona, Spain}
\author{S.~Allam}
\affiliation{Fermi National Accelerator Laboratory, P. O. Box 500, Batavia, IL 60510, USA}
\author{S.~Allen}
\affiliation{Department of Physics, Stanford University, 382 Via Pueblo Mall, Stanford, CA 94305, USA}
\author{J.~Annis}
\affiliation{Fermi National Accelerator Laboratory, P. O. Box 500, Batavia, IL 60510, USA}
\author{S.~Avila}
\affiliation{Instituto de Fisica Teorica UAM/CSIC, Universidad Autonoma de Madrid, 28049 Madrid, Spain}
\author{D.~Bacon}
\affiliation{Institute of Cosmology and Gravitation, University of Portsmouth, Portsmouth, PO1 3FX, UK}
\author{K.~Bechtol}
\affiliation{LSST, 933 North Cherry Avenue, Tucson, AZ 85721, USA}
\affiliation{Physics Department, 2320 Chamberlin Hall, University of Wisconsin-Madison, 1150 University Avenue Madison, WI  53706-1390}
\author{A.~Bermeo}
\affiliation{Department of Physics and Astronomy, Pevensey Building, University of Sussex, Brighton, BN1 9QH, UK}
% ...

\collaboration{DES Collaboration}

\thanks{For comments or questions please contact:\\ des-publication-queries@listserv.fnal.gov}%

\date{\today}

\begin{abstract}
We perform a joint analysis of the counts and weak lensing signal of redMaPPer clusters selected from the Dark Energy Survey (DES) Year 1 dataset.  Our analysis uses the same shear and source photometric redshifts estimates as were used in the DES combined probes analysis.
Our analysis results in surprisingly low values for $S_8 =\sigma_8(\Omegam/0.3)^{0.5}= 0.65\pm 0.04$,
driven by a low matter density parameter, $\Omegam=0.179^{+0.031}_{-0.038}$, with $\sigma_8-\Omegam$ posteriors 
in $2.4\sigma$ tension with the DES Y1 3x2pt results, and in $5.6\sigma$ with the \planck\ CMB analysis.
These results include the impact of post-unblinding changes to the analysis, which did not improve the level of consistency with other data sets compared to the results obtained at the unblinding.
The fact that multiple cosmological probes (supernovae, baryon acoustic oscillations, cosmic shear, galaxy clustering and CMB anisotropies), and other galaxy cluster analyses all favor significantly higher matter densities suggests the presence of systematic errors in the data or an incomplete modeling of the relevant physics. Cross checks with X-ray and microwave data, as well as independent constraints on the observable--mass relation from SZ selected clusters, suggest that the discrepancy resides in our modeling of the weak lensing signal rather than the cluster abundance. Repeating our analysis using a higher richness threshold ($\lambda \ge 30$) significantly reduces the tension with other probes, and points to one or more richness-dependent effects not captured by our model.  
\end{abstract}

\maketitle



\section{Introduction}
\label{sec:intro}
The flat $\Lambda$CDM model, despite its apparent simplicity---six parameters suffice to define it---has proven able to describe a wide variety of observations, from the low to the high redshift Universe. Despite its successes, however, the two dominant components of the Universe in this model---the Cold Dark Matter (CDM) and the Cosmological Constant ($\Lambda$)---lack a fundamental theory to connect them with the rest of physics.
Ongoing (e.g. the Dark Energy Survey (DES)\footnote{https://www.darkenergysurvey.org}, Hyper Suprime-Cam\footnote{http://hsc.mtk.nao.ac.jp/ssp/},
Kilo-Degree Survey\footnote{http://kids.strw.leidenuniv.nl/index.php}
eRosita\footnote{http://www.mpe.mpg.de/eROSITA}, South Pole Telescope (SPT)\footnote{https://pole.uchicago.edu/}, Atacama Cosmology Telescope (ACT)\footnote{https://act.princeton.edu/})  and future surveys (e.g. Euclid\footnote{http://sci.esa.int/euclid/}, Large Synoptic Survey Telescope\footnote{https://www.lsst.org/},
WFIRST\footnote{https://wfirst.gsfc.nasa.gov/}) aim to further test the \LCDM paradigm
, as well as the mechanism that drives the cosmic acceleration, be it a cosmological constant, some form of dark energy, or a modification of General Relativity.
Lacking a fundamental theory to test, one way to shed light on the latter is by looking at the evolution of cosmic structures over the past few Gyr, when the dark energy becomes dominant, and searching for  discrepancies between the observables in the low-redshift Universe and the predictions for said observables derived from the high-redshift Universe as measured through observations of the Cosmic Microwave Background (CMB) anisotropies \citep[e.g.][]{wmap9,Planck2018}.

The Dark Energy Survey is a six-year survey that mapped $5000\,{\rm deg}^2$ of the southern sky in five broadband filters, $g$, $r$, $i$, $z$, $Y$, between August 2013 and January 2019, using the 570 megapixel Dark Energy Camera \citep*[DECam;][]{decam2015} mounted on the 4m Blanco telescope at the Cerro Tololo Inter-American Observatory (CTIO).
DES was designed with the primary goal of testing the \LCDM model and studying the nature of dark energy through four key probes: cosmic shear, galaxy clustering, clusters of galaxies, and Type Ia supernovae.

Galaxy clusters have long proven to be a valuable cosmological tool: arising from the highest peaks of the matter density field, their abundance and spatial distribution are sensitive to the growth of structures and cosmic expansion \citep[see e.g.][for reviews]{Allen2011,Kravtsov2012}. More specifically, the cluster abundance constrains the parameter combination $\sigma_8(\Omegam/0.3)^\alpha$, where $\Omegam$ is the mean matter density of the Universe, $\sigma_8$ is the present-day rms of the linear density field in spheres of $8 \hMpc$ radius, and $\alpha$ ranges between $\sim\!0.2-0.5$ depending on the characteristics of the survey.
The evolution of the cluster abundance can thus be used to measure the growth rate of cosmic structure, which in turn constrains dark energy and modified gravity models \citep[e.g.][]{Burenin2012,Mantz2015,Cataneo2015,Bocquet2018}.

At present, cluster abundance studies at all wavelengths are limited by their ability to calibrate the relation between halo mass and the observable used as a mass proxy.
Among the different techniques to calibrate the observable--mass relation, the weak lensing signal, based on the distortion of background galaxy images due to the gravitational lensing of intervening clusters, is the current gold standard \citep[e.g.][]{Mantz2015,Murata2017,Bocquet2018,HST2018}. 
Still, many sources of systematic uncertainty affect this type of measurement, including shear and photometric redshift biases, halo triaxiality, miscentering, and projection effects, each of which contribute a significant fraction of the total error budget 
\citep[e.g.][]{Melchior2017,medezinksietal18,miyatakeetal18,desy1wl}
As we will discuss later, this is especially true for the optically-selected cluster sample adopted in this work, for which the systematic error represents $\sim\!60\%$ of the total error budget on mass estimates.

In this study we combine cluster abundances and weak-lensing mass estimates derived from data collected during the first year of observation of DES to simultaneously constrain cosmology and the observable--mass relation. 
Our optically-selected catalog is built using the red sequence Matched-filter Probabilistic Percolation cluster finder algorithm \citep[redMaPPer;][]{Rykoff2014}. For mass estimates, we rely on updated results of the stacked weak lensing analysis of \cite{desy1wl}, which include a new calibration of the selection effect bias\footnote{We use the term "selection effect bias" to refer to the bias introduced by the cluster finder for preferentially selecting clusters with properties that correlate with the lensing signal at fixed mass.}.
The latter has been studied by means of numerical simulations by Wu et al. (in preparation) to validate the systematic bias correction adopted in \cite{desy1wl}. The results of this analysis, which started before the unblinding but have been finalized only after, show that selection effects have a $\sim\!20-30\%$ impact on stacked weak lensing mass measurements, a much larger effect compared to the $\sim\!4\%$ correction estimated in \cite{desy1wl} combining simulations \citep{dietrichetal14} and analytic estimates \citep{Simet2016}.

This analysis follows the methodology described in \cite{costanzietal18b}, in which we develop our pipeline using the \redmapper\ SDSS cluster catalog.
This analysis was performed blind to the cosmological parameters to avoid confirmation bias.  However, the large tension between our original unblinded results and multiple cosmological probes, including \planck\ CMB \citep{Planck2018}, and especially the DES 3x2pt \citep{des17} results,  motivated a careful review of our handling of systematics. This led us to revisit our estimates of the selection effects bias and, in turn, to re-analyse and update our results post-unblinding.
The analysis presented in the main text of the paper make use of this post-unblinding correction, and we will refer to it as the \textit{unblinded} analysis. For completeness, the cosmological results obtained at the unblinding (\textit{blinded} analysis, hereafter) are presented in appendix \ref{sec:blinded_res}.
As discussed in the paper, the post-unblinding correction, while reducing by 2$\sigma$ the preferred $\sigma_8$ value, does not improve the consistency of our posteriors with either the \planck\ CMB or the DES 3x2pt results. 

This paper is organized as follows: In Section \ref{sec:datasets} we provide an overview of the DES Y1 data products used in this work. Section \ref{sec:datavector} presents the two data vectors---cluster abundance and mean weak-lensing mass estimates---employed for the cosmological analysis. Section \ref{sec:model} describes our theoretical model to predict cluster counts and mean cluster masses, and thus derive cosmological and observable--mass relation parameter constraints. We present our results and address their consistency with other probes in Section \ref{sec:res}, while we discuss their implication in Section \ref{sec:disc}. Finally, we summarize and draw our conclusions in Section \ref{sec:concl}.

\section{Data}
\label{sec:datasets}
\begin{figure}
 	\includegraphics[width= \linewidth]{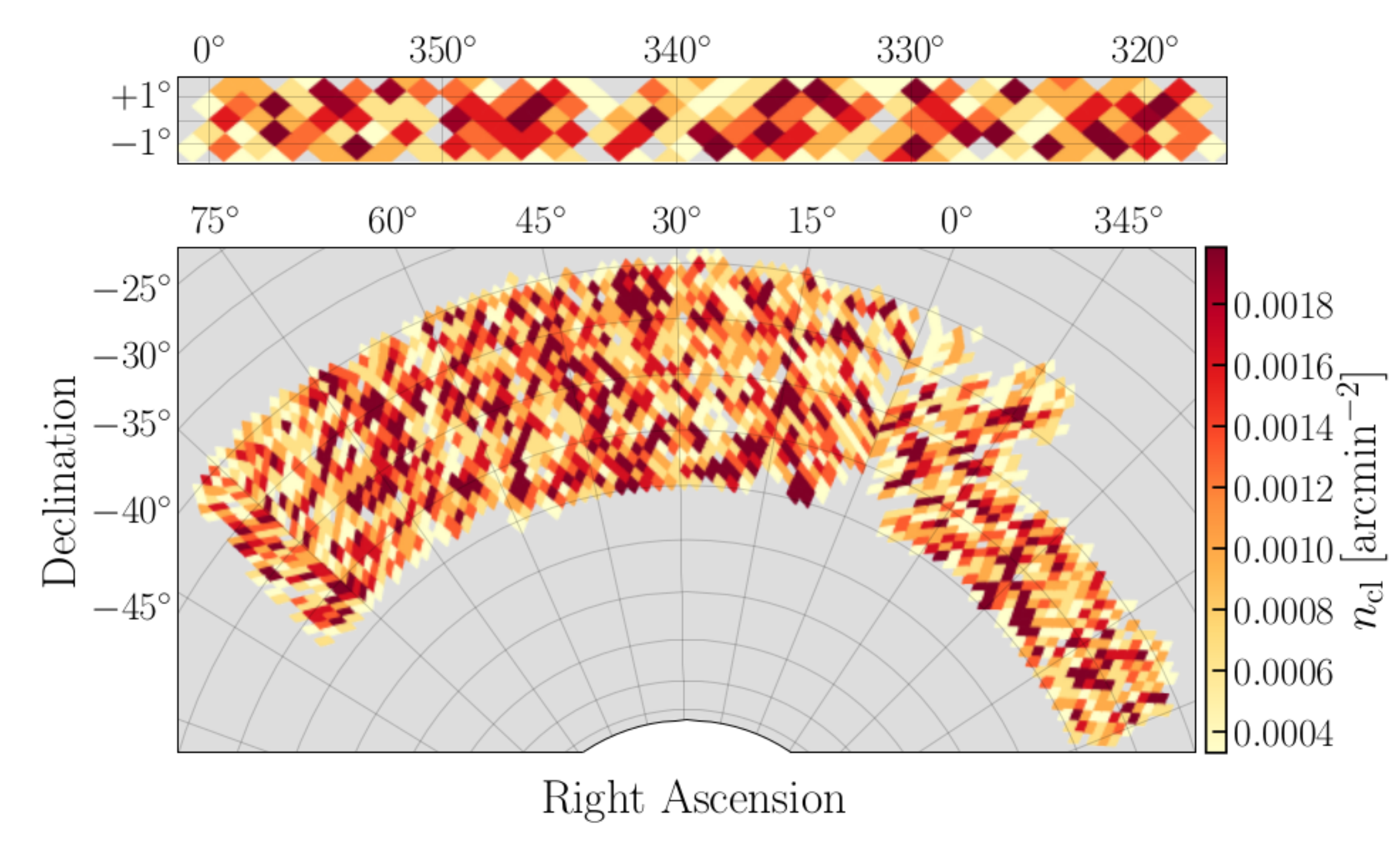}
	\caption{The DES Y1 \redmapper\ cluster density over the two non-contiguous regions of the Y1 footprint: the Stripe 82 region (116 deg$^2$; \textit{upper} panel) and the SPT region (1321 deg$^2$; \textit{lower} panel). 
	}
	\label{fig:footprint}
\end{figure}

In this work we use data collected by the DECam during the Year One (Y1) observational season, running from August 31, 2013 to February 9, 2014, which covers $\sim$1800 ${\rm deg}^2$ of the southern sky in the $g$, $r$, $i$, $z$ and $Y$ bands \citep{Y1gold}. Of the $\sim$1800  square degrees observed in Y1, $\sim$17$\%$ of them are excluded from the analysis due to a series of veto masks, vetting bright stars, bright nearby galaxies, globular clusters, and the Large Magellanic Cloud. The final DES Y1 footprint is shown in Figure \ref{fig:footprint}, and covers approximately $1500$ ${\rm deg}^2$ split in two non-contiguous regions: a larger region ($1321$ ${\rm deg}^2$; \textit{lower} panel) overlapping the footprint of the South Pole Telescope Sunyaev-Zel'dovich Survey \citep{spt2011}, and a smaller area (116 ${\rm deg}^2$; \textit{upper} panel), which overlaps the Stripe-82 deep field of the Sloan Digital Sky Survey \citep[SDSS,][]{s82}.

In sections \ref{sec:photo-cat}--\ref{sec:photoz_systematics} we summarize the main data products used in this work, and refer the reader to the relevant papers for further details.

\subsection{The DES Y1 Photometric Catalog}
\label{sec:photo-cat}

Photometry and `clean' galaxy samples are based on the Y1A1 Gold Catalog \citep{Y1gold}, the DES science-quality photometric catalog produced from Y1 data to enable cosmological analyses. This data set includes a multi-band photometric object catalog as well as maps of survey depth, foreground masks, and star--galaxy classification. Galaxy fluxes are measured using the multi-epoch, multi-object fitting (MOF) procedure described in \cite{Y1gold}. The typical $10\sigma$ limiting magnitude inside $2''$ diameter apertures for galaxies in Y1A1 Gold using MOF photometry is $g \simeq 23.7$, $r \simeq  23.5$, $i \simeq 22.9$, and $z \simeq 22.2$. Due to its shallow depth and significant calibration uncertainty, the Y band photometry was used in neither the \redmapper\ cluster finder nor for shape and photometric redshift measurements.

To build our cluster catalog, we rely on a subset of high-quality objects selected from the Y1A1 Gold catalog. First, we reject all objects classified as catalog artifacts, i.e. objects lying in regions having unphysical colors, astrometric discrepancies, or PSF model failures \citep[Section 7.4][]{Y1gold}. The sample is further refined via the {\tt MODEST\_CLASS} classifier, which was developed with the primary goal of selecting high-quality galaxy samples \citep[Section 8.1][]{Y1gold}. Finally, only galaxies that are brighter in the $z$ band than the local $10\sigma$ limiting magnitude are included in the galaxy catalog used by the \redmapper\ cluster finder.

\subsection{Cluster Catalog and Associated Systematics}
\label{sec:redmapper}

Our analysis relies on the DES Y1 \redmapper\ cluster catalog.
\redmapper\ is a photometric cluster finding algorithm that identifies galaxy clusters
as overdensities of red-sequence galaxies \citep{Rykoff2014}.  The algorithm has been extensively vetted against X-ray and Sunyaev-Zel'dovich (SZ) catalogs \citep{rozorykoff14,sadibekovaetal14,rozoetal15c,Saro2015}.  Incremental algorithmic updates are presented in \cite{Rozo2015b}, \cite{rykoffetal16}, and \cite{desy1wl}. Here, we present only a brief summary of the most salient features of the DES Y1 \redmapper\ catalog.  For further details on the algorithm, we refer the reader to the original work by \cite{Rykoff2014}. 

The DES Y1 \redmapper\ clusters are selected as overdensities of red-sequence galaxies in the DES Y1 photometric galaxy catalog. \redmapper\ counts the excess number of red-sequence galaxies  brighter than a specified luminosity threshold $\Lmin(z)$ within a circle of radius $R_\lambda = 1.0\ \hinv\ \Mpc (\lambda/100)^{0.2}$.  This number of galaxies is called the {\it richness}, and is denoted as $\lambda$.  We use all clusters of richness $\lambda\geq 20$ in the present analysis. The catalog is locally volume limited in that we use the survey depth to determine the maximum redshift $\zmax(\hatn)$ at which galaxies at our luminosity threshold are still detectable in the DES at $10\sigma$. Galaxy clusters are included in the volume-limited catalog if the cluster redshift $z \leq \zmax(\hatn)$.  The cluster survey footprint is mildly redshift dependent.  It is defined as a follows: a point $\hatn$ in the sky at redshift $z$ is included in the survey volume if a cluster at that redshift and position is masked by at most 20\% by the galaxy mask. The above criteria, along with the recovered redshift distribution of the \redmapper\ clusters, are used to generate a large random cluster catalog to characterize the survey volume.

A total of 7066 galaxy clusters are included in the DES Y1 \redmapper\ volume-limited catalog. We remove 32 ${\rm deg}^2$ corresponding to 10 non-contiguous deep fields for supernovae science, bringing down the total number of clusters to 6997.  We further  restrict ourselves to the redshift interval $z\in[0.2,0.65]$, which reduces the number of galaxy clusters to 6504. \redmapper\ performance below redshift $z=0.2$ is compromised by the lack of $u$-band data, while there are relatively few galaxy clusters in the catalog above redshift $z=0.65$, making it a convenient upper limit for calculating binned abundances.  Figure~\ref{fig:footprint} shows the footprint of the DES Y1 \redmapper\ cluster survey.  For illustration purposes only, we show the cluster density for clusters of richness $\lambda\geq 5$. The clusters with $\lambda<20$ have not been used for any other purpose in this analysis.

Galaxy clusters are centered on bright cluster galaxies, but not necessarily on the brightest cluster galaxy.  The \redmapper\ algorithm iteratively self-trains a filter that relies on galaxy brightness, cluster richness, and local galaxy density to determine candidate central galaxies.  The algorithm centers the cluster on the most likely candidate central galaxy.

Turning to our characterization of systematic uncertainties in cluster finding, we note that, at a fundamental level,  cluster catalogs should provide three measures for a cluster: 1) a sky location (center), 2) a cluster redshift estimate, and 3) an observable that serves as a proxy for mass.  We briefly summarize the DES Y1 \redmapper\ performance in each of these categories:

\bf Cluster centering: \rm  The centering efficiency of the \redmapper\ algorithm is studied using X-ray imaging by  \cite{zhangetal19}.  That work demonstrates that the fraction of correctly centered \redmapper\ clusters is $\fcen=0.75 \pm 0.08$.  The distribution of radial offsets for miscentered clusters relative to the true cluster center is modeled as a Gamma distribution with a characteristic length scale $R_{\rm mis}= \tau R_\lambda$, where $R_\lambda$ is the cluster radius assigned by \redmapper, and $\tau=0.17 \pm 0.04$. While the X-ray matched clusters are strongly biased to high richness, the authors do not find a significant richness dependence of their results.

\bf Photometric redshift estimation: \rm The DES Y1 \redmapper\ photometric redshifts are unbiased at the $|\Delta z| \leq 0.003$ level, and have a median photometric redshift scatter $\sigma_z/(1+z)\approx 0.006$ \citepalias[see Figure 3 in][]{desy1wl}. The photometric redshift uncertainties are estimated directly from the photometric data, and are rescaled to match the observed dispersion in spectroscopic cluster redshifts.   The photometric redshift errors are both redshift and richness dependent.  The redshift dependence is modeled using a polynomial of order ten, with the coefficients for the polynomial fit independently for each richness bin. 

Here, we assume the photometric cluster redshifts are unbiased, and we assume a perfect characterization of the photometric redshift scatter.  That is, we do not marginalize over our uncertainty in the scatter in the photometric cluster redshifts.  In light of other sources of systematic uncertainty in our analysis---in particular source photometric redshift uncertainties---we are confident that this approximation is sufficient. 

\bf Assigning a mass proxy (Richness Estimation): \rm If richness is a good mass proxy, then richer clusters should be more massive.  As evidenced by \cite{desy1wl}, this is indeed the case, with the mean mass of galaxy clusters scaling as $\approx \lambda^{1.3}$.  \cite{rozorykoff14} demonstrated that the \redmapper\ richness was the lowest scatter optical mass tracer among those available at the time of that study.  Nevertheless, the scatter in mass at fixed richness for \redmapper\ clusters is large.  Moreover, because of the coarse line-of-sight resolution achievable with broad-band photometric survey data, photometric cluster catalogs such as \redmapper\ will be susceptible to projection effects \citep[e.g.][]{Cohn2007}.  Indeed, there is now ample observational evidence confirming this expectation \citep{farahietal16,zuetal17,buschwhite17}.  As emphasized by \cite{Erickson2011}, a detailed quantitative characterization of the impact of projection effects is necessary to derive unbiased cosmological constraints from photometric cluster samples.  In this work, we forward-model the impact of projection effects on the DES Y1 cluster sample as described in \cite{costanzietal18}.  This modeling accounts not only for projection effects, but also for the masking of clusters by larger systems during the percolation step of the cluster  finding.\footnote{Percolation refers to removing from the candidate cluster member list galaxies that were blended into richer systems along the line of sight.}  

\subsection{Shear Catalog and Associated Systematics}
\label{sec:shearcat}
The weak-lensing analysis of \cite{desy1wl} relies on the galaxy shape catalogs presented in \cite{y1shapes}. In DES Y1, shape measurements have been performed with two independent pipelines, \metacal\  \citep{SheldonMETA, HuffMETA} based on {\sc NGMIX} \citep{sheldon2015}, and \IMthree \citep{im3shape}. Both codes passed a series of tests that show them to be suitable for cosmological studies.  However, for the stacked weak lensing analysis of \cite{desy1wl}, only the \metacal shape catalog has been used due its larger effective source density (6.28 ${\rm arcmin}^{-2}$). \metacal measures shapes by simultaneously fitting the galaxy images in the $r$, $i$, $z$ bands with a 2D Gaussian model convolved with the point-spread functions (PSF) appropriate to each exposure.

Galaxy shape estimators are subject to various sources of systematic errors. For a stacked shear analysis, the dominant source of uncertainty is a multiplicative bias, i.e., an over- or under-estimation of gravitational shear as inferred from the mean tangential ellipticity of lensed galaxies.  \metacal uses a self-calibration technique to de-bias shear estimates \citep*{y1shapes}. Specifically, each galaxy image is deconvolved from the estimated PSF, and a small positive and negative shear is applied to the two ellipticity components of the deconvolved image. The resulting images are then convolved once again with a symmetrized version of the PSF, and an ellipticity is estimated for these new images. This procedure allows one to estimate the response of the shape measurement to gravitational shear from the images themselves. An analogous technique is employed to calibrate shear biases due to selection effects.  This involves measuring the mean response of the ellipticities to the selection, and then repeating the selections on quantities measured on artificially sheared images.
The effectiveness of the \metacal self-calibration has been addressed in \cite{y1shapes} by means of simulated images generated with the \textsc{GALSIM} package \citep{Rowe2015} using high-resolution images of the COSMOS field processed to mimic the actual noise and PSFs of the DES Y1 data. 
From this analysis they obtained a Gaussian prior on the multiplicative bias of $ 0.012 \pm 0.013$, and found no evidence of a significant additive bias term. Among all the sources of multiplicative bias investigated---including errors due to the use of multi-epoch data, leakage of stellar objects into the galaxy sample, and errors in the modeling of the PSF---blending is the only component with a net bias.  The other sources are consistent with zero bias, although they contribute to the bias uncertainty.

\subsection{Photometric Redshift Catalog and Associated Systematics}
\label{sec:photoz_systematics}

Photometric redshifts of source galaxies were estimated using the template-based BPZ algorithm \citep{Benitez2000,Coe2006}.  Systematic uncertainties in the recovered redshifts were calibrated in a variety of different ways, including cross-matching to COSMOS galaxies, cross-correlation redshifts (\citealp*{gattietal18}; \citealp{davisetal17}), and through the redshift dependence of the shear signal of foreground galaxies of known redshift \citep*{pratetal18}.  The former two were combined in \cite{Y1pz} to arrive at the final systematic error budget for the source photometric redshifts.  We emphasize that all three methods resulted in mutually consistent calibrations.

The results of \cite{Y1pz} do not directly translate into a calibration of the systematic error associated with photometric redshift estimates in the cluster mass calibration analysis because of differences in how the data are used.
Specifically, rather than relying on a tomographic analysis of source galaxies, the cluster mass calibration effort in \cite{desy1wl} rescaled the shear signal of each galaxy into the corresponding density contrast variable $\Delta\Sigma$.  This allowed us to trivially combine the lensing signal of all sources to construct an estimate of the excess surface density profile ($\Delta\Sigma$) of the clusters. \cite{desy1wl} used the same COSMOS-matching algorithm of \cite{Y1pz} to calibrate the systematic uncertainty in the amplitude of the recovered weak-lensing profile due to photometric redshift uncertainties.   The principal sources of error in this calibration are the cosmic variance associated with the small area of the COSMOS field and uncertainties in connecting the COSMOS measurements to the source galaxy sample, which result in a $2\%$ systematic uncertainty in the amplitude of $\Delta\Sigma$. Here, we make the conservative assumption that this uncertainty is perfectly correlated across all cluster redshifts.   The resulting systematic uncertainty in the amplitude of the mass--richness relation of \redmapper\ clusters from this effect is 2.6\%.

\section{Data Vector and Error Budget}
\label{sec:datavector}

The DES Y1 data vector for the cluster abundance analysis comprises:
\begin{enumerate}
\item the number of galaxy clusters in bins of richness and redshift, and
\item the average mass of the galaxy clusters in said bins.
\end{enumerate}
We detail below how the data vectors and the associated covariance matrices are constructed, and characterize the associated sources of systematic uncertainty.

\subsection{Cluster Abundances and Uncertainties}
\label{sec:NC_err}
We bin the galaxy clusters in three redshift bins spanning the range $z\in[0.2,0.65]$ and four richness
bins spanning the range $\lambda\in[20,\infty]$.  
The richness selection threshold aims to avoid large fractional uncertainties in cluster richness due to Poisson sampling 
while the redshift range sampled is driven by the available photometric data: our bluest filter is $g$, which restricts our analysis to redshifts $z\geq 0.2$, while the depth of the data is such that there are few clusters past $z=0.65$.  Table~\ref{tab:abundances} collates the number of galaxy clusters in each of our richness and redshift bins, as labeled.  The binning scheme employed in this work is driven by the weak-lensing analysis of \cite{desy1wl}, which necessitates somewhat broad bins to achieve high signal-to-noise measurements of the weak-lensing profile of the galaxy clusters.  A byproduct of this choice is that the number of galaxy clusters in each bin is large; our least populated bin contains 91 galaxy clusters.  

The uncertainty in the cluster abundance is modeled as the sum of a Poisson component, a sample variance contribution associated with the unknown density contrast of the DES Y1 survey region as a whole \citep{hukravtsov03,Hu2006}, and a miscentering component.  We note that while the Poisson term of the likelihood is strictly non-Gaussian, the high occupancy number of all of our bins ensures that the Gaussian approximation to the Poisson likelihood is a good approximation. 

Sample variance is calculated using the technique of \cite{hukravtsov03}.  Briefly, the number density fluctuations in the cluster sample takes the form $\delta_N = b\delta_V$, where $b$ is the bias of the clusters in a given richness/redshift bin, and $\delta_V$ is the mean matter fluctuation within the appropriate DES Y1 survey volume (there is one such random variable for each redshift bin). The cluster bias as a function of mass is calculated using the fitting formula of \cite{Tinker2010}. The survey mask is approximated as spherically symmetric about the azimuthal axis.  In conjunction with this mask,  the redshift intervals for each of the bins defines a survey volume, and $\delta_V$ is the volume-averaged density contrast $\delta$.  The associated covariance can be readily calculated in terms of the linear matter power spectrum. We also account for the covariance between neighboring redshift bins.  For additional details, we refer the reader to Appendix A in \cite{costanzietal18b}. Our covariance matrix is explicitly model dependent: we compute both Poisson and sample variance contributions at each point in the chain, and we account for the determinant term of the covariance matrix in the likelihood.  We have verified that holding the covariance matrix fixed results in nearly identical posteriors.
At high richness, the Poisson contribution dominates, with sample variance becoming increasingly important at low richness \citep{hukravtsov03}.

Cluster miscentering tends to bias low our richness estimates and induces covariance amongst neighboring richness bins \citep[e.g.][]{zhangetal19}. Rather than forward modeling this effect we directly correct our observed data vector for it. The correction and the covariance matrix associated with miscentering are estimated as follows: starting from a halo catalog, we assign richness to each halo according to the model of \cite{costanzietal18}. We then randomly miscenter every halo in the catalog  following the miscentering model of \cite{zhangetal19}, and recompute the cluster abundance data vector.  The procedure is iterated $10^3$ times, and we use these realizations to derive the correction factors --- obtained as the mean of the ratios between the number counts in richness/redshift bins including or not the miscentering effect --- and the corresponding covariance matrix.
The uncertainty associated with cluster miscentering in the abundance function ($\approx 1.0-1.5\%$) is  sub-dominant to the Poisson and sample variance contributions in all richness and redshift bins (see table~\ref{tab:abundances}).
Note that miscentering only mixes neighboring richness bins at the same redshift; there is no covariance between different redshift bins due to miscentering. 


\begin{table*}
  \caption{Number of galaxy clusters in the DES Y1 \redmapper\ catalog for each richness and redshift bin.
  Each entry takes the form $N (N) \pm \Delta N\ {\rm stat}\ \pm \Delta N\ {\rm sys}$. The numbers between parenthesis correspond to the number counts corrected for the miscentering bias factors (see section \ref{sec:NC_err}). The first error bar corresponds to the statistical uncertainty in the number of galaxy clusters in that bin, and is the sum of a Poisson and a sample variance term.  The systematic error is due to miscentering errors in the \redmapper\
  catalog (see text for details).}
  \label{tab:abundances}
    \begin{tabular}{cccc}
		$\lambda$ & $z\in[0.2,0.35)$ & $z\in[0.35,0.5)$ & $z\in[0.5,0.65)$ \\ \hline
		$[20,30)$ & 762 (785.1) $\pm$ 54.9 $\pm$ 8.2 & 1549 (1596.0) $\pm$ 68.2 $\pm$ 16.6 & 1612 (1660.9) $\pm$ 67.4 $\pm$ 17.3\\
		$[30,45)$ & 376 (388.3) $\pm$ 32.1 $\pm$ 4.5 & 672 (694.0) $\pm$ 38.2 $\pm$ 8.0 & 687 (709.5) $\pm$ 36.9 $\pm$ 8.1\\
		$[45,60)$ & 123 (127.2) $\pm$ 15.2 $\pm$ 1.6 & 187 (193.4) $\pm$ 17.8 $\pm$ 2.4 & 205 (212.0) $\pm$ 17.1 $\pm$ 2.7\\
		$[60,\infty)$ & 91 (93.9) $\pm$ 14.0 $\pm$ 1.3 & 148 (151.7) $\pm$ 15.7 $\pm$ 2.2 & 92 (94.9) $\pm$ 14.2 $\pm$ 1.4\\
    \end{tabular}
\end{table*}


\subsection{Cluster Masses and Uncertainties}
\label{sec:wlmass}

The mean mass of the galaxy clusters in each richness and redshift bin is estimated through a stacked weak-lensing analysis \citep{desy1wl}. Briefly, we use the DES Y1 \metacal\ shear catalog \citep{SheldonMETA, HuffMETA} to estimate the shear for each cluster--source pair. This shear is turned into an estimate of the projected mass-density contrast $\Delta\Sigma$ using the inverse critical surface density $\Sigma_{{\rm crit}}^{-1}$. The latter depends on both the lens and source redshifts.  For the source redshift, we use the redshift probability distribution for the source as estimated using the BPZ code \citep{Benitez2000}.  The uncertainty in the overall lensing amplitude $\avg{\Sigma_{{\rm crit}}^{-1}}$ is calibrated by matching the sources in color--magnitude space to COSMOS galaxies with 30-band photo-$z$s \citep{laigleetal16}.
In addition, we evaluate the correction to the weak-lensing profiles due to the contamination of the source catalog by cluster members (boost factor) by measuring how interlopers distort the photometric redshift distribution of the source catalog towards the cluster cores.  For details, we refer the reader to \cite{Varga2018} (see also \citealp{Gruen2014}).  
The statistical uncertainties of the recovered weak-lensing profiles are characterized using a semi-analytic covariance matrix that is validated through comparisons to jackknife estimates of the variance.  The covariance matrices account for shape noise, cosmic variance, scatter in the richness--mass relation, scatter in the concentration--mass relation, and scatter in halo ellipticities \citep{Gruen2015,desy1wl}.  The covariance matrix on the boost factor profiles are jackknife estimates, but these uncertainties have a negligible impact on the mass posteriors.  

We simultaneously fit the recovered weak lensing $\Delta\Sigma(R|M)$ profile along with the corresponding boost factor data to arrive at the final posteriors for the mean mass.  The theory prediction for $\Delta\Sigma(R|M)$ is obtained by projecting an analytic model of the halo--mass correlation function.
In our fit we only consider data in the radial range $R\in [0.2,30]\ \Mpc$. For each redshift and richness bin considered we vary both the halo concentration and halo mass.
Model biases due to our choice of analytic model and the selection effect correction adopted in the \textit{unblinded} analysis are calibrated using numerical simulations. For further details, we refer the reader to section 5.4 of \cite{desy1wl} and appendix \ref{sec:unblnd}.
Table \ref{tab:posterior_masses} collects the mean mass estimates and associated errors adopted in the \textit{unblinded} analysis.


\begin{table*}
\caption{Mean mass estimates for DES Y1 \redmapper\ galaxy clusters in each redshift bin.  The reported quantities are $\log_{10}( {\rm M})$ where masses are defined using a 200-mean overdensity criterion ($M_{\rm 200m}$). The masses are measured in $\hinv {\rm M}_\odot$ and include the selection effect correction discussed in Appendix \ref{sec:unblnd}.  The first error bar refers to the statistical error in the recovered mass,  while the second error bar corresponds to the systematic uncertainty.}
  \label{tab:posterior_masses}
    \begin{tabular}{cccc}
		$\lambda$ & $z\in[0.2,0.35)$ & $z\in[0.35,0.5)$ & $z\in[0.5,0.65)$ \\ \hline
		$[20,30)$ & 14.036 $\pm$ 0.032 $\pm$ 0.045 & 14.007 $\pm$ 0.033 $\pm$ 0.056 & 13.929 $\pm$ 0.048 $\pm$ 0.072\\
		$[30,45)$ & 14.323 $\pm$ 0.031 $\pm$ 0.051 & 14.291 $\pm$ 0.031 $\pm$ 0.061 & 14.301 $\pm$ 0.041 $\pm$ 0.086\\
		$[45,60)$ & 14.454 $\pm$ 0.044 $\pm$ 0.050 & 14.488 $\pm$ 0.044 $\pm$ 0.065 & 14.493 $\pm$ 0.056 $\pm$ 0.068\\
		$[60,\infty)$ & 14.758 $\pm$ 0.038 $\pm$ 0.052 & 14.744 $\pm$ 0.038 $\pm$ 0.052 & 14.724 $\pm$ 0.061 $\pm$ 0.069\\
   \end{tabular}
\end{table*}


We note that the lensing profile $\Delta\Sigma(R)$ from the data requires an assumed cosmological model to transform angular separations into radial distances and to transform redshifts into angular diameter distances.  In addition, the two-halo term of the weak-lensing profile requires that we specify the clustering amplitude of the dark matter.  Within the context of a flat $\Lambda$CDM cosmological model, this implies that the recovered weak-lensing masses are sensitive to the matter density parameters $\Omega_{\rm m}$, the Hubble parameter $h$, and the clustering-amplitude parameter $\sigma_8$.  The Hubble-parameter dependence can be readily absorbed into the masses by quoting masses in units of $h^{-1}\ M_\odot$.  We approximate the dependence of the recovered masses as linear in $ \Omega_{\rm m}$ and $\ln (10^{10} A_s)$.  The coefficients of this dependence are evaluated numerically by computing the best-fit masses along a grid of values in $\ln (10^{10} A_s)$ and $\Omega_{\rm m}$, and fitting the resulting data in each bin with a line. The mean slopes obtained with this procedure are: $\de \log(M)/\de \Omega_{\rm m} = -0.40$ and $\de \log(M)/\de \ln (10^{10} A_s) = -0.015$.  We have verified that this approximation is accurate at better than the 2\% level in each bin, easily sufficient for our purposes (see Table~\ref{tab:posterior_masses}).  When iterating over the cosmological parameters in our analysis we explicitly account for the above cosmological dependence using this linear approximation. 


\subsection{Systematic Error Budget}
\label{sec:mass_sys}

Cluster cosmology has long been limited by systematic uncertainties in cluster mass calibration.  
This remains true today, and will likely remain so for the foreseeable future.  
We summarize the observational systematics that we have accounted for in our analysis.  Where quoted, the numbers refer to the uncertainty in the amplitude of the mass--richness relation, and are taken directly from Table 6 in \cite{desy1wl}, except as noted below.  Multiplicative shear and photometric redshift biases are assumed to be perfectly correlated across all richness and redshift bins.  Centering is not assumed to be perfectly correlated across all bins. 
The systematic errors we have accounted for are:
\begin{enumerate}
\item Multiplicative shear bias: 1.7\% Gaussian (\citep{y1shapes}, see section \ref{sec:shearcat}). 
\item Photometric redshift bias of the source galaxy population: 2.6\% Gaussian \citep[see section 4.3 of][]{desy1wl}. 
\item Cluster centering: $\leq 1\%$ \citep[see section 5.2 of][]{desy1wl}.  We forward model the impact of cluster miscentering on the weak-lensing profile, marginalizing over the priors derived by \cite{zhangetal19} and von der Linden et al. (in preparation).  

\item Modeling systematics: $2\%$ Gaussian \citep[see section 5.4 of][]{desy1wl}. Inaccuracies in our model of the halo--mass correlation function result in biased mass inferences from the weak lensing data.  These biases and their uncertainty are calibrated using numerical simulations.

\item Selection effect bias. Systematics which introduce correlation between cluster richness and lensing signal could bias our mass estimates. In \cite{desy1wl} we accounted for such bias using an analytical estimate of the impact of halo triaxiality and projection effects on weak lensing mass measurements (see their section 5.4.2). These estimates proved to be significantly smaller than our own, more recent determination using numerical simulations (see Appendix \ref{sec:unblnd} for details).  This simulation analysis lowers the recovered weak-lensing masses in a richness and redshift dependent way, with typical shifts being $\approx 20\%$--$30\%$.  The analysis presented in the main text of the paper (\textit{unblinded} analysis) adopts the selection effect corrections derived in Appendix \ref{sec:unblnd}.  We conservatively assume the correction to be uncertain at half its amplitude, leading to an $\approx 13\%$ systematic uncertainty on mass.  This uncertainty accounts for $\approx 60\%$ of our final error budget on the mass estimates.
\end{enumerate}


\section{Theoretical Model}
\label{sec:model}

Our theoretical model is the same as that described in detail in \cite{costanzietal18b}.  For this reason, here we only provide a summary of our method.  

The expectation value of the number counts and mean masses of the \redmapper\ galaxy clusters in a given richness and redshift bin are given by
\begin{eqnarray}
\avg{N} & = & \int_0^\infty d\ztrue \int_{\zmin}^{\zmax} d\zob \int_{\lmin}^{\lmax}d\lob\ \\ \nonumber
 & & \hspace{0.4in} \avg{n|\lob,\ztrue} \frac{dV}{d\ztrue} P(\zob|\ztrue) \\
\avg{M} & = & \frac{1}{\avg{N}} \int_0^\infty d\ztrue \int_{\zmin}^{\zmax} d\zob \int_{\lmin}^{\lmax}d\lob\ \\ \nonumber
 & & \hspace{0.4in} \avg{nM|\lob,\ztrue} \frac{dV}{d\ztrue} P(\zob|\ztrue) .
\end{eqnarray}
In the above expressions, $\lmin$ and $\lmax$ are edges of the richness bins, while $\zmin$ and $\zmax$ are the edges of the photometric redshift bins.  The quantities $\avg{n|\lob,\ztrue}$ and $\avg{nM|\lob,\ztrue}$ are the comoving space density of clusters and the mass weighted comoving densities, respectively. The term $dV/d\ztrue$ is the survey volume per unity redshift.  These various quantities are given by
\begin{eqnarray}
\avg{n|\lob,\ztrue} & = & \int_{0}^\infty dM\ \frac{dn}{dM} P(\lob|M,\ztrue) \\
\avg{nM|\lob,\ztrue} & = & \int_{0}^\infty dM\ \frac{dn}{dM}M P(\lob|M,\ztrue) \\
\frac{dV}{dz} & = & \Omega_{\rm mask}(z) cH^{-1}(z)\chi^2(z)
\end{eqnarray}
where $\Omega_{\rm mask}(z)$ is the survey area as a function of redshift, $H(z)$ is the Hubble parameters as a function of redshift, $\chi(z)$ is the comoving distance to redshift $z$, and $dn/dM$ is the halo mass function.  The above expression assumes a flat cosmology.  The survey area is computed as described in \cite{costanzietal18b}, and is nearly constant  up to redshift $z=0.5$, dropping to $\approx 50\%$ of the total survey area at $z\approx 0.63$.  Uncertainties in the survey area as a function of redshift are below 1\%, and do not contribute to our error budget.

As noted earlier in section~\ref{sec:redmapper}, we assume the photometric redshift probability distributions are known. The halo mass function is modeled using the \cite{Tinker2008} halo mass function, but allowing for power-law deviations that are calibrated using numerical simulations.  Specifically, we assume the mass function is specified by
\begin{equation}
\frac{dn}{dM} = \left( \frac{dn}{dM} \right)_{\rm Tinker} \left[ s \ln \left( \frac{M}{M_*} \right) + q \right].
\label{eq:qs}\end{equation}
The parameters $s$ and $q$ are fit to the {\sc Aemulus} simulations \citep{aemulus1}, which are also used to characterize the associated uncertainties in the parameters $s$ and $q$ (see table~\ref{tab:parameters}).  Our cosmological posteriors are marginalized over these uncertainties. 
Following \cite{costanzietal18b}, we do not include additional uncertainties due to the impact of baryonic physics on the halo mass function.  This assumption is well justified because of our choice of halo mass definition: for halos with $M\gtrsim 10^{14}\ h^{-1} M_\odot$, the radius $R_{200m}$ is sufficiently large that the baryonic redistribution within a halo due to cooling and feedback processes has a negligible impact on the mass within $R_{200m}$ \citep{Cui2014,Velliscig2014,Bocquet2016}.

The key remaining ingredient is the model for the richness--mass relation $P(\lob|M)$.  Our model is described in \cite{costanzietal18}, which was custom built for this analysis. Briefly, the intrinsic richness--mass relation is modeled using a conventional halo model parameterization, with $\ltrue = \lcen + \lsat$ where $\lcen$ and $\lsat$ are the number of central and satellite galaxies respectively.  $\lcen$ is assumed to be a deterministic function of mass, with $\lcen=1$ for $M\geq \Mmin$ and $\lcen=0$ otherwise. 
$\lsat$ is a random variable with an expectation value
\begin{equation}
\label{eqn:RMR}
\avg{\lsat|M,z} = \left( \frac{M-\Mmin}{ M_1 - \Mmin} \right)^\alpha \left( \frac{1 + z}{1 + z_*} \right)^\epsilon 
\end{equation}
where $M_1$ is the characteristic mass at which a halo of mass $M$ has on average one satellite galaxy, and the pivot redshift is set equal to the mean redshift of the sample $z_*=0.45$.  Note that the above formula ensures that only halos with central galaxies can have satellite galaxies. 
To allow for super-Poisson halo occupancies at high mass, we model $P(\ltrue|M)$ as the convolution of a Poisson and a Gaussian distribution, where the scatter of the latter is simply $\sint\avg{\lsat|M,z}$. 
For numerical reasons, we approximate this convolution using a skew-normal distribution.  For details, see \cite{costanzietal18b}, particularly Appendix B.  We note that because of the Gaussian component of $P(\ltrue|M)$, a large width may result in negative richness values.  These are interpreted as a finite probability of having $P(\ltrue=0)$, where the probability $P(\ltrue)$ is set to the integral of the Gaussian model below $\ltrue=0$.  In other words, negative $\ltrue$ values are considered halos with no satellite galaxies (and therefore no galaxy overdensity).  We investigate the sensitivity of our cosmological conclusions to our model for $P(\ltrue|M)$ in section \ref{sec:robustness}.  

The observed richness $\lob$ is a noisy measurement of $\ltrue$.  Four distinct sources of noise on $\lob$ are: 1) random errors associated with magnitude errors and background subtraction of uncorrelated structures; 2) projection effects; 3) percolation effects and 4) miscentering effects. The modeling of first three effects is the focus of our work in \cite{costanzietal18}.  In that work, we demonstrate that projection effects follow an exponential distribution, while photometric uncertainties and background subtraction lead to a Gaussian error.  Percolation effects modulate the richness of masked halos by a multiplicative factor that is uniformly distributed between $0$ and $1$, and the fraction of clusters that suffer from percolation effects is a decreasing function of richness. 
Parameters governing these distributions are determined by a semi-empirical method applied to halos in synthetic light-cone maps derived from N-body simulations \citep{DeRose2019BuzzardFlock}.  DES \redmapper\ data is used to calibrate a projection kernel that is used as a weight function applied to the simulated halos.  Using sightlines that target halos of specific intrinsic richness and redshift, a weighted sum of the richness of halos along the line of sight is used to estimate the component of $P(\lob|\ltrue,z)$ arising from two-halo and higher spatial correlations. These same simulations are used to calibrate the purely geometric impact of percolation.  The photometric and background subtraction noise is measured by injecting artificial clusters in the data.  The end result is a calibrated distribution $P(\lob|\ltrue,z)$ describing the impact of observational uncertainties and projection effects on the DES Y1 \redmapper\ cluster sample. Further details of this calibration are presented in Appendix~\ref{app:lob_calibration}. 

At this point we have described all the necessary ingredients for calculating the expectation value of our observable vector.  We model the likelihood function as a Gaussian distribution, which requires that we further specify the associated covariance matrix.  As described in section~\ref{sec:datavector}, the covariance matrix for the abundance reflects Poisson, sample variance, and miscentering uncertainties.  This covariance matrix is varied in parameter space, and we explicitly account for the term involving the determinant of the covariance matrix in our likelihood function. The covariance matrix for the recovered weak-lensing masses reflects the semi-analytic covariance matrix characterizing the weak-lensing data, and explicitly accounts for systematic uncertainties in the recovered weak-lensing masses.  All the systematic uncertainties, except the one associated with selection effects, are assumed to be correlated across richness and redshift bins. The lack of covariance in the selection effects correction allows for the selection effects to vary as a function of richness and redshift. 

Our analysis assumes no covariance between the number counts and the recovered mean masses in bins. However, it is reasonable to expect that an increase in projections will give rise to both an increase in the number counts, and an increase in the weak-lensing mass, e.g. due to the effects modeled in Appendix \ref{sec:unblnd}. 
Improved simulations and synthetic sky catalogs will allow us to simultaneously model coupled systematic effects within the data vector of counts and mean weak lensing masses. However, large, mass-independent positive correlations between the abundance and weak-lensing masses are ruled out as the resulting covariance matrix stops being positive definite.
In particular, assuming the element of the cross-covariance matrix to be given by $r\sigma_{\rm NC}\sigma_{\rm M_{\rm WL}}$, as $r$ increases, the determinant of the covariance matrix decreases, eventually becoming negative at $r\approx 0.15$. Adopting a ``large'' mass-independent correlation coefficient (compared to its maximum possible value above) of $r=0.125$ has only a minor impact on our cosmological posteriors, and does not impact any of the conclusions in the discussion below.


\subsection{Model and Data Summary}
\label{sec:cliff_notes}

We provide a short, bullet-point summary of our data and model below. 
Our data can be summarized as follows:
\begin{itemize}
\item Our data vector is the DES Y1 \redmapper\ cluster counts and weak-lensing masses.
\item The covariance matrix of the cluster counts is due to Poisson noise, sample variance, and cluster miscentering.
\item The covariance matrix of the weak-lensing data is dominated by the impact of selection effects on the weak-lensing profile of the galaxy clusters.  The next most important contribution is source photometric redshift uncertainties.  The remaining uncertainties are cluster miscentering, lensed galaxy source dilution, and multiplicative shear biases.
\item We assume no covariance between cluster counts and weak-lensing masses.
\end{itemize}

Our model can be summarized as follows:
\begin{itemize}
\item Cluster counts are modeled as a convolution of the \cite{Tinker2008} mass function with a richness--mass relation.
\item We characterize and account for possible deviations from the Tinker mass function using a suite of numerical simulations.
\item The intrinsic richness of a galaxy cluster $P(\ltrue|M)$ is a convolution of Poisson noise with a Gaussian scatter of fixed relative width.   
\item The impact of projection effects and observational uncertainties is forward modeled in the counts \citep*{costanzietal18}. There are no nuisance parameters associated with this calibration in our likelihood model.
\item Based on numerical simulation estimates we do not assign a systematic error budget to the halo mass function due to baryonic feedback.  
\item Based on the fact that the concentration parameter is allowed to float independently in each richness/redshift bin used in the stacked weak lensing analysis, we do not assign a systematic error to the recovered weak-lensing masses due to baryonic effects.  
\item Systematic biases (and their uncertainties) due to the use of an analytic halo model for the halo--mass correlation function are calibrated using numerical simulations.
\end{itemize}

Appendix~\ref{app:validation} applies our methodology to a simulated data set in order to validate the cosmological pipeline.



\begin{figure*}
\begin{center}
    \includegraphics[width= \textwidth]{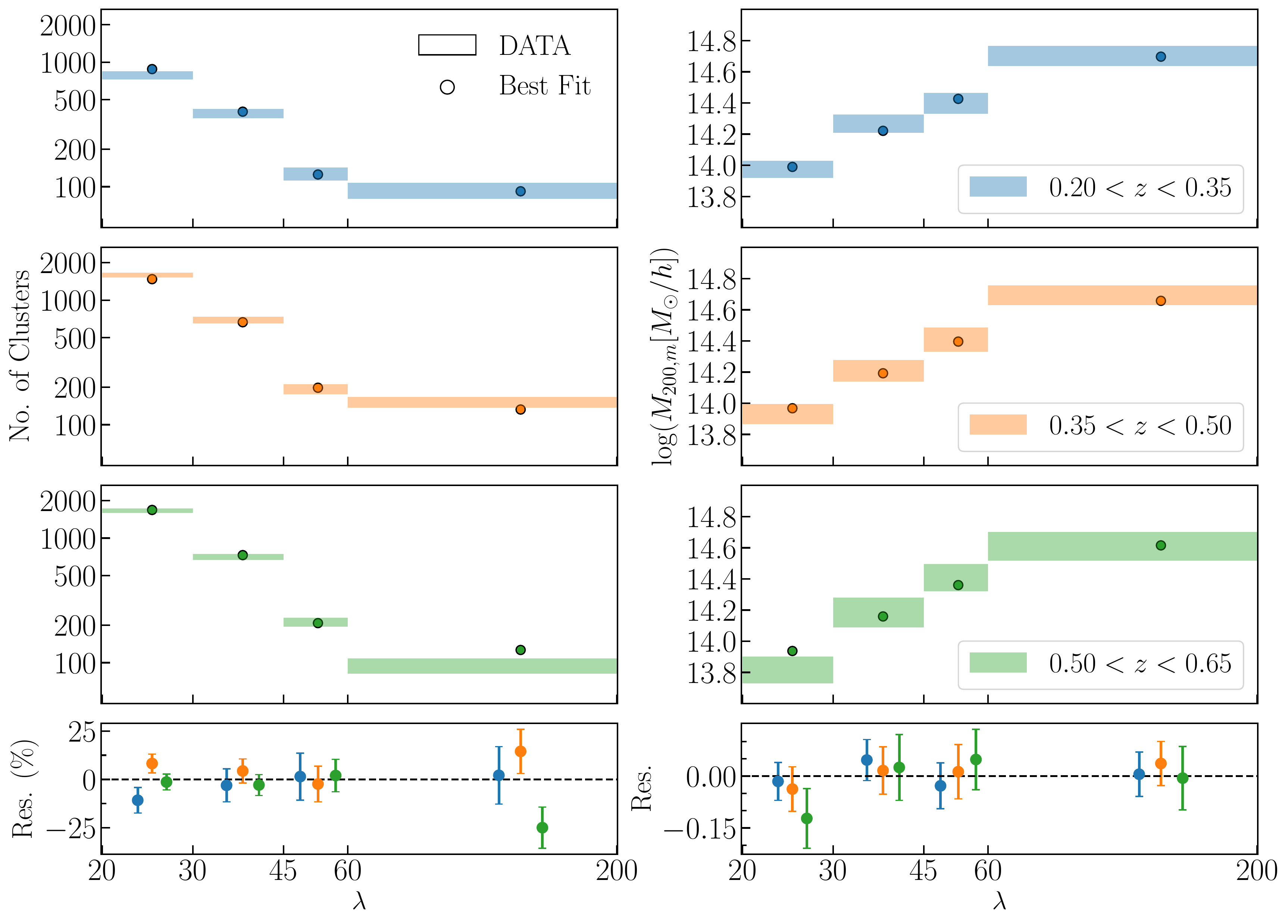}
\end{center}
\caption{Observed (\textit{shaded areas}) and best-fit model (\textit{dots}) for the cluster number counts (\textit{left}) and mean cluster masses (\textit{right}) as a function of richness for each of our three redshift bins. The \textit{y} extent of the data boxes is given by the square root of the diagonal terms of the covariance matrix. The bottom panel shows the residual between the data and our best-fit model.  All points have been slightly displaced along the richness axis to avoid overcrowding.}
\label{fig:datavec}
\end{figure*}


\section{Results}
\label{sec:res}

This analysis has been performed blind following the blinding and unblinding protocol outlined in Appendix \ref{sec:blinding}. After unblinding, a 2.3$\sigma$ and 6.7$\sigma$ tension in the $\sigma_8 - \Omega_m$ plane was found with DES 3x2pt \citep{des17} and {\it Planck} CMB data \citep{Planck2018}, as well as a larger than $3.5\sigma$ tension with BAO measurements \citep{BAO6dF,BAOSDSSMain,BAOBoss} and supernovae data \citep{SNpanth2017} (see appendix \ref{sec:blinded_res} for details). In the attempt to trace back the source of the tension, we found two clear but minor bugs, neither of which had a substantial impact on our posteriors.  We also discovered the impact of selection effects on weak lensing in simulations was significantly larger than originally expected (see Appendix \ref{sec:unblnd}), leading us to revise the estimate of the impact of selection effects on the cluster masses. Below, we present the results for the \textit{unblinded} analysis, which include the selection effects bias estimates from Appendix \ref{sec:unblnd}. These corrections increased the size of the error ellipse from DES Y1 clusters, but, as discussed below, significant tension with {\it Planck} and DES 3x2pt remains.  If not specified otherwise, we assume a flat $\Lambda$CDM cosmological model with three degenerate species of massive neutrinos ($\Lambda$CDM+$\sum m_\nu$). The parameter posteriors are estimated using the {\tt emcee} package \citep{emcee} which implement the affine-invariant Monte Carlo Markov Chain sampler of \cite{Goodman2010}.

\subsection{Goodness of Fit}
\label{sec:chi2}
Figure~\ref{fig:datavec} shows the abundance (left) and weak-lensing masses (right) of the DES Y1 \redmapper\ clusters as a function of the cluster richness for three separate redshift bins along with the corresponding best-fit model expectations. The measurements and associated uncertainties are shown as colored boxes, while the dots correspond to the best-fit model from our posteriors.  The bottom panel shows the residual between the data and our best-fit model for each of the three redshift bins under consideration, as labeled. For clarity, the points are slightly spread along the richness axis to avoid overcrowding. The $\chi^2$ of our best-fit model is $22.33$.

\begin{figure}
\begin{center}
    \includegraphics[width=0.45 \textwidth]{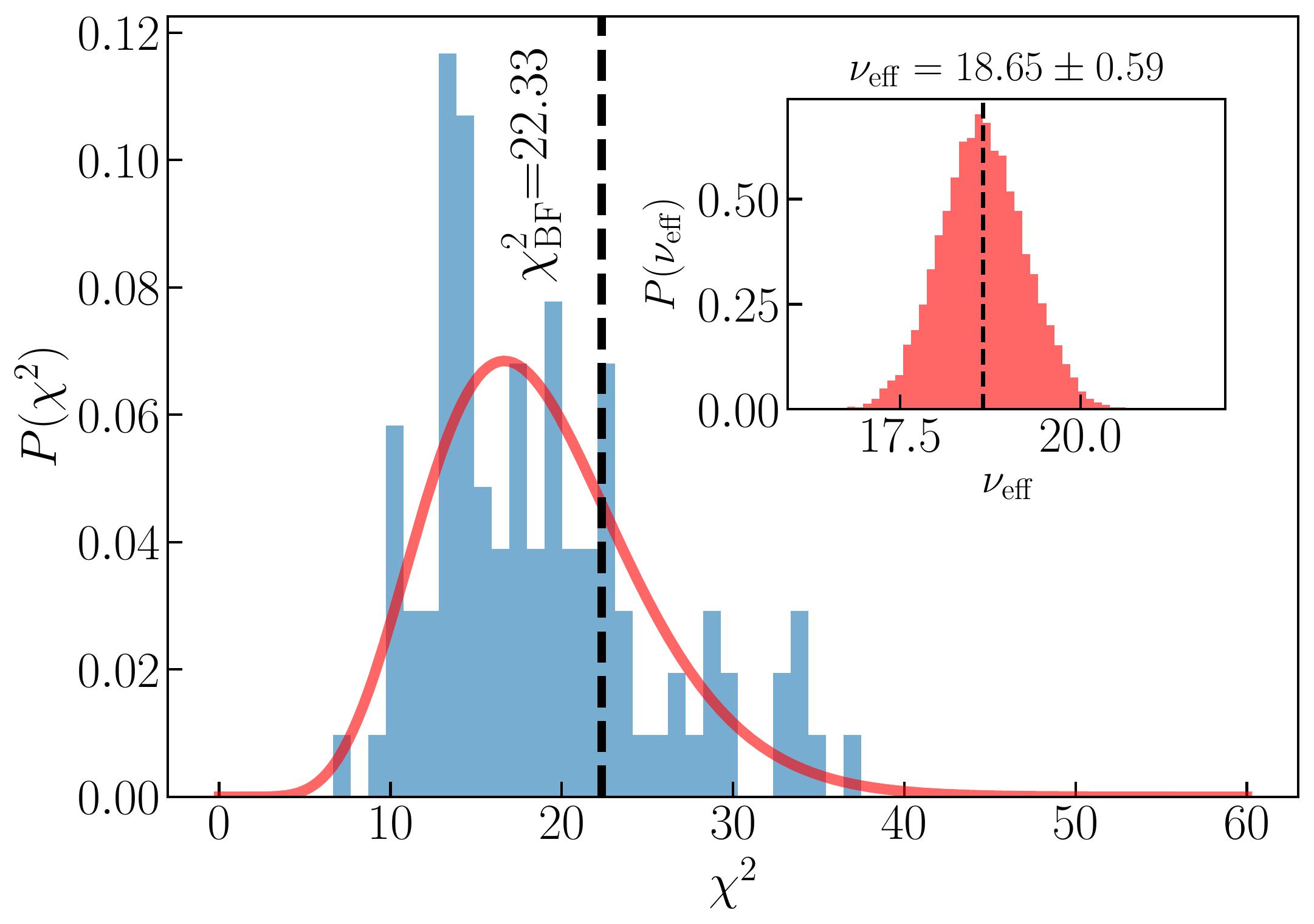}
\end{center}
\caption{Goodness-of-fit analysis. The \textit{blue} histogram shows the distribution of the best-fit $\chi^2$ values recovered from $100$ mock data realizations generated from the best-fit model of the data. The \textit{red} histogram in the inset plot shows the posterior distribution for the effective number of degrees of freedom obtained by fitting a $\chi^2$ distribution to the above $100$ $\chi^2$ values. The \textit{red solid} line represents the $\chi^2$ distribution for the best-fit model ($\nu_{\rm eff}=18.65$), while the vertical \textit{dashed} line corresponds to the $\chi^2$ value of the data.}
\label{fig:chisq}
\end{figure}

We assess the goodness of fit by generating $100$ realizations of our best-fit model data vectors adopting our best-fit covariance matrix, and fitting each in turn in order to arrive at the distribution of best-fit $\chi^2$ values of our mock-realizations.  The distribution is fit using a $\chi^2$ distribution, for which we find that the effective number of degrees of freedom is $\nu_{\rm eff}=18.65 \pm 0.60$.  The distribution of $\chi^2$ values in our simulated data, as well as the $\chi^2$ value in the real data, is shown in Figure~\ref{fig:chisq}. As evident from the figure, our model is a good fit to the data, with a probability to exceed of $0.25$.


\begin{table*}
    \centering
    \footnotesize
    \caption{Model parameters and parameter constraints from the joint analysis of \redmapper\ DES Y1 cluster abundance and weak-lensing mass estimates. In the third column we report our model priors: a range indicates a top-hat prior, while $\mathcal{N}(\mu,\sigma)$ stands for a Gaussian prior with mean $\mu$ and variance $\sigma^2$. The fourth column lists the modes of the $1$-d marginalized posterior along with the $1$-$\sigma$ errors. Parameters without a quoted value are those for which the marginalized posterior distribution is the same as their prior.}
    \label{tab:parameters}
   \begin{tabular}{lcccc}
    \hline 
	Parameter			&	Description	& Prior &	Posterior	\\
    \hline \vspace{-3mm}\\
	$\Omega_m$		& Mean matter density 						& $[0.0,1.0]$ &$ 0.179^{+0.031}_{-0.038}$ 		\vspace{0.5mm} \\
        $\ln(10^{10} A_s)$		& Amplitude of the primordial curvature perturbations	& $[-3.0,7.0]$ &$ 4.21 \pm 0.51$ 		\vspace{0.5mm} \\
    $\sigma_8$		& Amplitude of the matter power spectrum	& $-$ &$ 0.85^{+0.04}_{-0.06}$ 		\vspace{0.5mm} \\
    $S_8=\sigma_8 (\Omega_m/0.3)^{0.5}$		& Cluster normalization condition	& $-$ &$ 0.65^{+0.04}_{-0.04}$ 		\vspace{0.5mm} \\
    $\log M_{min} [{\rm M}_\odot /h]$	& Minimum halo mass to form a central galaxy& $(10.0,14.0)$ &$ 11.13\pm 0.18$	\vspace{0.5mm} \\
    $\log M_1 [{\rm M}_\odot /h]$		& Characteristic halo mass to acquire one satellite galaxy &$\log(M_1/M_{\rm{min}}) \in [\log(10),\log(30)]$ &$ 12.37 \pm 0.11 $\vspace{0.5mm} \\
    $\alpha$		& Power-law index of the richness--mass relation 				& $[0.4,1.2]$ &$ 0.748 \pm 0.045$ 		\vspace{0.5mm} \\
    $\epsilon$		& Power-law index of the redshift evolution of the richness--mass relation &$ [-5.0,5.0] $&$  -0.07 \pm 0.28$ \vspace{0.5mm} \\
    $\sigma_{intr}$	& Intrinsic scatter of the richness--mass relation				& $[0.1,0.5]$ &$ <0.325$		\vspace{0.5mm} \\
    $s$				& Slope correction to the halo mass function& $\mathcal{N}(0.047,0.021)$ &$ -$	\vspace{0.5mm} \\
    $q$				& Amplitude correction to the halo mass function& $\mathcal{N}(1.027,0.035)$ &$ -$	\vspace{2.0mm} \\
   $h$ & Hubble rate & $\mathcal{N}(0.7,0.1)$   & $0.744 \pm 0.075$ \vspace{0.5mm}  \\
   $\Omega_b h^2$ & Baryon density & $\mathcal{N}(0.02208, 0.00052)$ & $-$ \vspace{0.5mm}  \\
   $\Omega_\nu h^2$ & Energy density in massive neutrinos & $[0.0006,0.01]$ & $-$  \vspace{0.5mm} \\
   $n_s$ & Spectral index & $[0.87, 1.07]$ & $-$ \vspace{0.5mm}  \\
   \hline \vspace{-3mm}\\   
    \end{tabular}
\end{table*}

\subsection{Cosmological Constraints from DES Y1 Cluster Data}
\label{sec:cosmo_result}

\begin{figure*}
\begin{center}
    \includegraphics[width=\textwidth]{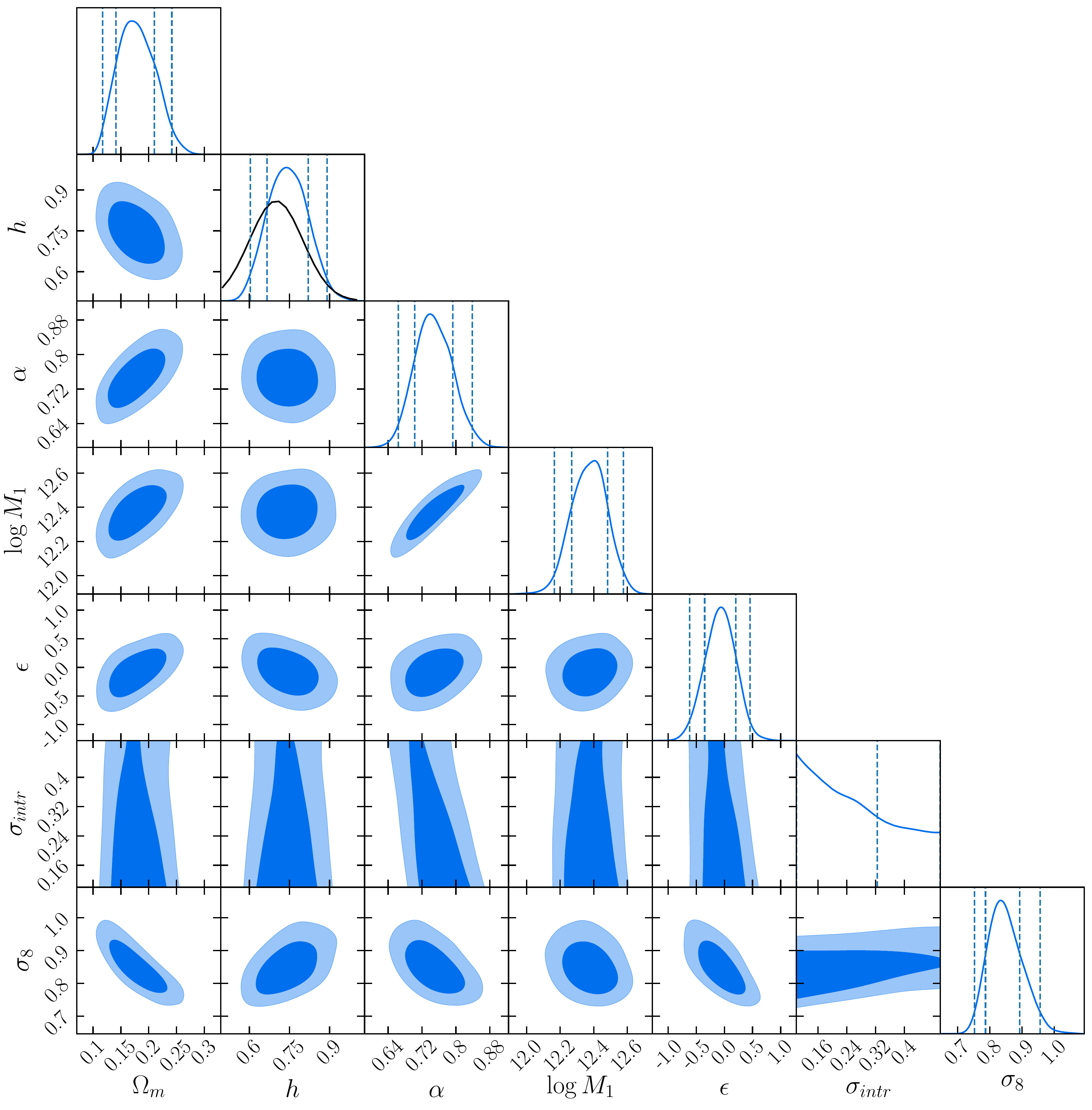}
\end{center}
\caption{Marginalized posterior distributions of the fitted parameters. The $2D$ contours correspond to the $68\%$ and $95\%$ confidence levels of the marginalized posterior distribution. The dashed lines on the diagonal plots correspond respectively to the $2.5$th, $16$th, $84$th and $97.5$th percentile of the $1$-d posterior distributions. The black line in the 1-d posterior plot of $h$ corresponds to the Gaussian prior adopted in the analysis. The description of the model parameters along with their posteriors are listed in Table \ref{tab:parameters}. Only parameters that are not prior dominated are shown in the plot.}
\label{fig:des_constraints}
\end{figure*}

Figure~\ref{fig:des_constraints} shows the posteriors of the parameters used to model the DES Y1 cluster cosmology data set.  The parameter $\Mmin$ is not shown because it is prior dominated.   All of our parameters, along with their corresponding priors and posteriors, are summarized in Table~\ref{tab:parameters}.

%
%

The only two cosmological parameters that are not prior dominated in our analysis are $\sigma_8$ and $\Omegam$.  Our posteriors for each of these are $\sigma_8 =0.85^{+0.04}_{-0.06}$ and $\Omegam = 0.179^{+0.031}_{-0.038}$  The corresponding cluster normalization condition is $S_8=\sigma_8(\Omegam/0.3)^{0.5} = 0.650 \pm 0.037$.

In addition, the posterior for the Hubble parameter $h=0.744 \pm 0.075$ is slightly improved relative to our prior, $h=0.7 \pm 0.1$. This improvement arises due to the mild sensitivity of number counts and mean cluster masses to $h$: a shift of $h$ tilts the slope of the number counts around the pivot point $\lambda \simeq 55$ while changing the amplitude of the mean mass--richness relation.  Despite the modest degeneracy of $h$ with $\Omegam$ and $\sigma_8$, we verified that adopting a flat prior on $h \in [0.55,0.90]$ (as in \citealt{des17}) does not affect the cosmological posteriors of $\Omegam$ and $\sigma_8$.

\begin{figure}
\begin{center}
    \includegraphics[width=0.475 \textwidth]{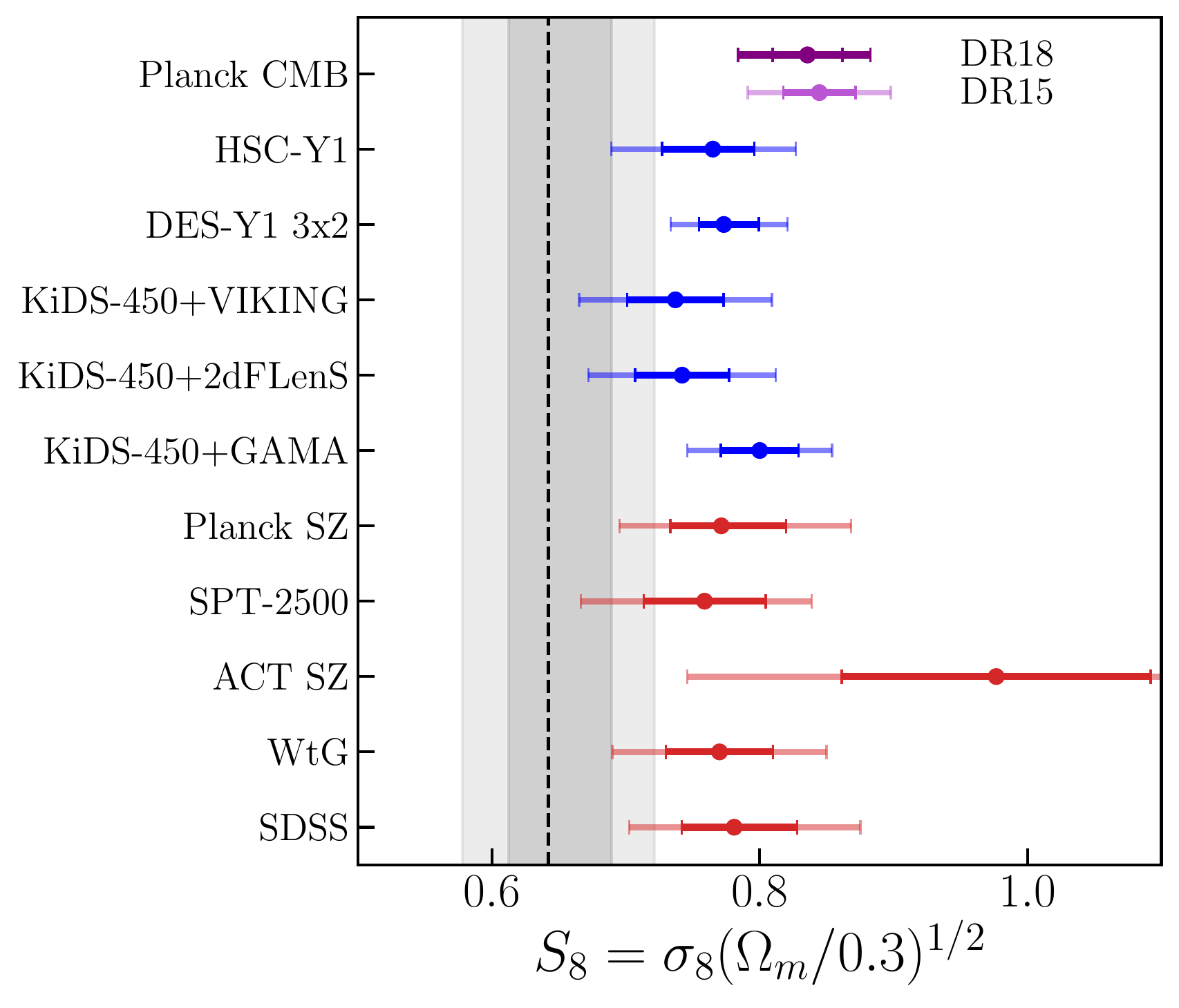}
\end{center}
\caption{Comparison of the $68\%$ (\textit{dark}) and $95\%$ (\textit{light}) confidence level constraints on $S_8$ derived from our baseline model (\textit{shaded gray} area) with other constraints from the literature: \textit{red} error bars for cluster abundance analyses, \textit{blue} error bars for weak lensing and galaxy clustering analyses and \textit{purple} for the CMB constraint. From the bottom to the top: SDSS from \cite{costanzietal18b}; WtG from \cite{Mantz2015}; ACT SZ from \cite{ACT2013} (BBN+H0+ACTcl(dyn) in the paper); SPT-2500 from \citep{Bocquet2018}; Planck SZ from \cite{PlanckSZ2016} (CCCP+$H_0$+BBN in the paper); KiDS-450+GAMA from \cite{kidsgama2018}; KiDS-450+2dFLens from \cite{kids2df2018}; KiDS-450+VIKING from \cite{kidsviking}; DES-Y1 3x2 from \cite{des17}; HST-Y1 from \cite{HST2018}; \Planck\ CMB from \cite{PlanckXIII2016} (DR15) and \cite{Planck2018} (DR18). Note that all the constraints but those from SDSS, DES-Y1 3x2, HSC-Y1 and Planck CMB have been derived fixing the total neutrino mass either to zero or to $0.06$ eV.}
\label{fig:S8}
\end{figure}

We compare our posterior on the parameter $S_8=\sigma_8(\Omegam/0.3)^{0.5}$ to that derived from a variety of different weak lensing and cluster abundance experiments in Figure~\ref{fig:S8}. This figure also compares our posterior in $S_8$ to that of \planck\ 2016 and \planck\ 2018. Our posterior is clearly lower than all other constraints shown, with the tension in $S_8$ relative to other low-redshift probes typically ranging from $1.5\sigma$ to $2.5\sigma$.  Notably, one of the largest tensions is with respect to the DES Y1 3x2pt analysis, at $2.9\sigma$.  We note that these tensions in $S_8$ were only slightly impacted by the post-unblinding corrections we adopted.  If we naively combine all nine low-redshift experiments assuming they are mutually independent, the DES Y1 cluster result has a $2\%$ probability of being a statistical fluctuation around their mean.  The difference becomes even stronger when considering Planck CMB results, for which the significance of the tension with $S_8$ reaches $4.0\sigma$.


\begin{figure}
\begin{center}
    \includegraphics[width=0.45 \textwidth]{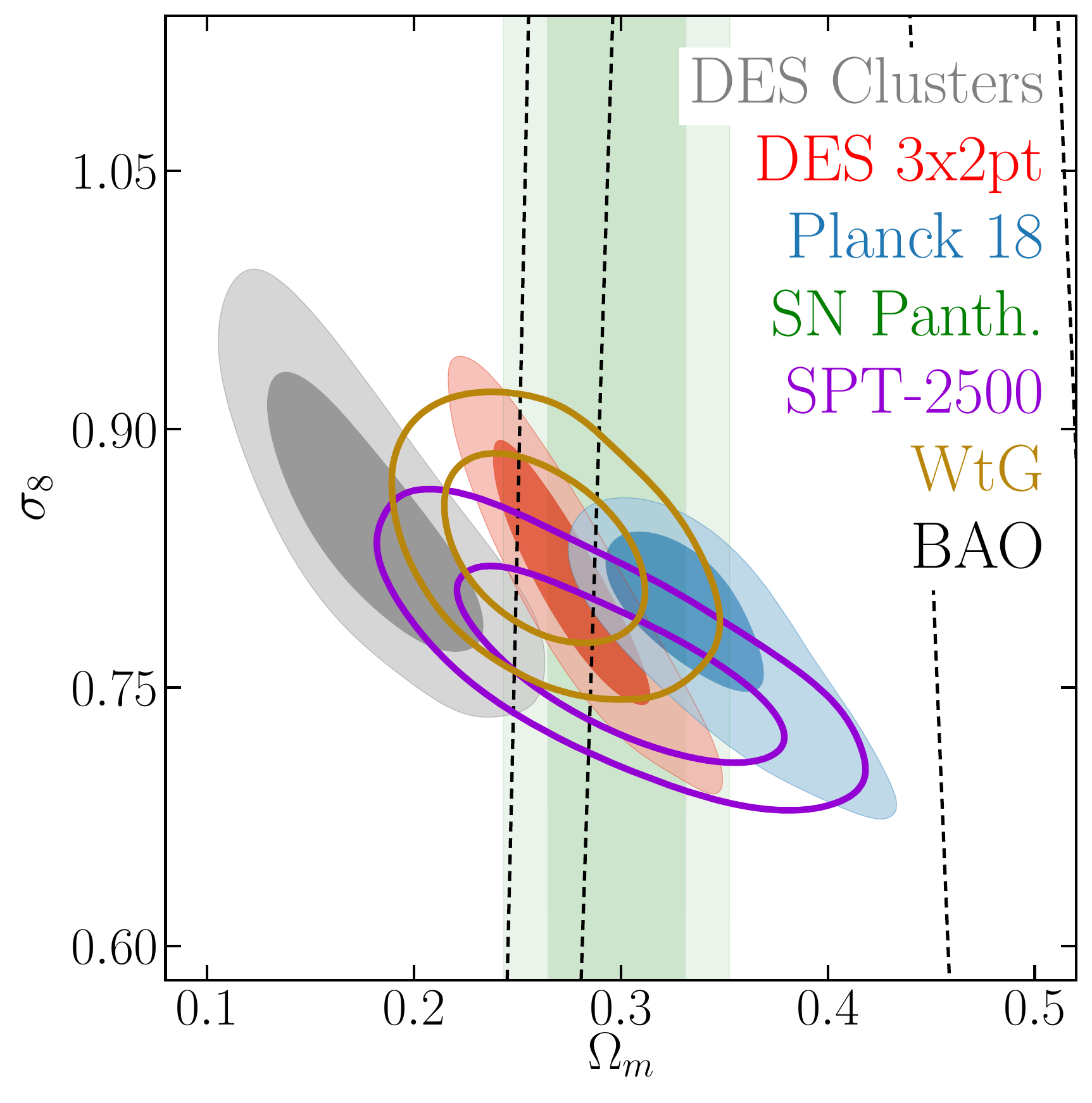}
\end{center}
\caption{Comparison of the $68\%$ and $95\%$ confidence contours in the $\sigma_8$-$\Omega_m$ plane derived from DES Y1 cluster counts and weak-lensing mass calibration (\textit{gray} contours) with other constraints from the literature: BAO from the combination of data from Six Degree Field Galaxy Survey \citep[6dF][]{BAO6dF}, 
the SDSS DR 7 Main galaxy sample \citep{BAOSDSSMain}, and the Baryon
Oscillation Spectroscopic Survey \citep[BOSS][]{BAOBoss} (\textit{black dashed} lines); Supernovae Pantheon \citep{SNpanth2017} (\textit{green} contours); DES-Y1 3x2 from \cite{des17} (\textit{red} contours); \Planck\ CMB from \cite{Planck2018} (\textit{blue} contours); SPT-2500 from \citep{Bocquet2018} (\textit{violet} contours); WtG from \citep{Mantz2015} (\textit{gold} contours). }
\label{fig:s8om}
\end{figure}


Figure~\ref{fig:s8om} compares the $68\%$ and $95\%$ confidence regions in the $\sigma_8-\Omegam$ plane derived from DES Y1 Cluster data to the DES 3x2pt statistics \citep{des17}, the Planck CMB DR18 \citep{Planck2018}, a combination of BAO measurements \citep{BAO6dF,BAOSDSSMain,BAOBoss}, Supernovae Pantheon data \citep{SNpanth2017}, and cluster counts analyses from  \cite{Mantz2015} and \cite{Bocquet2018} (respectively WtG and SPT-2500 in the figure).  As is evident from the figure, the $S_8$ tension is due to the low $\Omegam$ value preferred by the DES Y1 cluster data set. Specifically, looking at the $\Omegam$ sub-space, our cluster posterior displays a $1.7\sigma$ tension with SPT-2500, $1.8\sigma$ tension with WtG, a $2.2\sigma$ tension with DES Y1 3x2pt, a  $3.0\sigma$ tension with SN data, a $3.3\sigma$ tension with BAO, and a $4.7 \sigma$ tension with Planck CMB. The corresponding tensions in the $\sigma_8-\Omegam$ plane are $1.1\sigma$ (SPT-2500), $1.7\sigma$ (WtG), $2.4\sigma$ (DES 3x2pt) and $5.6\sigma$ (\Planck).\footnote{Here consistency between two data sets $A$ and $B$ is established by testing whether the hypothesis $\bm{p}_A - \bm{p}_B = 0$ is acceptable  \citep[see method `3' in][]{charnocketal17},  where $\bm{p}_A$ and $\bm{p}_B$ are the model parameters of interest as constrained by data sets $A$ and $B$, respectively.}
The fact that all other cosmological probes, including those using the same DES data employed in this work, return significantly higher values for the matter density than ours suggests the presence of unexpected systematics or physics in our analysis. We will comment on the possible origin of this tension in Section \ref{sec:disc}. Due to the inconsistencies between the DES Y1 cluster data and internal and external probes we do not perform any joint analysis of cluster data with other data sets.

One intriguing possibility to consider is whether the tensions seen in Figure~\ref{fig:s8om} could be reduced within the context of a different cosmological model.  We have run chains assuming a $w$CDM+$\sum m_\nu$ model with a flat prior $w\in[-2,-1/3]$ for the equation of state of the dark energy.  We find that these models do not improve the agreement between DES clusters and the remaining data sets.

\subsection{Robustness Tests}
\label{sec:robustness}

\begin{figure*}
\begin{center}
    \includegraphics[width=0.65 \textwidth]{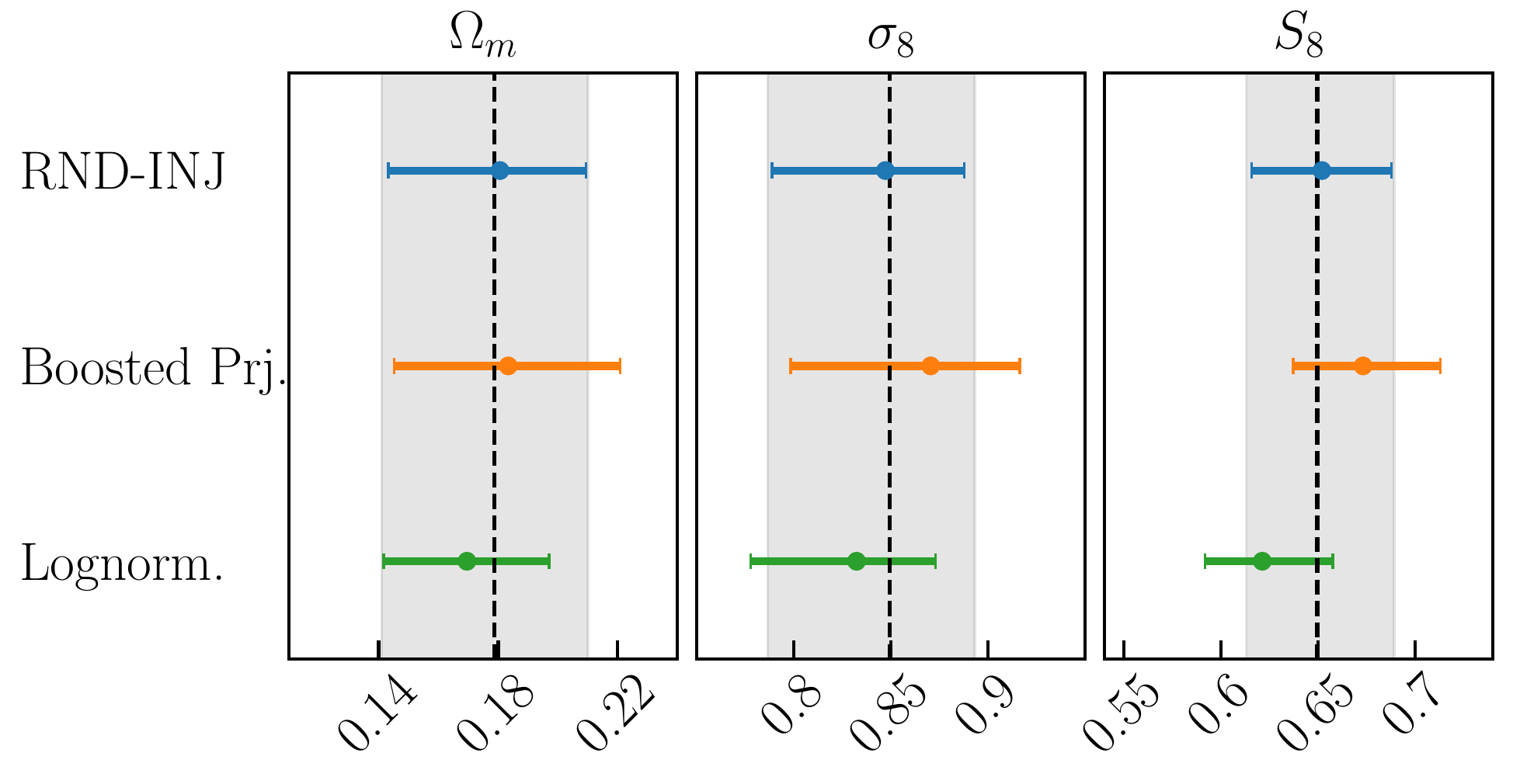}
    \includegraphics[width=0.30 \textwidth]{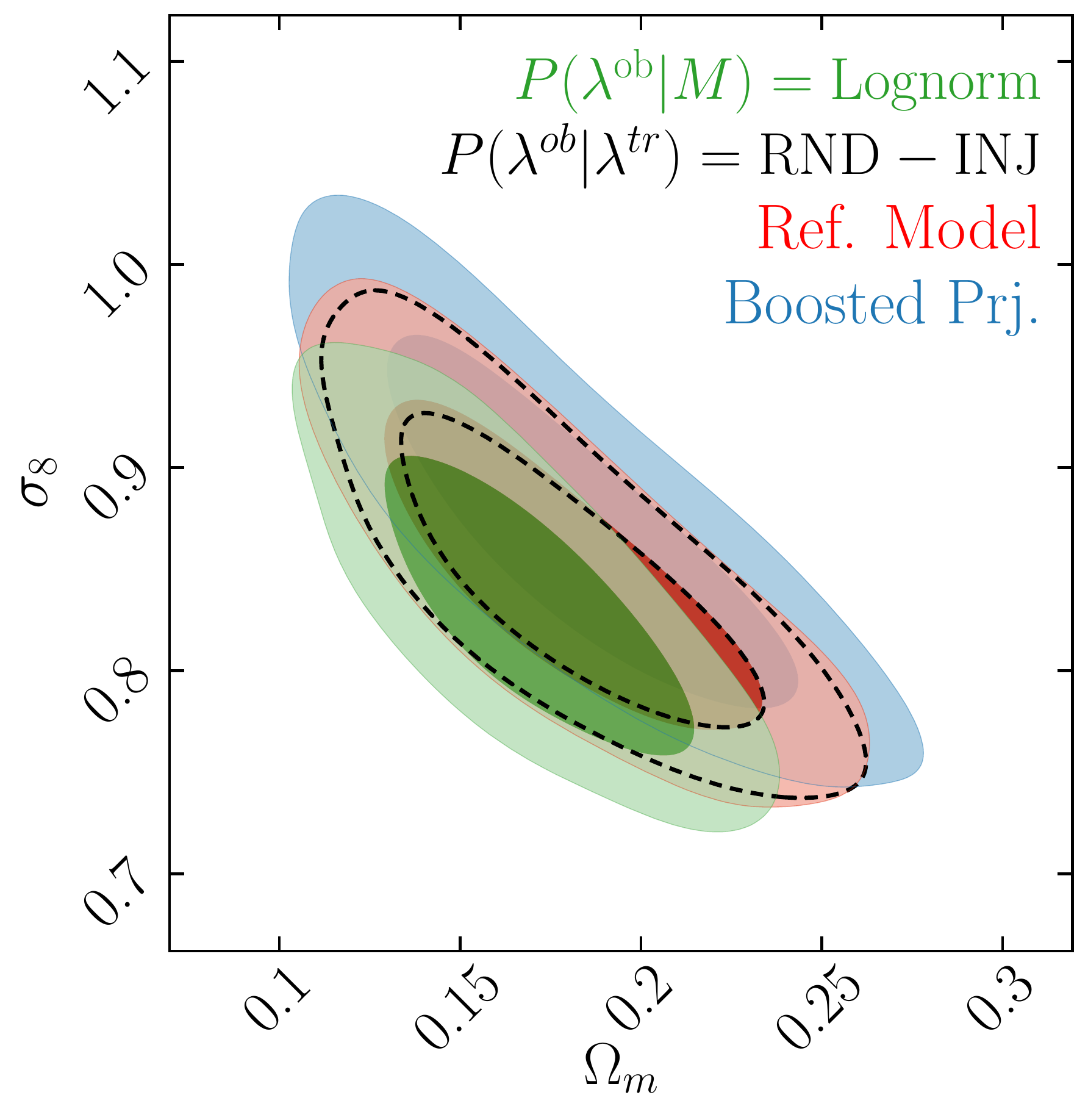}
\end{center}
\caption{\textit{Left} panel: Comparison of the $68\%$ confidence regions for $\Omegam$, $\sigma_8$ and $S_8$ derived from different model assumptions . The shaded area corresponds to the constraints derived using our reference model, while the dashed vertical line is the mean of our fiducial posterior. The model assumptions we consider are, from top to bottom, the random injection, boosted projection, and lognormal models described in Section \ref{sec:res:RMR}. \textit{Right}: Same as \textit{left} panel in the $\Omegam$-$\sigma_8$ plane.
}
\label{fig:robustness}
\end{figure*}

Of special interest to us is the robustness of our cosmological posteriors to our choice of theoretical model. To test for robustness we consider three different modifications to our fiducial model for the richness--mass relation, which in turn affect the expectation values for the number counts and mean cluster masses. 
These are: 
\begin{enumerate}
\item A random-point injection model, in which projection effects are estimated assuming clusters are randomly located throughout the sky.  This provides a firm lower limit on projection effects.  We consider this an extreme model (i.e. we know clusters live in highly clustered regions of the Universe).
\item A model with boosted projection effects, in which $P(\lambda\ob |\lambda\true)$ is calibrated doubling the magnitude of projection effects relative to our fiducial model.  We expect this model provides an upper limit on the effect that an underestimation of projection effects could have on cosmological posteriors.
\item A model in which $P(\lambda\ob |M)$ is a log-normal, the mean richness--mass relation is a power law and the intrinsic scatter is mass dependent; note that in this case we do not include our model for $P(\lambda\ob|\lambda\true)$, and all the scatter due to observational noise and projection effects is absorbed by the $\sint$ parameter.
\end{enumerate}
As detailed in appendix \ref{sec:blinding}, these models were selected and tested before unblinding. We thus repeated these tests for the \textit{unblinded} analysis finding consistent effects on the parameter posteriors to those obtained in the \textit{blinded} analysis.
Figure~\ref{fig:robustness} shows how our cosmological posteriors of the \textit{unblinded} analysis change for each of these different model assumptions.
As noted above, we consider model (i) to be extreme and (ii) to provide a conservative upper limit on the amplitude of projection effects, and use them to define a $2\sigma$ systematic error in our cosmological parameters associated with the projection-effect calibration.
That is, we estimate the systematic uncertainty in our cosmological posteriors as half the difference between the recovered parameters in these models and our fiducial model.  These systematic errors are negligible compared to our posteriors, and will therefore be ignored from this point on.

Similarly, the central values of our cosmological posteriors when using model (iii) are within the one-sigma posterior of our reference model.  We include this model here for comparison purposes, since previous analyses have relied on power-law log-normal models \citep[e.g.][]{Rozo2010,Murata2017}. 

Appendix~\ref{sec:unblnd} details further tests of the parameterization of the richness--mass relation performed after unblinding.  The summary of those results is consistent with our conclusions above: the adopted form of the richness--mass relation does not have a large impact on the cosmological posteriors derived from our analyses.


\subsection{Constraints on the Richness--Mass Relation}
\label{sec:res:RMR}

\begin{figure*}
\begin{center}
    \includegraphics[width=\textwidth]{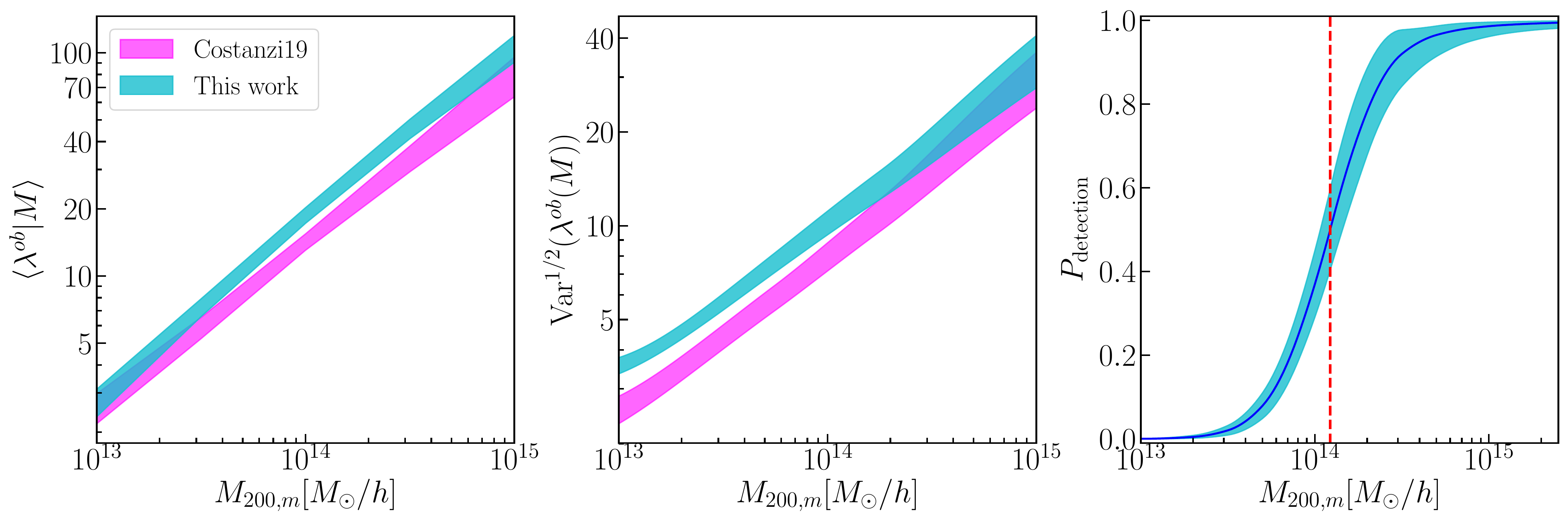}
\end{center}
\caption{Observable--mass relation and mass-selection function of the redMaPPer DES Y1 catalog assuming our reference richness--mass relation model (Eq. \ref{eqn:RMR}) at the mean sample redshift $z=0.45$. {\it Left} panel: Expectation value for the observed richness as a function of mass. {\it Central} panel: Scatter of $\lambda\ob$ -- $\Var^{1/2}(\lambda\ob|M)$ -- as a function of mass. 
{\it Right} panel: Detection probability as a function of cluster mass. The \textit{dashed} vertical line corresponds to the mass at which the detection probability is $50\%$  ($\log M_{50\%}= 14.09 [M_{\odot}/h])$. 
The \textit{blue} area corresponds to the $68\%$ confidence interval derived for the different quantities in this work. For comparison, the results of \cite{costanzietal18b} (\textit{magenta}) for the \redmapper\ SDSS catalog are shown in the two left panels. All results are corrected for the systematic richness offset between the SDSS and DES catalogs, and account for the expected redshift evolution between $z=0.22$ (SDSS) and $z=0.45$ (DES) as determined by the best-fit model to the DES data.}
\label{fig:lob_m_rel}
\end{figure*}

Figure~\ref{fig:lob_m_rel} shows the posterior of the richness--mass relation of the DES Y1 \redmapper\ galaxy clusters.  The left panel shows the expectation value of the richness--mass relation, $\avg{\lambda\ob|M}$ at the mean sample redshift $z=0.45$.  The central panel shows the variance in richness at fixed mass, $\Var(\lambda\ob|M)$, again at the mean sample redshift.  It is important to emphasize that the shape of the variance as a function of mass is intrinsic to our fiducial model: while we have a single scatter parameter $\sint$, which is mass independent, our model for both the intrinsic richness--mass relation and projection effects results in a mass-dependent variance. Finally, the right panel of Figure~\ref{fig:lob_m_rel} shows the probability that \redmapper\ will detect a halo of mass $M$ as a cluster with more than 20 galaxies.  The mass at which the detection probability is $50\%$ is $M_{200m}=1.2\times 10^{14}\ \hMsun$.

Figure~\ref{fig:lob_m_rel} also compares our posteriors to those of our analysis of the SDSS \redmapper\ cluster sample \citep{costanzietal18b}.  For the purposes of this comparison, we cross match low-redshift DES clusters with SDSS clusters, and correct the SDSS richnesses for the systematic richness offset of 0.93 between SDSS and DES \citepalias[Eq. 67 in][]{desy1wl}.  Further, we correct our SDSS result for the expected redshift evolution from $z=0.23$---the mean redshift of the SDSS \redmapper\ clusters---to our chosen pivot point of $z=0.45$ using the best-fit value for the evolution parameter $\epsilon$ from the DES chain. 
While the slopes of the richness--mass relations are in agreement between the two analyses, the DES data prefers a larger value for the amplitude. This difference is explained by the selection effect bias correction applied to the weak-lensing mass estimates (see Appendix~\ref{sec:unblnd}): while the mass estimates in \cite{desy1wl} were consistent with those of SDSS \redmapper\ clusters \citep{Simet2016}, our selection effect correction lowered the DES Y1 masses by $\sim 20\%$ relative to our analysis in \cite{desy1wl}.  By the same token, the variance as a function of mass is similar between the two analyses, but shifted to lower masses in this work because of the selection effects correction.  We note, however, that the selection effects characterized in this work should also impact the SDSS constraints. That is, we expect the SDSS richness--mass relation shown above to be biased low by $\approx 15\%$.

\begin{figure}
\begin{center}
    \includegraphics[width=0.45 \textwidth]{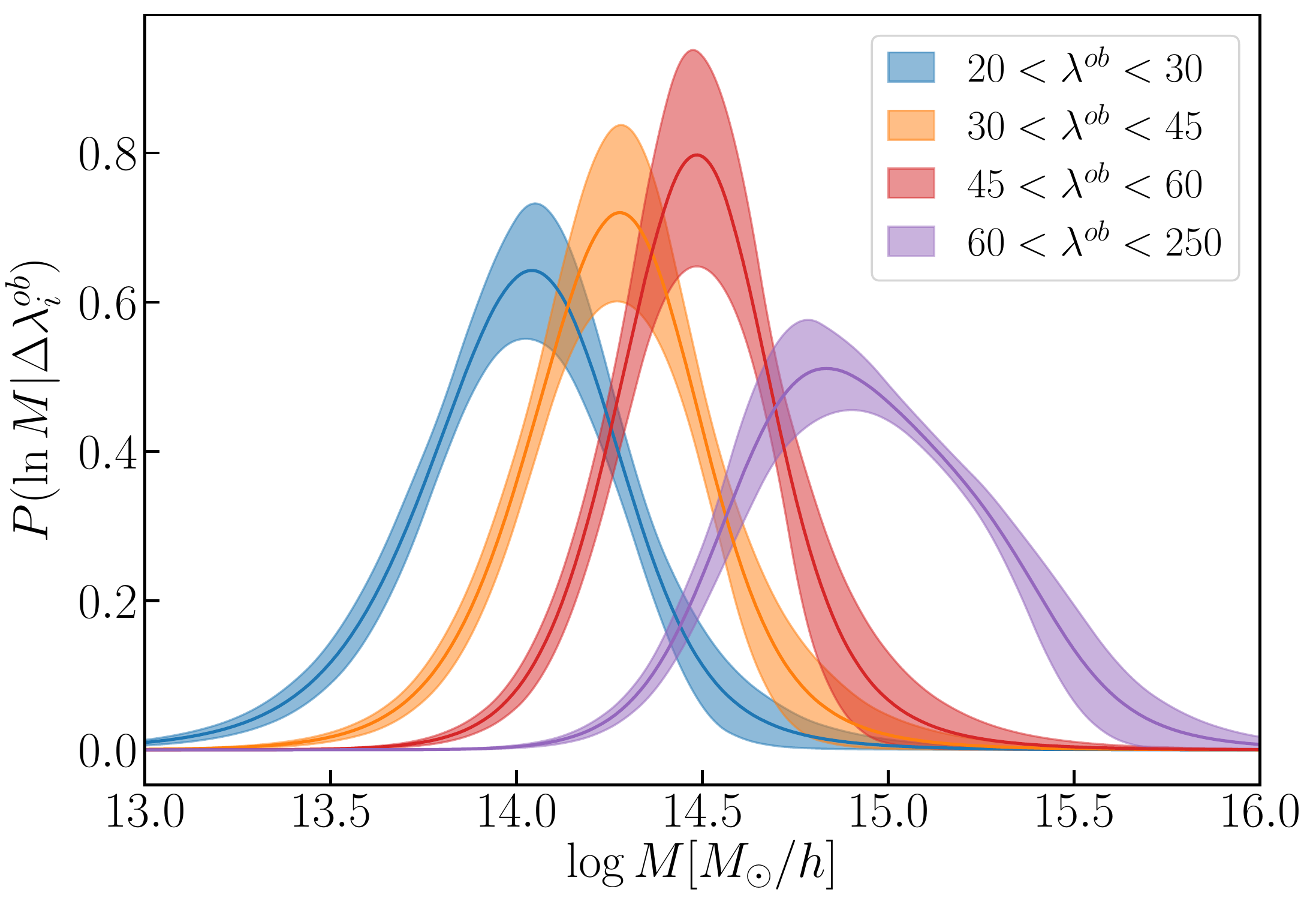}
\end{center}
\caption{Distribution of halo mass for clusters in each of the four richness bins employed in this work at median redshift $z=0.45$, as labeled.  The width of the bands correspond to the 68\% confidence interval of the distribution as sampled from our posterior.}
\label{fig:mdist}
\end{figure}

Figure~\ref{fig:mdist} shows the mass distribution for each of our four richness bins at a redshift $z=0.45$, as constrained through our posteriors.  Integrating over these distributions, we can recover the mean mass of the \redmapper\ galaxy clusters of a given richness.
This mean mass is shown with a \textit{blue} band in Figure~\ref{fig:m_rel_comp}. 
From the combination of DES Y1 cluster counts and weak-lensing mass estimates we constrain the mean mass at the pivot richness $ \lambda\ob=40$ to $\log \langle  M| \lambda\ob \rangle = 14.252 \pm 0.026$. As before, the selection effect bias correction applied in this work lowered our masses by $\sim20\%$, leading to a mismatch between our results and that presented in \cite{desy1wl}: $\log (M_0 [\Msun/h]) = 14.334 \pm 0.022$. Remarkably, the $\approx 6\%$ precision in the posterior masses is similar to the uncertainty quoted in \cite{desy1wl}, despite the large systematic uncertainty we have added to the weak lensing masses.   This demonstrates that the inclusion of cluster count data offsets the factor of $\sim 2$ larger uncertainty in mass  due to the uncertain calibration of selection effects in our final results. However, the calibration of the scaling relation through number counts data is made at the expense of more relaxed cosmological constraints.  For the same reason, this posterior would likely relax in extended cosmological models such as $w$CDM.

\section{Discussion}
\label{sec:disc}

\subsection{What Drives the Tension Between DES Clusters and Other Probes?}

The internal consistency of the other DES probes,
along with their consistency with external cosmological probes, rule out the possibility that the tension observed with the DES Y1 clusters data is driven by observational systematics affecting the DES data (e.g. photometry or shear calibration). Thus,
the tension between our results and other cosmological probes provides strong evidence that at least one aspect of our theoretical model is incorrect: either the cosmological model assumed is wrong (\LCDM+$\sum m_\nu$ and $w$CDM+$\sum m_\nu$), our interpretation of the stacked weak lensing signal as mean cluster mass is incorrect, or our understanding of the richness--mass relation and/or selection function is flawed. The interpretation of our results as evidence for the first is unlikely: it would require our analysis to be correct, while all other cosmological experiments
would need to have large, as of yet undiscovered systematics. 
Turning to our understanding of the richness--mass relation, we have verified (section~\ref{sec:robustness}) that our cosmological conclusions are robust to the form of the richness--mass relation adopted within the uncertainty suggested by numerical simulations and data. As discussed below, while additional observational tests will be critical to further validate it, currently available multi-wavelength data already disfavour the possibility that an unmodeled systematic in $P(\lambda\ob|M)$ could fully account for the bias in our cosmological posteriors.
Given the surprisingly large impact of selection effects in simulations, and that these effects have only been calibrated with one set of simulations, it appears likely that it is our understanding of selection effects on the weak-lensing signal where our model fails. 

\begin{figure}
\begin{center}
    \includegraphics[width= 0.42 \textwidth]{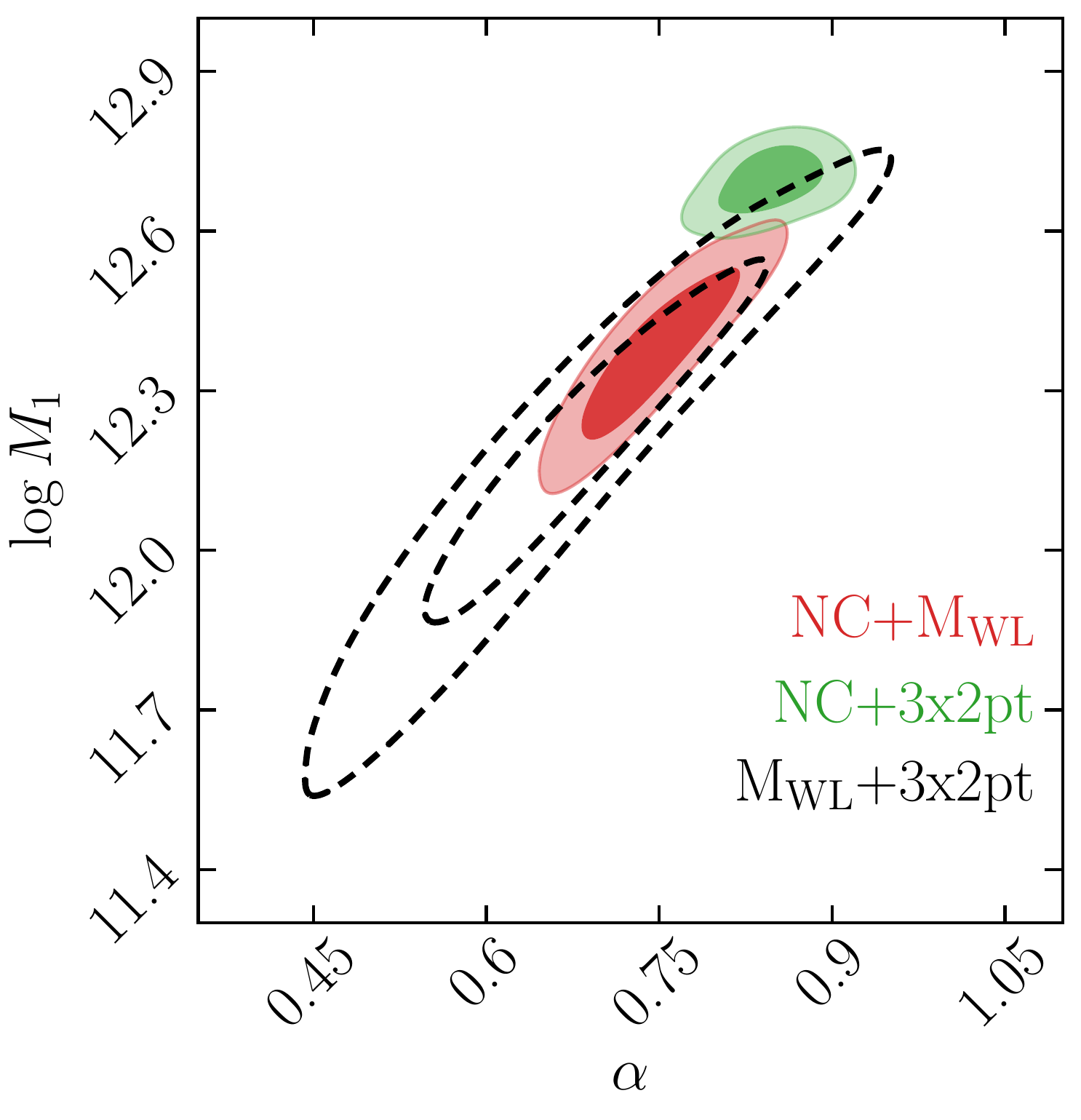}
\end{center}
\caption{$68\%$ and $95\%$ confidence contours from the combination of cluster counts data and DES Y1 3x2pt cosmology (\textit{green}) and the combination of weak-lensing mass estimates and DES Y1 3x2pt cosmology (\textit{black}). For comparison also shown in \textit{red} our reference model results from the combination of cluster counts and weak lensing data.
Not shown in the plot are the $\sint$ posteriors since the parameter is not constrained without the inclusion of number counts data (${\rm M}_{\rm  WL}$+3x2pt), whereas we recover the reference model posterior in the NC+3x2pt case.
}
\label{fig:3x2cosmo}
\end{figure}

To study possible unmodeled systematics in our data, we separately reanalyze either the number counts or the weak-lensing mass data, adopting as priors the cosmological posteriors derived from the DES 3x2pt analysis \citep[][]{des17}. By doing so, we can compare the posteriors of the richness--mass relation derived using each of our two types of cluster observables independently.  The result of this exercise is shown in Figure \ref{fig:3x2cosmo}. \textit{Green} contours are derived from the combination of number counts data and DES Y1 3x2pt priors, while the \textit{black dashed} contours combine the Y1 3x2pt priors with the cluster mass data only.  Also shown in \textit{red} for comparison are our reference model posteriors obtained from the combined analysis of number counts and weak lensing data.

As expected, in both cases the cosmological posteriors are dominated by the DES Y1 3x2pt priors, while the richness--mass relation parameters are constrained by either the cluster counts or the weak-lensing mass data alone. It is clear from Figure~\ref{fig:3x2cosmo} that the posteriors for the richness--mass relation derived from either of the cluster observables assuming a DES 3x2pt cosmology are only marginally consistent with one another. In particular, the abundance data prefer a steeper slope and a larger normalization for the richness--mass relation compared to the weak lensing data. This is not unexpected: had they been consistent, we would have expected the DES 3x2pt cosmology to be contained within our joint cosmological posterior.  The marginal consistency reflects the fact that our posteriors are only marginally consistent ($2.4\sigma$) with the DES 3x2pt cosmology constraints.
Interestingly, \cite{murata2019} found a similar trend between the slope preferred by either weak lensing data or cluster abundance when analysed separately for the first-year HSC data set in a Planck cosmology. However, a direct comparison with our results is not feasible due to the different richness definition and richness--mass relation adopted in their work.


\begin{figure*}
\begin{center}
    \includegraphics[width= \textwidth]{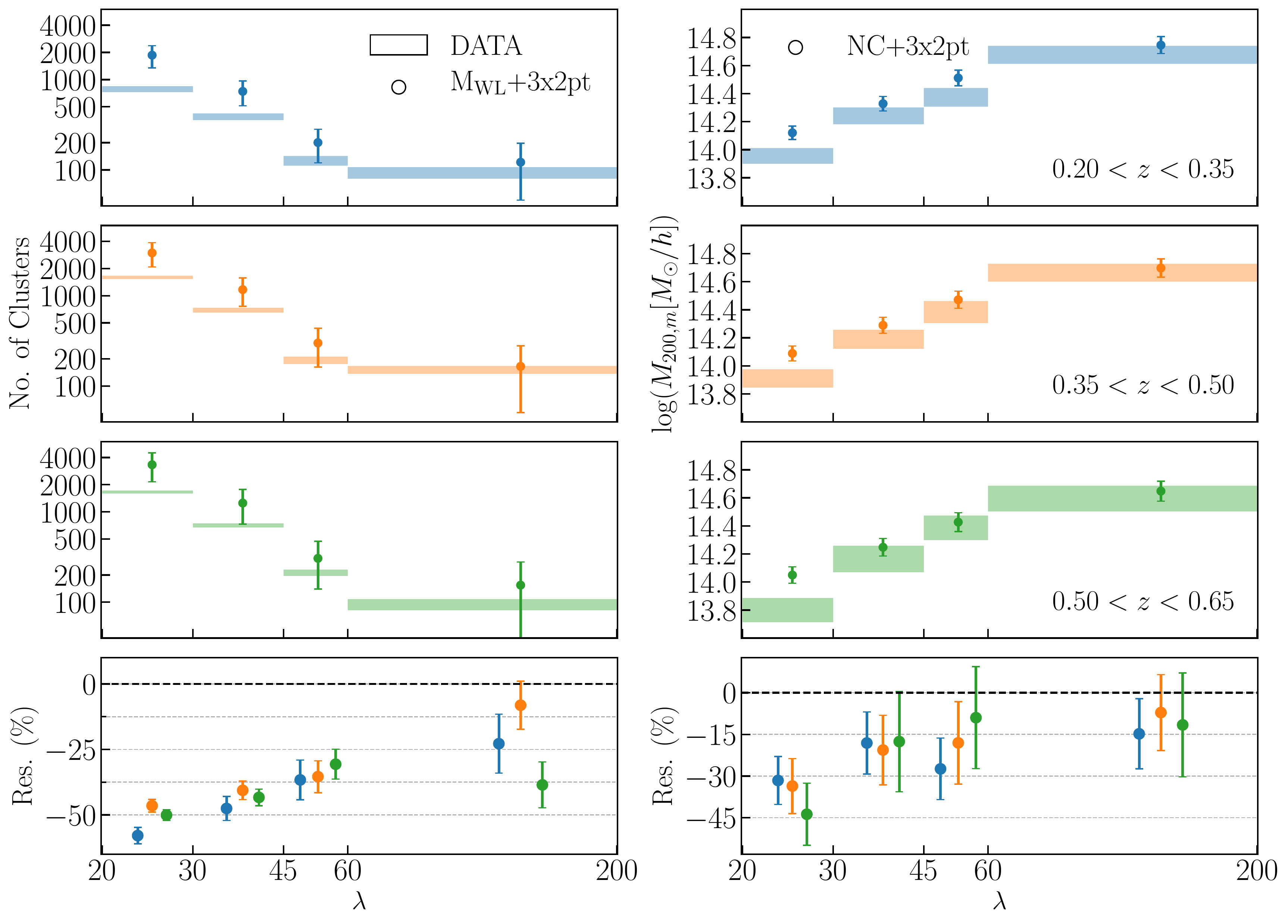}
\end{center}
\caption{Comparison of the observed data vectors (\textit{shaded areas}) with the number counts predicted from the combination of weak-lensing mass estimates and DES Y1 3x2pt cosmology (\textit{left} panel), and mean masses predicted from the combination of Y1 number counts data and DES Y1 3x2pt cosmological priors (\textit{right}). The \textit{y} extent of the \textit{shaded areas} correspond to the error associated with the data. The error bars on the predicted number counts and mean masses represent one standard deviation of the distribution derived sampling the corresponding MCMC chain. The lower panel shows the percent residual of the predictions to the data vectors, where the error bars refer to data vector uncertainties.}
\label{fig:datavec_comp}
\end{figure*}

We may now use the posteriors of the richness--mass relation derived using one observable (cluster counts or cluster masses) to predict the complementary observable.  This allows us to determine which aspects of the data are driving the tension in Figure~\ref{fig:s8om}.  Figure \ref{fig:datavec_comp} shows the comparison of our data vectors (\textit{shaded areas}) with our two predictions based on the complementary data set combined with DES 3x2pt priors (\textit{filled circles with error bars}). 

We see that the assumption that our recovered cluster masses and 3x2pt cosmology are correct implies that the \redmapper\ catalog is highly incomplete. Specifically, \redmapper\ should be $\sim 50\%$ incomplete at low richness, and between $10\%-40\%$ incomplete in the highest richness bin.  The \redmapper\ catalogs have been extensively vetted over the years, and such a large incompleteness, especially at high richness, is unlikely.  For instance, $100\%$ of the SPT and Planck SZ clusters within the DES Y1 footprint and below redshift $0.65$ are detected by \redmapper.  Extensive cross checks with both SPT cluster samples at $z>0.25$ \citep{bleemetal15,Bleem2019} and X-ray cluster samples at $0.1<z<0.35$ \citep{hollowoodetal18}  have so far failed to identify a single instance of a clear non-detection of a galaxy cluster due to \redmapper\ algorithmic failures.
In short, while there is still some room for a small fraction of undetected clusters at low richness, the level of incompleteness in the number counts required at $\lambda \gtrsim 40$ by our weak lensing cluster masses in a 3x2pt cosmology is unfeasible.


The right panels of Figure~\ref{fig:datavec_comp} compare the cluster masses predicted by the cluster counts assuming a 3x2pt cosmology to the masses estimated using weak lensing.  We find that the weak-lensing masses are low relative to the predicted masses based on the cluster number counts using the 3x2pt cosmology, with the difference ranging from $\sim 10\%$ percent in the highest richness bins to $\sim 30-40\%$ in the lowest richness bins.
In other words, the \it slope \rm of the recovered mass--richness relation from our weak lensing analysis appears to be biased high, a point to which we will return below. 

\begin{figure}
\begin{center}
    \includegraphics[width=0.45 \textwidth]{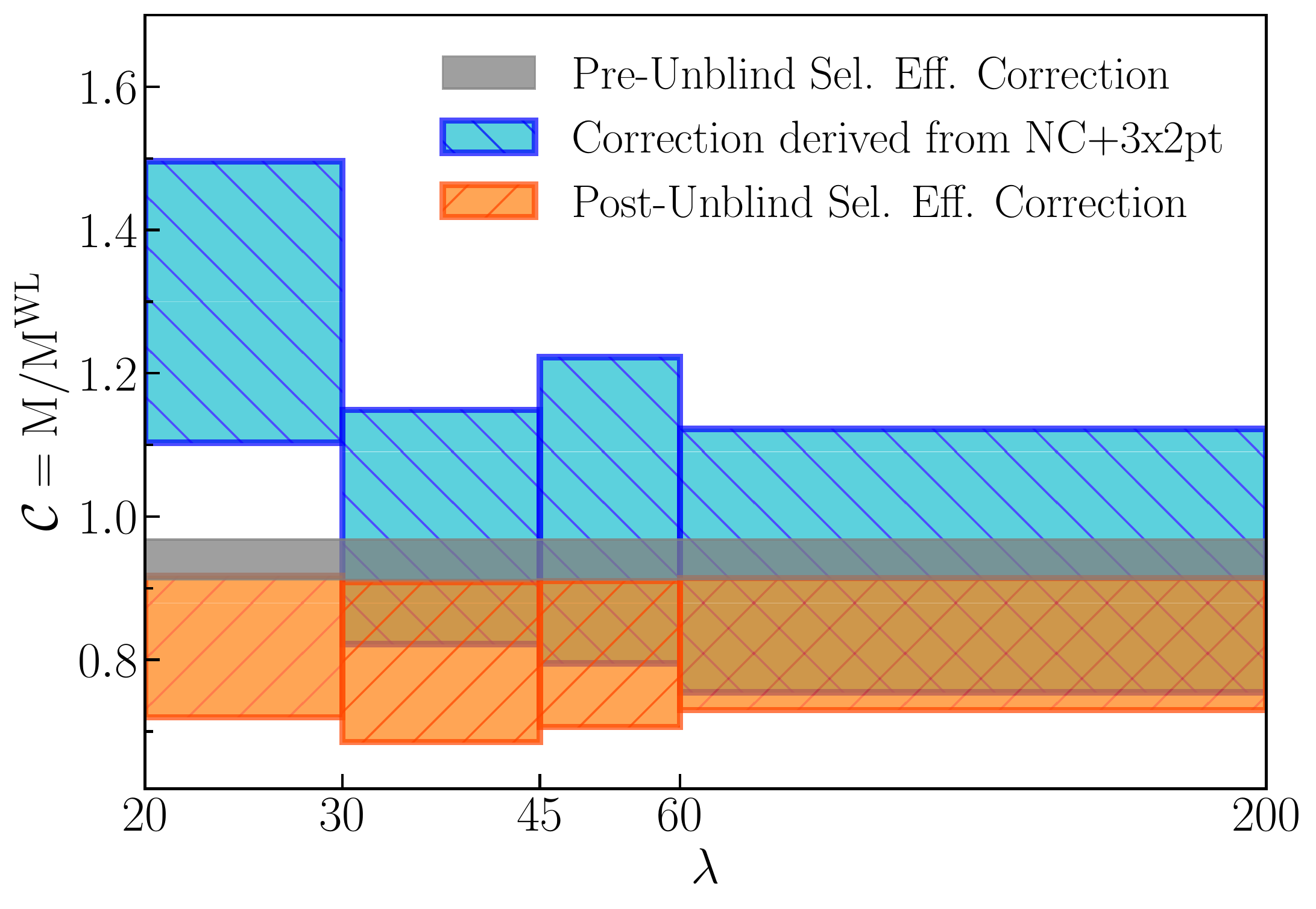}
\end{center}
\caption{\textit{Cyan bars}: Mean correction required to reconcile the weak-lensing mass estimates from \cite{desy1wl} -- without the triaxiality and projection effects corrections -- with the mean masses predicted by the combination of  Y1 cluster counts and 3x2pt cosmology. Also over-plotted the projection and triaxiality effects correction estimated analytically in  \cite{desy1wl} and adopted pre-unblinding (\textit{gray band}), and the selection effect correction adopted post-unblinding (\textit{orange bars}). The \textit{y} extent of the bars represent the $68\%$ confidence interval; the \textit{cyan} bars are estimated as the ratio of the masses predicted by randomly sampling the NC+3x2pt chain, and the ``raw'' weak-lensing masses randomly drawn from their posterior distribution.}
\label{fig:m_correction_comp}
\end{figure}

With the exception of our lowest richness bins, the difference between our predicted and observed weak-lensing masses can be reconciled within the systematic uncertainty associated with the selection effects corrections. 
It is interesting that interpreting the tension in terms of selection effect bias requires lowering the amplitude of the selection effect correction derived in Appendix~\ref{sec:unblnd} to a level comparable to our pre-unblinding analytical estimates. 
This is shown most clearly in Figure~\ref{fig:m_correction_comp}, in which we compare the correction to the ``raw'' weak-lensing masses necessary to reconcile the weak-lensing data with the number counts within the context of a DES Y1 3x2pt cosmology (\textit{cyan bars}) with the selection effect correction applied to the data (\textit{orange bars}).  There are two key takeaways from this figure: 1) the simulation-based estimates of the impact of selection effects appear to over-correct the weak-lensing masses, with the original analytical estimates being closer to what we would expect given a DES 3x2pt cosmology and the observed cluster counts, and 2) remarkably, a DES 3x2pt cosmology requires that we \it increase \rm the recovered weak-lensing masses in our lowest richness bins by $\approx 30\%$ to be consistent with our number counts.  The fact that the weak-lensing masses of the low richness clusters are biased low is counter to our {\it a priori} expectations.

\subsection{What are Possible Solutions?}

If we interpret our results as due to an offset between the recovered weak lensing masses and true mean cluster masses, 
Figure~\ref{fig:m_correction_comp} poses a remarkably difficult challenge. First, in order to match the DES 3x2pt expectation, the resulting bias must be richness dependent.  This immediately rules out traditional weak-lensing systematics---e.g. source photometric redshifts and/or multiplicative shear biases---since these systematics give rise to coherent shifts in the recovered masses across all richnesses. 
It is also worth noting that in addition to our own weak-lensing analysis, \cite{baxteretal18} and \cite{raghunathanetal19} used CMB lensing signal around DES clusters to determine the amplitude of the mass--richness relation, finding results consistent with our own.
This further strengthens the case that the weak-lensing signal is being measured correctly, but that its interpretation in terms of mean true mass is potentially problematic.

Perhaps the biggest challenge that Figure~\ref{fig:m_correction_comp} poses is the fact that while the ``raw'' weak-lensing masses are biased high at high richness (as expected), at low richness the weak-lensing masses are biased low by a very large amount. Since projection effects and cluster triaxiality tend to boost richness and weak-lensing masses in concert---leading to raw weak-lensing masses that are biased high---Figure~\ref{fig:m_correction_comp} suggests that these systematics are incapable of reconciling the weak lensing and abundance data within the context of a DES 3x2pt cosmology.  

The above argument assumes that projection effects act primarily as a form of noise that boosts the richness and weak-lensing masses of existing clusters, but one might wonder whether projection effects are better thought of as creating ``false detections'' in which ``clusters'' are really a string-of-pearls type arrangement, with no especially massive halo along the line of sight. One way to think of such projections is as very large non-Gaussian tails in the richness--mass relation toward high richness. From Figure~\ref{fig:robustness}, we see that doubling the amount of projection effects in our galaxy clusters moves our cosmological posteriors towards the DES 3x2pt model. However, a further increase of the amplitude of projection effects will not correspond to an additional relaxation of the tension with DES 3x2pt: the benefit of lowering the predicted mean cluster masses will be counterbalanced by the worse fit to the abundance data due to the predicted larger number of clusters.

More quantitatively, we assess the capability of a large contamination fraction to relieve the tension with 3x2pt as follows: we consider a model in which a fraction $f_{\rm cont}$ of the detected clusters is contributed by line-of-sight projections with effectively zero weak-lensing mass. To account for this systematic, we re-scale the predicted number counts and weak lensing masses by $1/(1-f_{\rm cont})$ and $(1-f_{\rm cont})$, respectively. Also, to account for a possible richness dependence we model the contamination fraction with a power law of the form: $f_{\rm cont}(\lambda\ob)= \Pi_0 (\lambda\ob/25)^\pi$. Finally, we fit for those parameters (along with all the others) combining cluster abundance and weak lensing data with DES 3x2pt cosmological priors, to derive the contamination fraction preferred by the our data sets in that cosmology. The fit results in a steeply decreasing contamination fraction ranging from $\sim 15\%$ in the lowest richness bin to $\sim 1\%$ in the highest richness bin. As expected, though, the model does not provide a good fit to the data in a 3x2pt cosmology, especially in the lowest richness bin where the predicted masses exceed the data by $15-30\%$. Specifically, repeating the analysis without including the cosmological priors and fixing the contamination fraction parameters to their best-fit values, we obtain cosmological posteriors which are still at $1.6\sigma$ tension with DES 3x2pt.
Importantly, a high fraction of false detection at low richness and redshift is also disfavoured by {\it Swift} X-ray follow up of $\lambda \approx 30$ clusters, in which all but one of $\approx 150$ low-richness ($\lambda \in [25,35]$ and $0.1<z<0.35$) SDSS \redmapper\ targets were X-ray detected (von der Linden et al, in preparation).

One systematic that might seem like a good candidate for explaining the bias in Figure~\ref{fig:m_correction_comp} is the impact of baryonic processes: baryonic feedbacks redistribute and expel mass from a galaxy cluster, leading to cluster counts and weak-lensing masses that are biased low relative to expectations from dark matter only simulations. Moreover, the effect would be stronger at low richness than at high richness, naturally producing a richness-dependent bias. However, results from hydrodynamical simulations disfavor this solution.  If the triaxiality and projection effects are roughly mass independent, as found in Appendix~\ref{sec:unblnd} and per our {\it a priori} expectations, then the amplitude of the baryonic feedback would be $\sim 30\%$ for clusters of richness $\lambda\approx 25$.  That is, baryonic feedback would need to expel nearly 30\% of the mass of a $\sim 10^{14} \Msun$ galaxy cluster, a fraction twice as large as its baryonic content ($f_b\simeq \Omega_b/\Omega_m\simeq 0.15$), a clearly unphysical proposition \citep[e.g.][]{Cui2014,Velliscig2014,Bocquet2016,Springel2017}. Similarly, \cite{hensonetal17}, using $M>10^{14}\Msun$ clusters extracted from a hydrodynamical simulation, found that the redistribution of mass due to baryonic feedback processes induces a $\sim 9\%$ bias on the recovered weak lensing mass, a factor of 3 times smaller than the bias required to reconcile our data sets in a 3x2pt cosmology. Moreover, we expect this bias to be further reduced in our analysis given that our fits allow the concentration parameter to vary with no informative priors in each bin, partially absorbing the effect of the mass redistribution.

Richness-dependent cluster miscentering suffers from much the same difficulty in explaining the observed discrepancy. While a systematic trend in cluster miscentering could introduce a richness-dependent bias in the recovered weak-lensing masses, it is hard to imagine miscentering giving rise to a 30\% under-estimate of the cluster mass. Such a correction would require a very high miscentering fraction at low richness, again in tension with {\it Swift} X-ray follow-up of low-richness SDSS \redmapper\ clusters (von der Linden et al., in preparation).

Cluster percolation has recently been identified as another possible source of systematic uncertainty \citep{garciarozo19}.  Excessive percolation could give rise to severe incompleteness in the low-richness bins, as we found was needed to reconcile our final weak-lensing masses with the cluster counts within the context of a DES 3x2pt cosmology.  If this were the case, then our percolation scheme must be overly aggressive. To test this, we reduce the percolation radius used from $1.5R_\lambda$ to $1.25R_\lambda$.  The corresponding change in the number of clusters is just under 1\%, far from what would be needed to reconcile the cluster lensing and number counts data in a 3x2pt cosmology.   We have also tested the impact of percolation on the weak lensing bias expected from numerical simulations, again finding a negligibly small impact.  

In short, we have thus far been unable to identify a systematic that can plausibly explain the tension between the weak lensing data and the cluster counts assuming a DES 3x2pt cosmology (Figure~\ref{fig:3x2cosmo}), particularly for our lowest-richness bins.

Interestingly, a lensing signal lower by $\sim20-40\%$ compared to predictions from galaxy clustering has been measured by \cite{Leauthaud2017} around BOSS CMASS massive galaxies at small scales ($M \sim 10^{13}\Msun$).  If the discrepancy in their measurement were somehow related to the low weak lensing mass of our low richness clusters, that would point towards a mass-dependent physical origin for the bias that ``turns on'' around $\approx 10^{14}\ \Msun$.


\subsection{Relation to Other Works}
\label{sec:other_works}
We have seen that the bias in the cosmological posterior shown in Figure~\ref{fig:s8om} can be fundamentally traced to the slope derived from our weak-lensing masses.  Figure 15 of \cite{desy1wl} compares the DES mass--richness relation to several other works in the literature.  All of these tend to have relatively large slopes, though the DES value is unusually large.  Two works in particular find slopes below unity: \cite{geachpeacock17} and \cite{Saro2015}.  Of these, \cite{geachpeacock17} has large error bars, so we will focus on the work by \cite{Bleem2019}, which is an update to the \cite{Saro2015} analysis.\footnote{The recent analysis of \cite{capasso2018} also results in a shallower slope of the mass--richness relation, but their analysis includes assumptions about X-ray scaling relations and the scatter of the richness--mass relation, which make it more difficult to interpret their results within the context of our analysis.} 

\begin{figure}
\begin{center}
    \includegraphics[width=0.45 \textwidth]{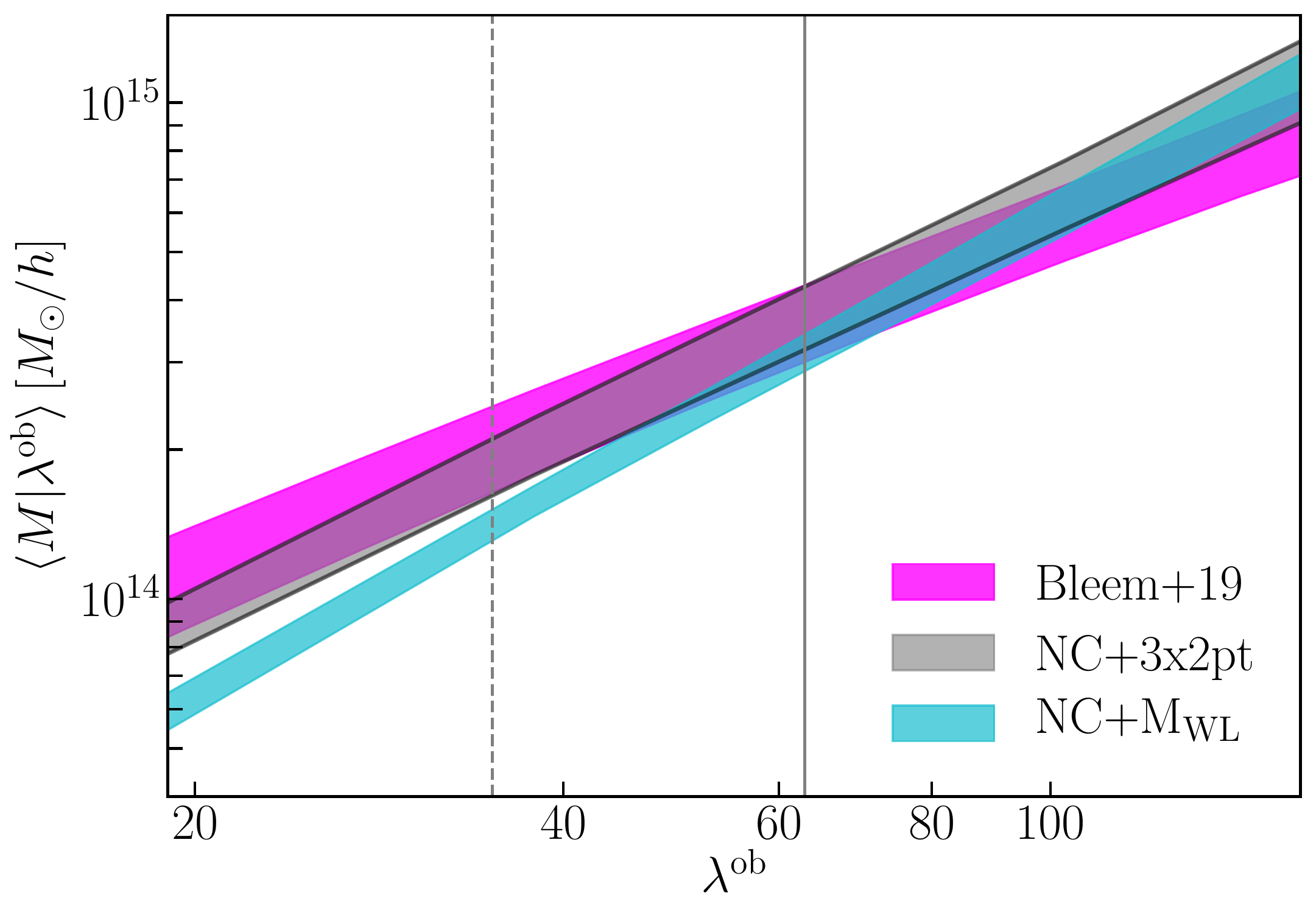}
\end{center}
\caption{Comparison of mass--richness relations at the mean sample redshift $z=0.45$. The \textit{cyan} and \textit{gray} bands show the M-$\lambda\ob$ relation derived in this work combing number counts data with weak-lensing mass estimates or DES Y1 3x2pt cosmological prior, respectively. For comparison, shown in \textit{magenta} the $\langle M | \lambda\ob \rangle$ relation from SPT SZ clusters \citep{Bleem2019}. The \textit{solid} ($\lambda\ob=35$) and \textit{dashed} ($\lambda\ob=63$) vertical lines correspond to the richnesses above which $68\%$ and $95\%$ of the SPT-SZ sample used in \cite{Bleem2019} is contained.}
\label{fig:m_rel_comp}
\end{figure}

We use the method of Section~\ref{sec:res:RMR} to derive the mass--richness relation as constrained using cluster abundances when assuming a DES Y1 3x2pt cosmology.  In Figure~\ref{fig:m_rel_comp} we compare this mass--richness relation (gray band) to that derived from our combined counts and weak-lensing analysis (cyan band), and to the mass--richness relation from \cite{Bleem2019} (magenta band).  The latter is obtained as follows. First, they cross-match clusters selected using the Sunyaev--Zel'dovich (SZ) effect as measured using the South Pole Telescope (SPT) so that each SPT cluster is assigned a richness.  Second, they assume a fiducial cosmology with $\sigma_8=0.8$ and $\Omega_m=0.3$.  Using the SPT selection function, the abundance of clusters as a function of SZ-signal constrain the cluster masses, which in turn leads to a constraint of the richness--mass relation.  In practice, this whole procedure is simultaneous and occurs at the likelihood level. It is worth noting that the SPT clusters typically have high richness values, with a median richness of 71.  Thus, the constraint shown in Figure~\ref{fig:m_rel_comp} at low richness is an extrapolation of their results.

The agreement between the \cite{Bleem2019} analysis and the posterior obtained by analyzing the optical cluster abundance assuming a DES 3x2pt cosmology is remarkable.  Given the similarity of the $S_8$ values---$S_8=0.782 \pm 0.027$ for DES 3x2pt and $S_8=0.8$ in the \cite{Bleem2019} analysis---this agreement implies that the optical and SPT abundances are compatible with each other, further strengthening the case that some unmodeled systematics reside
with the interpretation of the stacked weak lensing signal as mean cluster mass rather than the modeling of the richness--mass relation.
In particular, assuming a large incompleteness or contamination fraction as discussed above would result, for the combination of abundance data and DES 3x2pt cosmology priors, in a slope inconsistent with the results of \cite{Bleem2019}.  Importantly, at $\lambda \gtrsim 40$ --- the richness range probed by the SPT sample --- the weak-lensing masses and \cite{Bleem2019} results overlap.
Consistent results are also obtained by Grandis et al., (in preparation), 
who use cross-matched \redmapper--SPT clusters with $\lambda>40$ and the SZ signal--mass relation derived from the cosmological analysis of the SPT 2500 deg$^2$ cluster sample \citep{Bocquet2018} to calibrate the richness--mass relation. Similarly to \cite{Bleem2019}, when extrapolating their results to low richnesses ($\lambda \lesssim 30$) the predicted cluster masses are $\sim 30\%$ larger compared to our weak lensing mass estimates, while the predicted number counts are consistent with the \redmapper\ abundance data.

Figure~\ref{fig:m_rel_comp} is a modern incarnation of an old problem.  \cite{planck_maxbcg} studied the scaling relation between the richness of maxBCG clusters \citep{koesteretal07} and the SZ signature of those clusters using \planck\ data.  They found both a large amplitude offset, and a large difference in the slope, relative to that predicted using weak-lensing masses.  \cite{rozoetal14} argued that the difference in amplitude was due primarily to the assumed \planck\ masses being biased low by $\approx 30\%$, and the weak-lensing masses being biased high by $\approx 10\%$.  The difference in slope was, at that time, not significant given the corresponding uncertainties.  This is related to the fact that, even though our analysis of the SDSS \redmapper\ sample \citep{costanzietal18b}
undoubtedly suffers from the same systematics as our DES analysis, our SDSS results are consistent with the DES 3x2pt cosmology analysis.  In other words, it is only because of the improved statistical constraining power of the DES that the ``high'' slope of the mass--richness relation derived using weak lensing is now clearly problematic.

\begin{figure}
\begin{center}
    \includegraphics[width=0.45\textwidth]{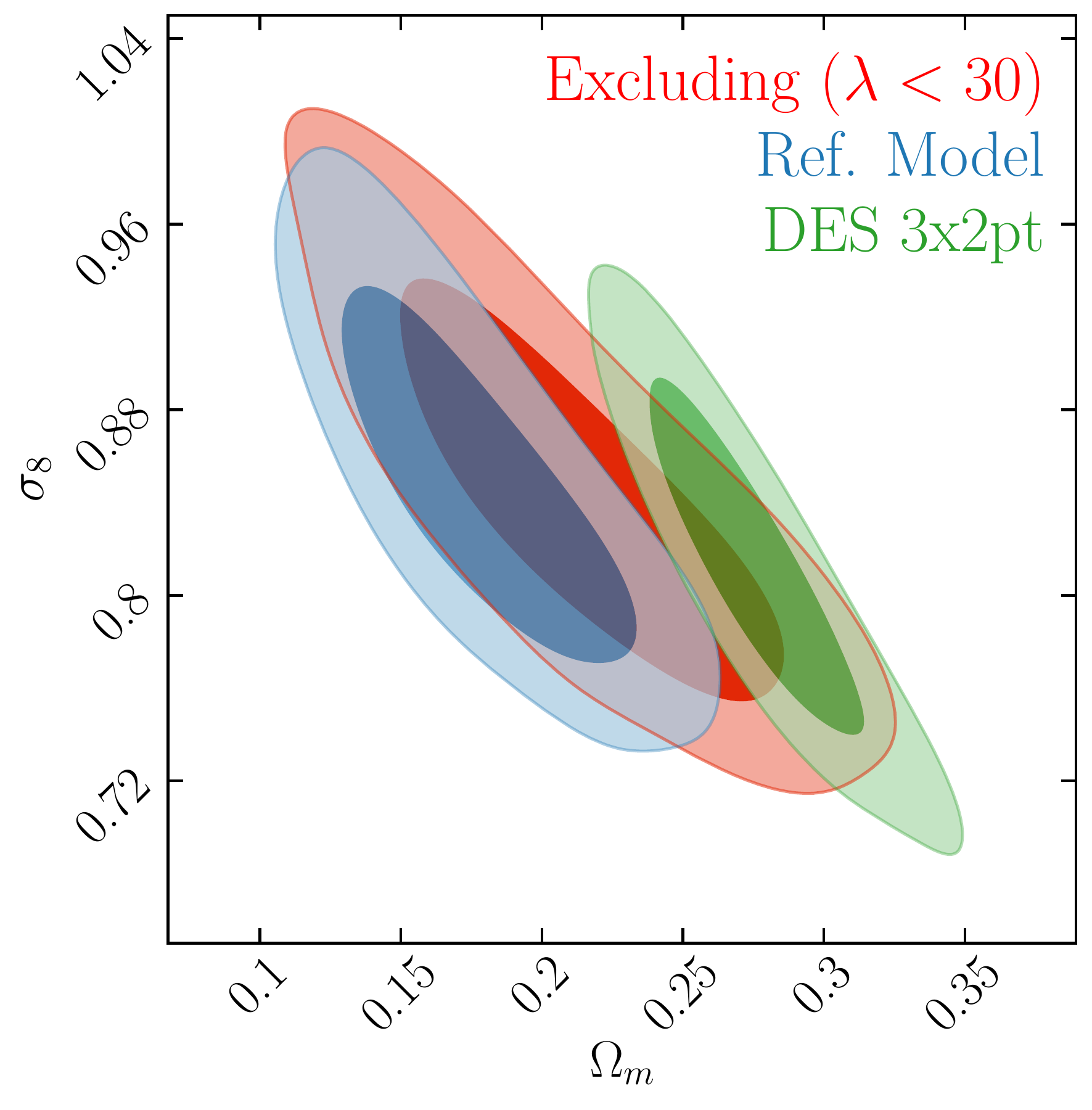}
    \includegraphics[width=0.46\textwidth]{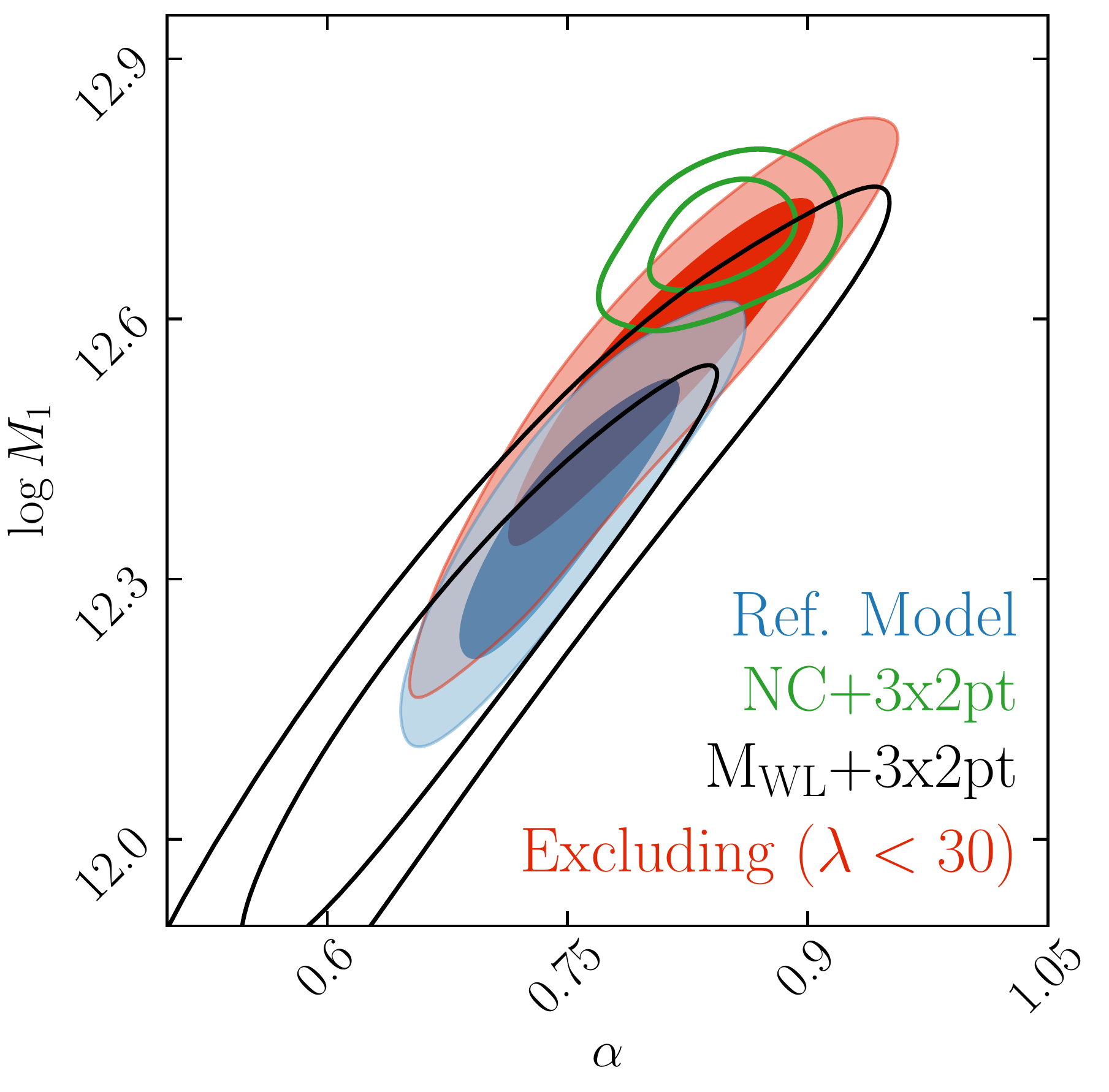}
\end{center}
\caption{Cosmological posteriors in the $\sigma_8$--$\Omegam$ (\textit{upper} panel) and $\log M_1$--$\alpha$ (\textit{lower} panel) plane for our fiducial analysis (\textit{blue}), and a new analysis in which we remove the lowest richness bins (\textit{red}). Removing the low richness bins shifts the posteriors towards larger $\Omegam$ values, bringing our analysis into agreement with the DES Y1 3x2pt cosmology analysis ($0.9\sigma$ tension; \textit{green} contours in the upper panel). Similarly, when excluding the low richness bins, the richness--mass relation posteriors move towards the region of the parameter space preferred by the combination of DES number counts and 3x2pt cosmological priors (\textit{green} contours in the lower panel).}
\label{fig:s8om_nolowrichness}
\end{figure}

To emphasize this point, we have rerun our analysis after dropping our lowest richness bin, making the mass range of our cluster sample more similar to that probed by X-ray and SZ selected catalogs.  The resulting posteriors are shown in Figure~\ref{fig:s8om_nolowrichness}. As we can see, dropping our lowest richness bins shifts our posteriors towards higher matter density, bringing our analysis into agreement ($0.9\sigma$) with the DES 3x2pt cosmology (\textit{upper} panel). 
On the other hand, the posteriors of the richness-mass relation move towards the region of the parameter space preferred by the combination of number counts and DES 3x2pt priors (\textit{lower} panel), and thus by the analysis of \cite{Bleem2019} using SZ selected clusters (see Figure~\ref{fig:m_rel_comp}). Moreover, if we use the results of this analysis to predict our observables in the lowest richness bins, we obtain predictions for the number counts which are consistent with the abundance data, while the predicted mean cluster masses are higher by $15-30\%$ than the weak lensing mass estimates. These results highlight the fact that most of the tension with the DES 3x2pt cosmology is driven by the $\lambda<30$ data, and that our weak lensing mass estimates for $\lambda<30$ and $\lambda>30$ are inconsistent with each other within our model when combined with abundance data.
Further removal of the next-lowest richness bin does not systematically shift the contours of the posterior. 
Aside from noting that our results are indeed especially sensitive to our lowest richness bin, Figure~\ref{fig:s8om_nolowrichness} makes a simple but important point: had we performed our analysis with fewer, more massive clusters -- analogously to previous abundance studies using X-ray and SZ selected clusters -- the underlying systematic that biased the cosmological posteriors in Figure~\ref{fig:s8om} would have remained undiscovered.  While this does not in any way demonstrate that clusters selected at other wavelengths will suffer from a similar systematic, 
it does open the possibility that such a systematic might exist also for low mass objects selected at other wavelengths. 

One intriguing possibility that arises from this discussion is the extent to which the biases uncovered in our analysis could be mitigated using different mass-calibration strategies.  For instance, in a recent work \cite{capasso2018} used dynamical information to calibrate the richness--mass relation of galaxy clusters using the CODEX cluster sample.  Encouragingly, they find a much shallower slope for the richness--mass relation, though their amplitude is in tension with ours and that of \cite{Bleem2019}. Of course, this does not negate the fact that the as-of-yet unidentified reason for discrepancy must be identified and understood, but it is encouraging to find that alternative methods of mass calibration may be less susceptible to the latter.  

Another possibility resides in the use of different mass proxies. A stellar mass based mass proxy, such as the one presented in \cite{Palmese2019} is expected to be less impacted by projection effects \citep{Bradshaw2019}. In future work, we plan on comparing results from these different mass proxies, which could help with shedding light on the unknown systematics found in this work.


\subsection{Correlated Scatter}

The analysis presented here is a ``backward'' analysis, in that one uses the weak lensing data to infer a cluster mass.  This is to be contrasted to a ``forward'' analysis, in which one forward-models the weak-lensing shear profile of galaxy clusters.  Forward analyses \citep[e.g.][]{Mantz2015,Murata2017,Bocquet2018} have traditionally assumed log-normal observable--mass relations, where the weak lensing signal is characterized by a weak-lensing mass $M_{\rm WL}$ that can correlate with the cluster-selection observable.  In the presence of correlated scatter, $P(M_{\rm WL}|M,\lambda) \neq P(M_{\rm WL}|M)$. Instead, the expectation value of $M_{\rm WL}$ is still a log-normal distribution, but the mean is given by \citep{evrardetal14}
\begin{equation}
    \avg{\ln M_{\rm WL}|\lambda} = \avg{\ln M|\lambda} + \beta r \sigma_{\rm M|WL}\sigma_{\rm M|\lambda} ,
\label{eq:wlbias}
\end{equation}
where $\beta$ is the slope of the halo mass function at the appropriate mass, and $r$ is the correlation coefficient between the weak-lensing mass and the cluster observable.  

Based on the above equations, it is easy to understand how the forward and backward modeling approaches are related.  In the backward modeling approach, we consider the ``correction term'' $\beta r \sigma_{\rm M|WL}\sigma_{\rm M|\lambda}$ to be an unknown for which we place priors based on numerical simulations.  When $r>0$, as expected from projection effects and triaxiality, this leads $M_{\rm WL}$ to be biased high.  

There are two points to emphasize here.  First, there is the simple equivalence of forward and backward modeling.  A ``forward model'' with the same assumptions as we have would result in identical cosmological posteriors.  Second, within the context of a log-normal model, Figure~\ref{fig:m_correction_comp} demonstrates that, under the assumption of the DES 3x2pt cosmology prior, the correlation coefficient between richness and weak-lensing mass must change as a function of mass, with $r>0$ at high mass (as expected), and $r<0$ at low mass.  What can give rise to such a trend in the correlation coefficient remains unknown.  Put another way, neither the ``direction'' of the  analysis, nor the adoption of a multi-variate log-normal model with correlated scatter, can resolve the tension in Figure~\ref{fig:s8om}.   


A second point of interest for forward modeling comes about because of the results shown in Appendix~\ref{sec:unblnd}.  In particular, Figure~\ref{fig:sel_eff_bias} demonstrates that selection biases may have strong scale dependence, and therefore cannot generally be modelled using a single ``weak-lensing mass bias''.  In a forward-model analysis, one should introduce the scale-dependent perturbations to the weak-lensing profiles, and marginalize over the amplitude of said perturbations.  While we fully expect that an effective ``weak-lensing mass'' suffices for now, we expect future cluster analyses will require an understanding of the scale-dependent impact of selection effects (or, within the context of a log-normal model, an understanding of the scale dependence of the weak-lensing scatter and correlation coefficient).

\section{Summary and Conclusion}
\label{sec:concl}

We have performed a cosmological analysis of the abundance and the weak lensing signal of the DES Y1 \redmapper\ clusters.  We summarize our findings below:

\begin{itemize}
    \item The cosmological posteriors of our unblinded analysis are in $5.6\sigma$ tension with \planck\ CMB, and $2.4\sigma$ tension with the DES 3x2pt cosmological analysis in the $\sigma_8-\Omegam$ plane.  This is driven by a low $\Omegam$ posterior that is in tension with all existing cosmological probes.  This finding is robust to the adopted cosmological and richness--mass relation model.
    \item The internal inconsistency of the DES Y1 cluster data with other DES probes rule out the possibility that the tension is driven by an observational systematic affecting the DES data.
    \item Cross checks of the \redmapper\ catalog with X-ray and SZ data  suggest that the abundance data and related modeling are not driving the tension but it is likely a consequence of an incorrect interpretation of the stacked  weak lensing signal of the DES \redmapper\ clusters.
    \item Low richness data ($\lambda \in [20,30]$) are the main driver of the tension with the DES 3x2pt cosmological results: dropping our lowest richness bin from the analysis removes the tension with DES 3x2pt ($0.9\sigma$). In particular, the weak lensing mass estimates for $\lambda<30$ push the slope and amplitude posteriors of the richness--mass relation towards lower values compared to the ones preferred by the combination of number counts and weak lensing data at $\lambda>30$, as well as by the analysis of \cite{Bleem2019} using SPT clusters ($\lambda \gtrsim 40$).
    \item Assuming our abundance data, modelling and DES 3x2pt results to be correct, we estimate the required bias in the observed weak-lensing masses by comparing the latter to the predicted masses assuming a DES 3x2pt cosmology and using the cluster counts to constrain the richness--mass relation. The relative mass offset we recover is richness dependent, corresponding to a steeper slope in the richness--mass relation compared to the one preferred by the weak lensing data. 
    \item Our understanding of how photometric cluster selection impacts the stacked lensing profiles of clusters might have a major role in the observed tension. However, at low richness, the necessary selection effect bias requires the raw weak-lensing masses of photometrically selected clusters to be biased \it low \rm relative to a mass-selected sample.  This is contrary to our {\it a priori} expectations, and we have not yet been able to identify a systematic that could give rise to such a selection effect.
    \item Interpreting our results within the context of correlated observables, our data implies that the correlation coefficient between richness and weak lensing is mass dependent, and changes sign in going from high mass clusters (positive correlation) to low mass clusters (negative correlation).  As noted above, this is very surprising. 
\end{itemize}

As discussed in section~\ref{sec:other_works}, hints of a richness-dependent bias in the weak lensing signal of galaxy clusters go as far back as \cite{planck_maxbcg}, but it is only with the improved statistical power of the DES that these biases have become statistically significant.  Understanding the origin of this systematic effect, and the degree to which it can be calibrated using multi-wavelength cluster data, is an absolute necessity for future photometric cluster cosmology analyses. 
Observational and simulation-based campaigns to study the relation of true cluster mass, observed richness, and weak lensing profiles, independent of the inherent limitations of purely photometric data, will shed light on the puzzles posed by DES Y1 cluster abundance and lensing data.


\section*{Acknowledgements}  
\label{sec:acknowledgements}

Funding for the DES Projects has been provided by the U.S. Department of Energy, the U.S. National Science Foundation, the Ministry of Science and Education of Spain, the Science and Technology Facilities Council of the United Kingdom, the Higher Education Funding Council for England, the National Center for Supercomputing Applications at the University of Illinois at Urbana-Champaign, the Kavli Institute of Cosmological Physics at the University of Chicago, 
the Center for Cosmology and Astro-Particle Physics at the Ohio State University,
the Mitchell Institute for Fundamental Physics and Astronomy at Texas A\&M University, Financiadora de Estudos e Projetos, 
Funda{\c c}{\~a}o Carlos Chagas Filho de Amparo {\`a} Pesquisa do Estado do Rio de Janeiro, Conselho Nacional de Desenvolvimento Cient{\'i}fico e Tecnol{\'o}gico and 
the Minist{\'e}rio da Ci{\^e}ncia, Tecnologia e Inova{\c c}{\~a}o, the Deutsche Forschungsgemeinschaft and the Collaborating Institutions in the Dark Energy Survey. 

The Collaborating Institutions are Argonne National Laboratory, the University of California at Santa Cruz, the University of Cambridge, Centro de Investigaciones Energ{\'e}ticas, 
Medioambientales y Tecnol{\'o}gicas-Madrid, the University of Chicago, University College London, the DES-Brazil Consortium, the University of Edinburgh, 
the Eidgen{\"o}ssische Technische Hochschule (ETH) Z{\"u}rich, 
Fermi National Accelerator Laboratory, the University of Illinois at Urbana-Champaign, the Institut de Ci{\`e}ncies de l'Espai (IEEC/CSIC), 
the Institut de F{\'i}sica d'Altes Energies, Lawrence Berkeley National Laboratory, the Ludwig-Maximilians Universit{\"a}t M{\"u}nchen and the associated Excellence Cluster Universe, 
the University of Michigan, the National Optical Astronomy Observatory, the University of Nottingham, The Ohio State University, the University of Pennsylvania, the University of Portsmouth, 
SLAC National Accelerator Laboratory, Stanford University, the University of Sussex, Texas A\&M University, and the OzDES Membership Consortium.

Based in part on observations at Cerro Tololo Inter-American Observatory, National Optical Astronomy Observatory, which is operated by the Association of 
Universities for Research in Astronomy (AURA) under a cooperative agreement with the National Science Foundation.

The DES data management system is supported by the National Science Foundation under Grant Numbers AST-1138766 and AST-1536171.
The DES participants from Spanish institutions are partially supported by MINECO under grants AYA2015-71825, ESP2015-66861, FPA2015-68048, SEV-2016-0588, SEV-2016-0597, and MDM-2015-0509, 
some of which include ERDF funds from the European Union. IFAE is partially funded by the CERCA program of the Generalitat de Catalunya.
Research leading to these results has received funding from the European Research
Council under the European Union's Seventh Framework Program (FP7/2007-2013) including ERC grant agreements 240672, 291329, and 306478.
We  acknowledge support from the Australian Research Council Centre of Excellence for All-sky Astrophysics (CAASTRO), through project number CE110001020.

This manuscript has been authored by Fermi Research Alliance, LLC under Contract No. DE-AC02-07CH11359 with the U.S. Department of Energy, Office of Science, Office of High Energy Physics. The United States Government retains and the publisher, by accepting the article for publication, acknowledges that the United States Government retains a non-exclusive, paid-up, irrevocable, world-wide license to publish or reproduce the published form of this manuscript, or allow others to do so, for United States Government purposes.

MC is supported by the ERC-StG 'ClustersXCosmo' grant agreement 716762. ER was supported by the DOE grant DE-SC0015975, by the Sloan Foundation, grant FG-2016-6443, and the Cottrell Scholar program of the Research Corporation for Science Advancement. This research used simulations that were performed resources of the National Energy Research Scientific Computing Center (NERSC), a U.S. Department of Energy Office of Science User Facility operated under Contract No. DE-AC02-05CH11231.


\appendix


\section{Calibration of the Distribution $P(\lob|\ltrue,\lowercase{z}\true )$}
\label{app:lob_calibration}

A key ingredient in our analysis is our characterization of noise in photometric richness estimates.  As discussed in section~\ref{sec:model}, we consider three distinct sources of noise in $\lob$:
\begin{enumerate}
\item A Gaussian random noise associated with photometric uncertainties, uncorrelated structures, and background subtraction.
\item An exponentially decaying additive contribution to the richness due to projection effects that is dominated by the contribution from correlated structures along the line-of-sight.
\item A multiplicative correction that removes galaxies from the cluster richness estimates of a small fraction of low mass systems due to the impact of percolation in the construction of the \redmapper\ cluster catalog. 
\end{enumerate}
The random noise can straightforwardly be calibrated from the data.  We use the matched filter used to detect \redmapper\ clusters to generate Monte Carlo realizations of our cluster model, and proceed to insert these artificial clusters into our data set.  We generate $10^4$ cluster realizations along a grid of cluster richness $\lambda\true$ and cluster redshift $z$.   Each of these clusters is placed at a random point within the survey footprint, and the magnitude of every galaxy in the simulated cluster is perturbed according to the effective survey depth in each band at the galaxy's location. We then estimate the richness of the galaxy clusters.  The distribution $P(\lambda\ob|\lambda\true)$ obtained in this way is very well approximated as a Gaussian, and the observational uncertainty on the posteriors of these parameters is negligible.  In this way, we fully characterize observational uncertainties due to photometric uncertainties, uncorrelated structures, and background subtraction. 

We characterize the impact of correlated large scale structure using the method developed in \citet*{costanzietal18}.   This model is intuitively very simple: when two clusters are aligned along the line of sight, the smaller of the clusters will get blended into the larger of the two systems, with some fraction of its galaxies being mistakenly assigned to the larger system.  The fraction of galaxies that the small cluster loses will depend on the separation along the line of sight between the two systems: if the separation is zero, the smaller cluster will be entirely subsumed within the larger system, while if the separation is large the two clusters will be easily distinguished from each other, so there will be no artificial projection effects.  Evidently, the critical input to this model is the calibration of how the strength of projection effects decreases with increasing cluster separation.  Note that the fraction of the cluster lost to projection effects must be unity at zero separation, zero at large separation, and must have a slope of zero at zero separation.  Consequently, we expect a priori that a simple Gaussian can succesfully describe this function.  

\begin{figure}
\begin{center}
    \includegraphics[width=0.45 \textwidth]{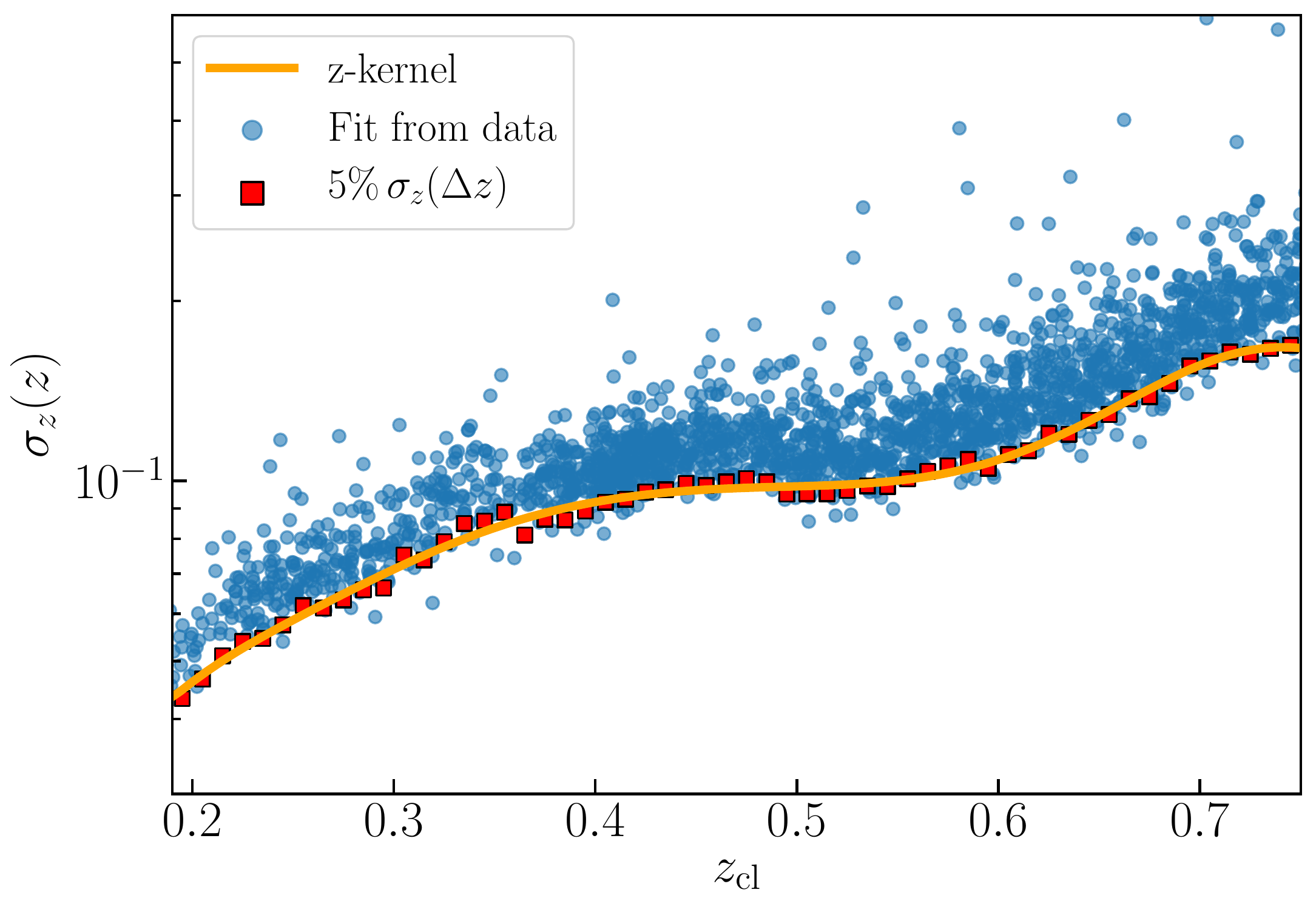}
\end{center}
\caption{The blue dots are the best-fit values for $\sigma_z$ obtained when fitting the curves $\lambda(z)$ for each cluster in the \redmapper\ cluster catalogue.
The {\it red squares} represent the $5$ percentile of the $\sigma_z$ distribution estimated in redshift bins of width $\Delta z=0.01$. The {\it solid orange} line shows the model for $\sigma_z(z)$ adopted for the analysis.}
\label{fig:z_kernel}
\end{figure}

We calibrate the separation dependence of projection effects by calculating the fractional decay of the cluster richness as a function of redshift, that is, the fraction of member galaxies of a cluster that \redmapper\ would assign to a putative cluster perfectly aligned with the former as a function of their separation in redshift.
This fractional decay is in fact well described by a Gaussian, enabling us to calibrate the width of this Gaussian as a function of cluster redshift.  Because this function should reflect only the intrinsic width of the red sequence and photometric errors, we did not expect this fraction to depend on cluster richness, an expectation that we explicitly confirmed.  We then calibrated the width of the Gaussian decay as a function of redshift in the DES data.  The resulting calibrated data is shown in Figure~\ref{fig:z_kernel}. Our best-fit model is a simple polynomial fit that successfully described our data.


We use our projection effects model to generate a synthetic data set as follows.  Starting from the DES Buzzard light cone simulation (DeRose et al. 2018, in prep, Wechsler et al. 2018, in prep), we assign to each halo an intrinsic richness $\lambda\true$.  We then rank order the halos by $\lambda\true$, and proceed to compute their projected richnesses using the projection effect model of \citet*{costanzietal18} as calibrated above.  Halos that contribute a fraction $f$ of their galaxies to a richer system along the line-of-sight have their own final richness decreased by a factor $1-f$, i.e. we enforce galaxy conservation.  The end result is a galaxy cluster catalog that includes both projection effects and percolations.  We use this simulated catalog to characterize both the characteristic richness enhancement due to projection effects which characterizes the exponential distribution of this noise, and the fraction of galaxy clusters that suffer from percolation effects (i.e. the fraction of clusters who lost some of their galaxies to richer systems along the line of sight).  Both of these effects are richness and redshift dependent: richer systems live in denser environments, which increases the importance of projection effects.  Likewise, systems at higher redshift are noisier, making it easier to blend systems together, and therefore increasing the impact of projection effects.  Finally, with regards to percolation, low richness systems are much more likely to suffer from percolation effects (the richest systems rarely have an even richer system along their line of sight).  

These trends are all very precisely measured in the simulation, and the corresponding observational uncertainties are negligible compared to the associated systematic uncertainties.  In particular, it should be obvious that the impact of projection effects is cosmology dependent: higher $\sigma_8$ and higher $\Omegam$ models will result in increased projection effects.  Fortunately, as demonstrated in \citet*{costanzietal18}, these differences are very nearly degenerate with the parameters of the intrinsic richness--mass relation, so the cosmological posteriors from our analysis are extremely robust to these types of effects.  Indeed, as we demonstrate in the main body of this text, even if we entirely neglect the impact of correlated structures along the line of sight, our cosmological posteriors are hardly affected.  

\begin{figure}
\begin{center}
    \includegraphics[width=0.45 \textwidth]{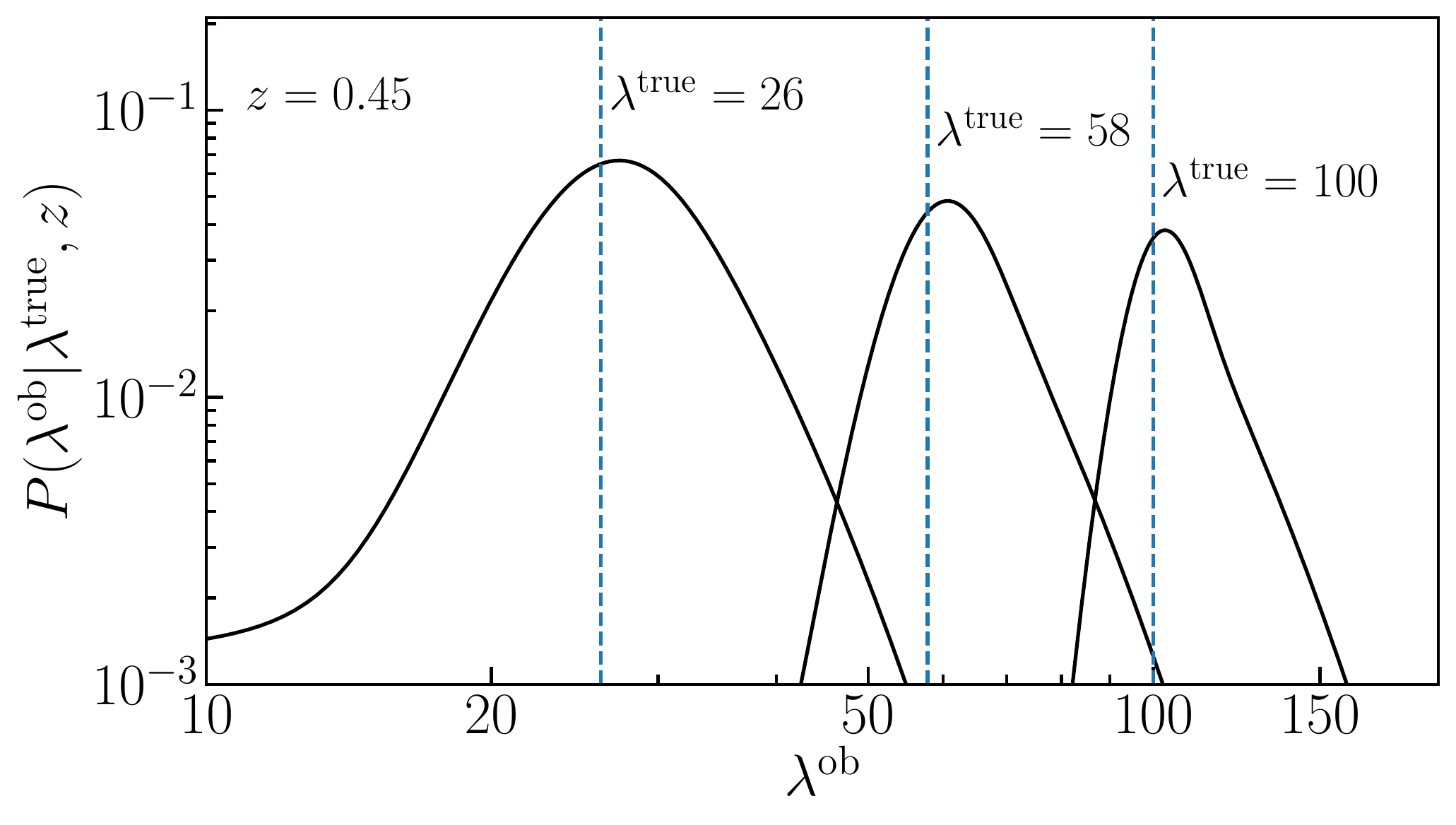}
\end{center}
\caption{$P(\lambda\ob|\lambda\true)$ distribution for clusters of true richness $\lambda\true=20$, $58$, and $100$ at the mean redshift of the sample $z=0.45$.
}
\label{fig:p_lob_ltr}
\end{figure}

Figure~\ref{fig:p_lob_ltr} shows our calibration of the distribution $P(\lambda\ob|\lambda\true)$ for clusters of richness $\lambda\true=20$, $58$, and $100$ at the mean redshift of the sample $z=0.45$. The Gaussian peak due to observational noise is evident, as is the long-tail to high richness due to projection effects.  The low tail at low richness is due to percolation.  

\section{Blinding and Unblinding Protocol}
\label{sec:blinding}

The DES analysis was performed blind, but followed a staged unblinding procedure.  In particular, we used the DES inference pipeline to analyze the SDSS \redmapper\ cluster catalog \citep{costanzietal18b}.  Both the SDSS and DES analyses were performed blind, but the unblinding of the SDSS analyses was part of the DES unblinding protocol, as described below.  This staged unblinding has both benefits and drawbacks.  The principal benefit is that ``unknown unknowns'' may be discovered and fixed after unblinding a ``weak'' data set (SDSS), enabling us to implement any necessary corrections before unblinding the ``strong'' (DES) data set.  The principal drawback is that this type of blinding is somewhat weaker than simply unblinding the ``strong'' data set.
However, we emphasize that the DES photometry, shear, and photo-$z$ catalogs are completely independent of the corresponding SDSS catalog.
In practice, no ``unknown unknowns'' were identified when unblinding the SDSS data sets, so the effective path through the DES unblinding protocol was identical to that of a simultaneous unblinding.

Our DES blinding protocol is as follows:
\begin{enumerate}
\item The cosmological parameters in the MCMC were randomly displaced before being stored.  The displacement was stored in a not-human-readable format (binary).
\item All modeling choices for both SDSS {\it and} DES were made before unblinding of the SDSS data sets.  Modeling choices were not allowed to change after unblinding of the SDSS data set.  
\item In addition to our fiducial model for the scaling relation, we considered one additional model for projection effects, namely random-point-injection.  Random-point-injection refers to the projection effects model calibrated by inserting galaxy clusters at random locations in the sky. This method obviously underestimates the impact of projection effects, so we take half of the difference in cosmological parameters between our fiducial model and this extreme projection effects model as the systematic uncertainty associated with modeling of projection effects.
\item All priors for both the SDSS and DES data sets were finalized before unblinding of the SDSS data set, with one critical exception: the prior on the intrinsic scatter parameter $\sint$.  In \cite{costanzietal18b}, we applied a prior $\sint \in [0.1,0.8]$.  At $\sint=0.8$, the model predicts that $\approx 11\%$ of massive clusters ($M\geq 10^{15}\ \msun$) do not host red-sequence galaxies ($\lambda\true=0$).  This seems implausible.   \cite{nohcohn12} studied the scatter in richness at fixed mass in numerical simulations in which galaxies were used to populate resolved halo substructures.  They then fit Gaussian distributions to their results.  Their best fit total fractional scatter in a maxBCG-like cluster catalog \citep{koesteretal07} was 0.37.\footnote{Tails due to projection effects were obvious, but we note that our model explicitly accounts for such tails.}  Note this is a total scatter, so $\sint$ must be strictly less than 0.37 in this simulation.  Moreover, since \redmapper\ is demonstrably superior to maxBCG \citep{rozorykoff14}, the above number should be pessimistic.  Based on this argument, we set for the DESY1 analysis the conservative upper limit $\sint \leq 0.5$.  This upper limit is also low enough that fluctuations that produce negative richnesses (formally non-detections) are sufficiently rare for them to be irrelevant for our study ($P(\lambda\true \leq 0) \leq 0.6\%$).\footnote{A Gaussian model is mathematically more convenient than a log-normal model, both because Poisson distributions are closer to a Gaussian distribution than to a log-normal distribution, and because the exponential model for projection effects is easy to convolve with a Gaussian.} 

Finally, the DES analysis includes an additional parameter, $\epsilon$, governing the redshift evolution of the intrinsic richness--mass relation (see Equation \ref{eqn:RMR}).

\item The weak-lensing masses of the clusters in each richness/redshift bin remained blind throughout the entire weak lensing analysis, which was completed before SDSS unblinding. No alterations of the lensing pipeline were made post-unblinding of the DES weak lensing data, except for a minor bug-fix that affected the boost factor correction of only one richness/redshift bin.  The change in mass was well below the uncertainty for that bin, and the bug was found and fixed before unblinding the cosmological constraints.  For details on our weak lensing calibration of the DES data set, we refer the reader to \cite{desy1wl}.
\item No comparison of our cosmological constraints to any other data sets were performed prior to unblinding of the DES data.  Our analysis in \cite{desy1wl} did compare the DES and SDSS weak-lensing masses, but this was only done after the DES masses were unblinded.
\end{enumerate}

Our unblinding protocol was defined by the set of requirements detailed below.
\begin{enumerate}
\item The SDSS analysis was unblinded, and ``unknowns unknowns" were either not found or addressed, as appropriate.
\item All non-cosmological systematics tests of the shear measurements were passed, as described in \cite{y1shapes}, and all priors on the multiplicative shear biases were finalized.
\item Photo-$z$ catalogs were finalized and passed internal tests, as described in \citet*{Y1pz}.
\item Our inference pipeline successfully recovered the input cosmology in a synthetic data set (see Appendix~\ref{app:validation}).
\item All planned DES-only chains (including alternative models) were run and satisfied the Gelman-Rubin convergence criteria \citep{Gelman1992} with $R-1 \leq 0.03$.
\item Since not explicitly included in the analysis, we demanded that the systematic uncertainty in our posteriors due to projection effects modeling---estimated as half the difference between the central values of the posteriors for our fiducial model and the random-point-injection model---were smaller than the corresponding statistical uncertainties.
\item We verified that the posteriors of all parameters which we expected would be well-constrained did not run into the priors within the 95\% confidence region when using a flat $\Lambda$CDM model. Parameters that are prior dominated are $M_{min}$, $\sigma_{intr}$, $s$, $q$, $h$, $\Omega_b h^2$, $\Omega_\nu h^2$, and $n_s$.  All of these were expected to be prior dominated, and all prior ranges were purposely conservative. Of these, the two that might be most surprising to the reader might be $M_{min}$ and $\sigma_{intr}$, as these parameters help govern the richness--mass relation.  However, notice that $M_{min}$ is the mass at which halos begin to host a single central galaxy; since our cluster sample is defined with the richness threshold $\lambda\geq 20$, the mass regime of halos which host a single galaxy is simply not probed by our data set.  Likewise, our data vector is comprised only of the mean mass of galaxy clusters in a given richness bin, a quantity that is largely independent of the scatter in the richness--mass relation \citepalias[see][which accounts for the modest scatter dependence as a systematic uncertainty in the recovered masses]{desy1wl}.\footnote{Interestingly, in the log-normal model the data does constrain the scatter parameter.}
\item Finally, this paper underwent internal review by the collaboration prior to unblinding.
All members of the DES cluster working group, as well as our internal reviewers, had to agree that our analysis was ready to unblind before we proceeded.
\end{enumerate}

\section{Blinded Analysis Results}
\label{sec:blinded_res}
After all of our unblinding requirements were satisfied, we proceeded to unblind our results. For the two cosmological parameters constrained by our data set we obtained for the \textit{blinded} analysis $\Omega_m=0.172^{+0.023}_{-0.029}$ and $\sigma_8=0.956^{+0.045}_{-0.056}$, corresponding to $S_8=0.720\pm0.032$.
Figure \ref{fig:sel_eff_comp} shows the resulting posteriors on the $\sigma_8 - \Omegam$ plane (\textit{blinded} analysis; \textit{gray}), along with the posteriors obtained from the \textit{unblinded} analysis (i.e. our reference results; \textit{red}), DES 3x2pt (\textit{blue}) and \Planck\ CMB (\textit{gold}). As can be seen from the figure, the \textit{blinded} analysis results are in clear tension with those derived by the other DES probes and \planck\ CMB ($2.3\sigma$ and $6.7 \sigma$ in the $\sigma_8-\Omegam$ plane, respectively). Driven by the low $\Omegam$ value recovered, a larger than $3.5 \sigma$ tension is also present with BAO measurements \citep{BAO6dF,BAOSDSSMain,BAOBoss} and supernovae data \citep{SNpanth2017}.

The $\chi^2$ of the best-fit model of the \textit{blinded} analysis is $38.35$. Based on the expected $\chi^2$ distribution (see section \ref{sec:chi2} for details) the model adopted in the \textit{blinded} analysis did not provide a good fit to the data ($\chi^2/\nu_{\rm eff}=38.35/18.65$). This was driven primarily by the offset between the predicted and observed abundances of galaxy clusters in our highest redshift and largest richness bin.

Given the large tension with the DES 3x2pt and \planck\ results, as well as with BAO, supernovae and other independent cluster count analyses, we attempted to trace back the source of the tension, whether it be an objective bug in the code and/or an unknown/underestimated source of systematic bias.

\section{Post-unblinding Tests and Selection Effect Calibration}
\label{sec:unblnd}

Two minor bugs were discovered in our pipeline post-unblinding. First, the projection effect correction adopted in \cite{desy1wl} was implemented with the wrong sign, and second we implement in our pipeline $\avg{\ln M|\lambda}$ rather than $\ln\avg{M|\lambda}$.  Fixing these bugs had only a minor impact on the cosmological posteriors.
Post unblinding, an independently coded version of our cosmological pipeline was completed.  The two pipeline codes were found to be in excellent agreement with each other, precluding the possibility of a bug in the code used to analyze the data (the bugs above came from the processing of the data).

To address possible model systematic biases we test a variety of different models for $P(\lambda\ob|M,z)$ besides those considered pre-unblinding, which include:
\begin{enumerate}
\item A model in which the intrinsic scatter of the richness--mass relation is allowed to be mass dependent.
\item A model in which the intrinsic scatter of the richness--mass relation is allowed to be redshift dependent.
\item A model in which the slope of the mass-richness relation $\alpha$ in Eq.~\ref{eqn:RMR} depends on mass: $\alpha(M)=\alpha_0+\alpha_M \log(M/10^{14.2})$.
\item A model in which the slope of the mass-richness relation $\alpha$ can evolve with redshift: $\alpha(z)=\alpha_0 [(1+z)/(1+z_*)]^{\alpha_z}$.
\end{enumerate}
None of the models tested seem to suggest a large systematic bias on cosmological posteriors related to model assumptions: the differences between $P(\lambda\ob|M)$ models are mainly accommodated by a shift of the richness--mass relation parameters.

Finally, as noted in the main text, we used numerical simulations to update our model for the impact of selection effects on the recovered weak-lensing mass of galaxy clusters.  This work was started before unblinding, but was only completed post-unblinding and found an effect in excess of previous literature results. We describe our calibration of selection effects below.

We ran \redmapper\ on 12 simulated Y1-like light-cones from the Buzzard Flock suite \citep{DeRose2019BuzzardFlock}. The synthetic data have been tuned to match the observed evolution of galaxy counts at different luminosities as well as the spatial clustering of the galaxy population of DES Y1 data. To avoid double counting of miscentering effects, \redmapper\ has been run fixing the cluster center on the dark matter halo center.
We thus computed the azimuthally averaged stacked mass density profile $\bar{\Sigma}(R)$ of the clusters in richness/redshift bins using the dark matter particle distribution.
Then, for each richness and redshift bin, we randomly selected 1000 halos from the simulations with the same mass and redshift distribution as the clusters in the bin. Finally, we  measured the stacked mass density profiles of mass-selected cluster samples and compare them to those obtained from the richness/redshift selected samples.

\begin{figure*}
\begin{center}
    \includegraphics[width=\textwidth]{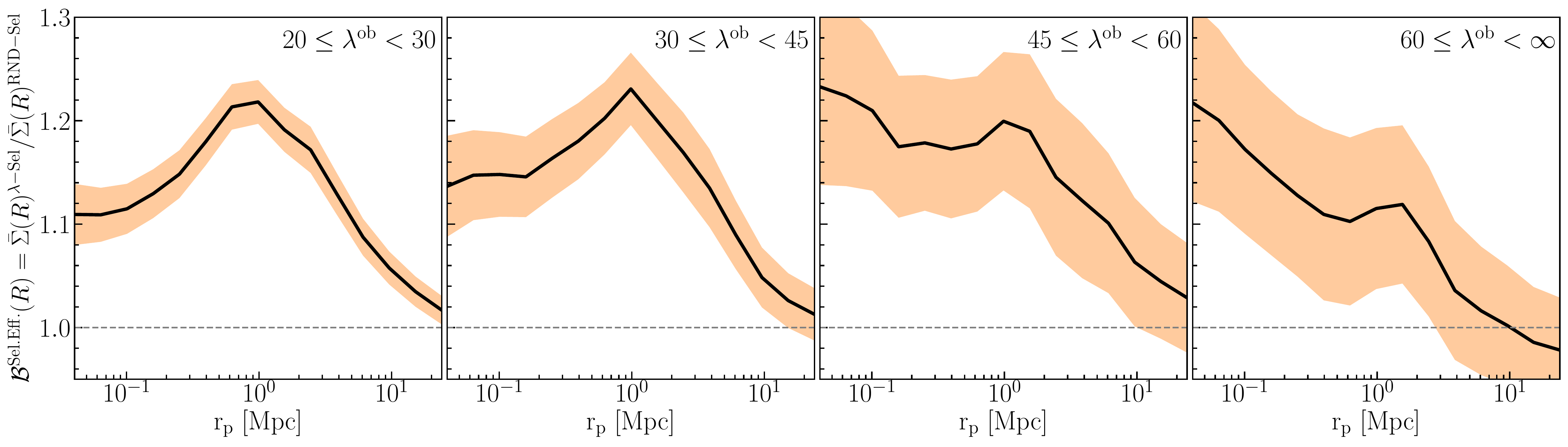}
\end{center}
    \caption{Selection effect bias on the stacked mass density profile derived from synthetic clusters in the redshift range $0.35<z<0.50$ for the four $\lambda\ob$ bins considered in the analysis. The bias is computed as the ratio of the stacked $\bar{\Sigma}(R)$ profiles measured in clusters selected by richness and clusters randomly selected from the simulations so as to match the mass and redshift distribution of the $\lambda\ob$-selected sample. The \textit{black} lines correspond to the means of the biases retrieved from 12 Y1-like simulations, while the shaded area represent one standard deviation of the mean.
}
\label{fig:sel_eff_bias}
\end{figure*}

Figure~\ref{fig:sel_eff_bias} shows the ratio of stacked mass density profiles of the \redmapper\ selected clusters to that of the mass-selected sample.  Note that the mass and redshift distribution of the two samples is identical by construction, thus any difference between the two is due to selection effects.  We find that \redmapper\ selected clusters have a weak lensing signal that is biased higher by $\approx 10-25\%$ over the relevant radius range than that of similar, purely mass/redshift selected clusters\footnote{During the finalization of this analysis similar findings  have been presented in the work of \cite{Sunayama20}} (see figure \ref{fig:sel_eff_bias}).
This indicates that, at a given halo mass, \redmapper\ preferentially selects halos with a boosted lensing signal compared to a random sample.
This bias is partially due to triaxiality and projection effects, and will be studied in greater detail in an upcoming publication (Wu et al., in preparation).
Specifically, matching our control samples not only by mass and redshift, but also by the $\sigma_z$ (our proxy for projection effects; see Appendix~\ref{app:lob_calibration}) and halo orientation distributions of the richness-selected sample, reduces the bias between the two samples by $\sim 50\%-100\%$ depending on the bin and radius considered.

\begin{figure*}
\begin{center}
    \includegraphics[width=\textwidth]{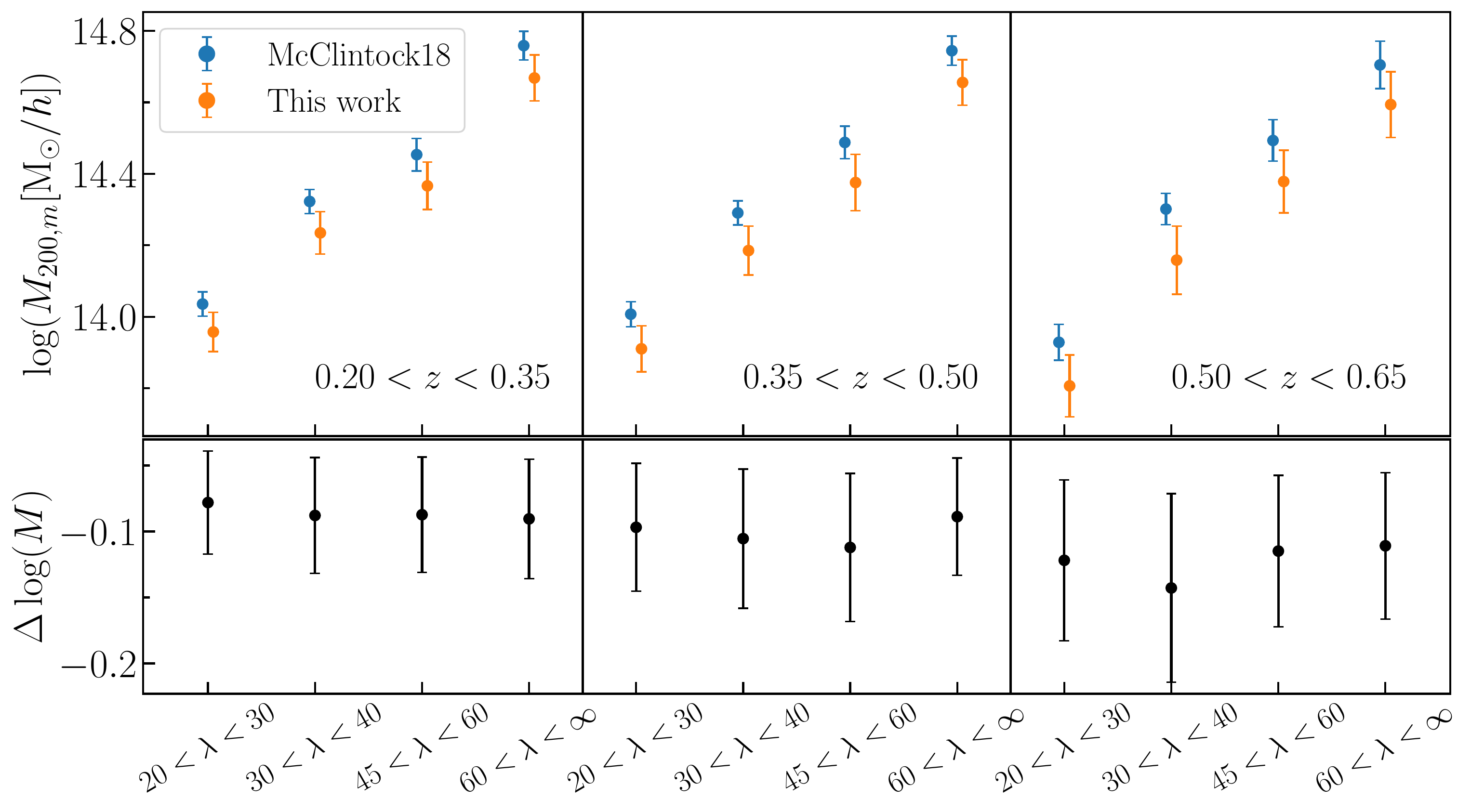}
\end{center}
    \caption{\textit{Upper} panel: Comparison of the weak-lensing masses derived in \cite{desy1wl} and the ones adopted in this work which include the selection effect bias correction and uncertainty. The inclusion of this systematic lowers the weak-lensing mass estimates by $20-30\%$ and increases the error budget by $50-60\%$ depending on the richness/redshift bin. \textit{Lower} panel: Difference of the log masses derived including or not the selection effect bias correction. The error bars correspond to the uncertainty associated with the selection effect bias estimated for each bin as half of the difference between the two mass estimates.
}
\label{fig:mass_comparison}
\end{figure*}

We re-analyzed the stacked weak lensing data including in the $\bar{\Sigma}(R)$ model \citepalias[Eq. 28 of][]{desy1wl} the multiplicative selection effect bias factor relevant for the bin considered: $ \mathcal{B}^{\rm Sel. Eff.}(R)=\bar{\Sigma}(R)^{\lambda{\rm -Sel}}/\bar{\Sigma}(R)^{\rm RND-Sel}$. The masses derived including this systematic are $\sim 20-30\%$ smaller compared to the previous results, with a larger bias for richer and high redshift clusters (see Figure \ref{fig:mass_comparison}). Since this systematic effect is still under investigation we add in quadrature to the re-fitted masses a conservative error equal to half of the difference between the old and the new results -- $\sigma^{\rm Sel.Eff.}=|M^{\rm new}-M^{\rm old}|/2$ -- that is, the absence or doubling of selection effects on weak lensing mass estimates is excluded at $2\sigma$.
While the mock \redmapper\ catalogs obtained from the Buzzard Flock suite are known to underestimate the richness of the clusters at fixed mass \citep{DeRose2019BuzzardFlock}, since the selection effect correction is calibrated on the relative bias of stacked lensing profiles of samples with the same mass distribution, we do not expect this to affect our results.
As a confirmation of the latter statement, we repeat the analysis on synthetic \redmapper\ clusters extracted from a Buzzard simulation adopting different assumptions for the red-sequence and clustering model, finding results consistent with the one above.
Nonetheless, additional tests on different synthetic data will be fundamental to further validate our findings and reduce the associated uncertainty.

The cosmological constraints in the main text adopt this systematic calibration, as noted in section~\ref{sec:mass_sys}.

\begin{figure}
\begin{center}
    \includegraphics[width= 0.45 \textwidth]{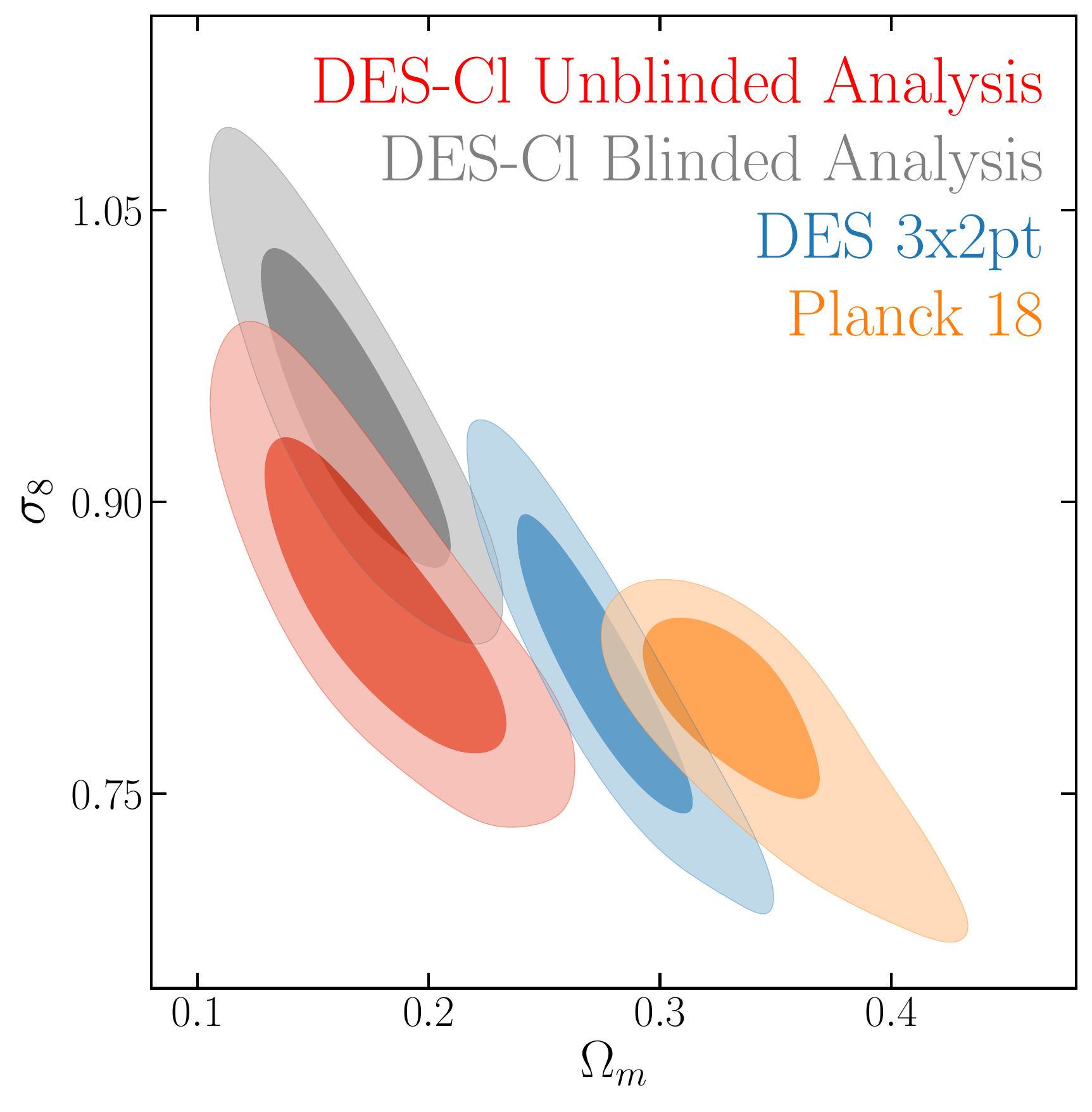}
\end{center}
    \caption{$\sigma_8$-$\Omegam$ posteriors from the DES Y1 cluster blinded analysis (\textit{gray}) and unblinded analysis (\textit{red}). The latter adopt the selection effect bias correction on the weak lensing mass estimates detailed in Appendix \ref{sec:unblnd}. Shown for comparison also are the posteriors obtained from the DES 3x2pt (\textit{blue}) and \planck\ CMB (\textit{gold}) analysis.  The smaller weak lensing masses recovered including the selection effect bias lead to a $\sim 2\sigma$ shift of the $\sigma_8$ posterior, while the larger systematic error associated to the masses entails a relaxation of the $S_8$ posterior of $\sim18\%$. As evident from the figure, the inclusion of the selection effect bias does not substantially change the level of tension with DES 3x2pt or \planck\ CMB results in the $\sigma_8-\Omegam$ plane.
}
\label{fig:sel_eff_comp}
\end{figure}

In figure \ref{fig:sel_eff_comp} we show the effects of the selection effect bias and associated systematic uncertainty on the DES Y1 cluster posteriors in the $\sigma_8-\Omegam$ plane. The $\sim 20\%$ lower weak-lensing mass estimates adopted in our reference analysis shift downwards by $\sim 2\sigma$ the $\sigma_8$ posteriors while leaving the $\Omegam$ posterior mostly unaffected (compare \textit{gray} and \textit{red} contours). Furthermore, the larger systematic error associated with the mass estimates causes the $S_8$ posteriors to relax by $\sim 18\%$. We note that the inclusion of the selection effect bias does not substantially affect the level of consistency of our results with DES 3x2pt or \planck\ CMB posteriors. We further stress that the \textit{gray} contours are shown here only to illustrate the effect of the selection bias, and should not be considered as possible alternative results of this analysis.


\section{Pipeline Validation}
\label{app:validation}

\begin{figure*}
\begin{center}
    \includegraphics[width=\textwidth]{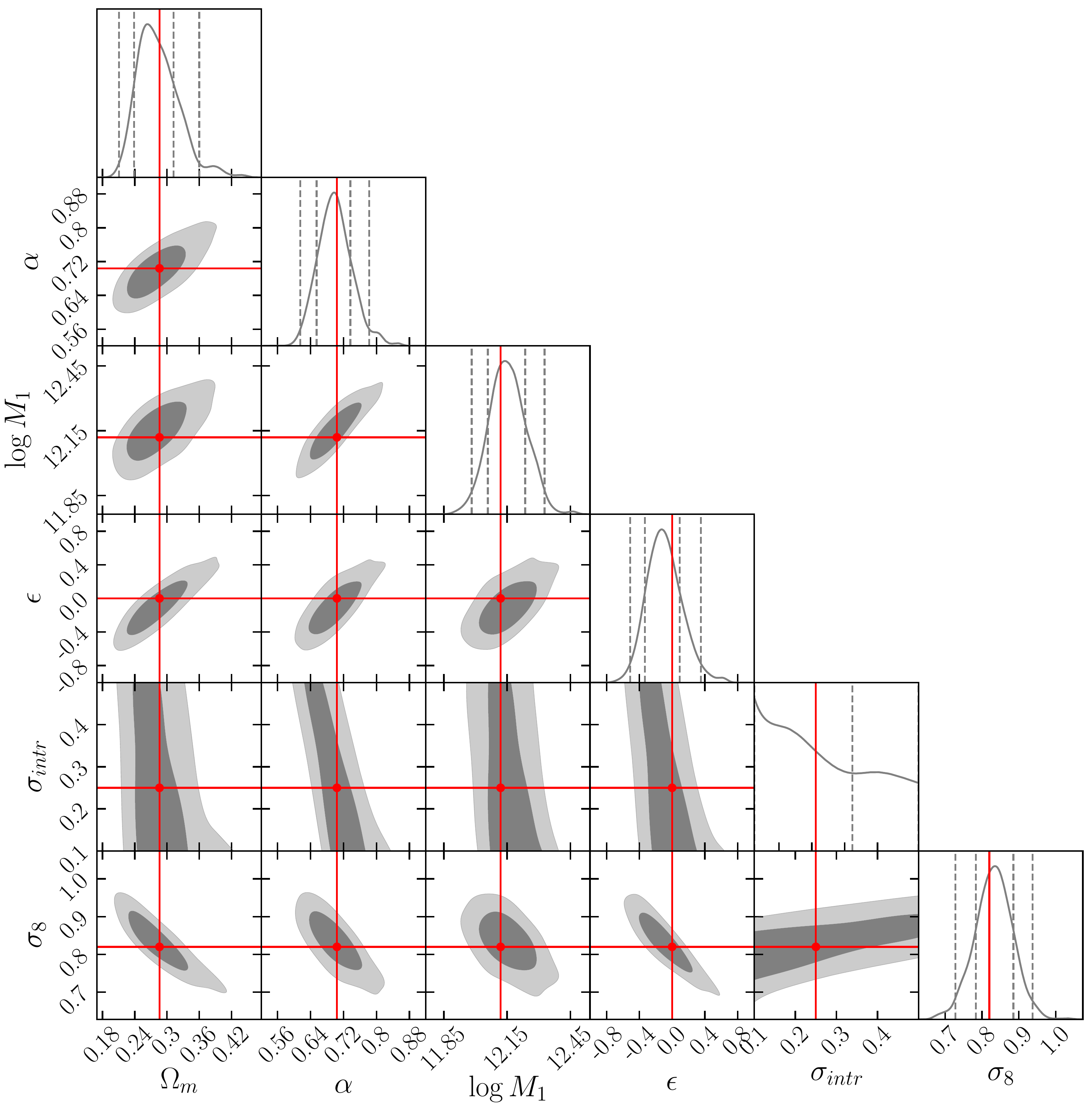}
\end{center}
\caption{ $68\%$ and $95\%$ confidence contours obtained running our pipeline on mock data. The input parameter values used to generate the simulation and the mock data catalog are shown in {\it red}. The {\it dashed} lines shown in the $1$-d marginalized distributions (diagonal of the triangle plot) correspond to the $0.025$, $0.16$, $0.84$ and $0.975$ quantiles of the distributions. This plot includes only the model parameters that are not prior dominated.}
\label{fig:mock_res}
\end{figure*}

We validate our analysis pipeline using the simulated cluster catalog described in Appendix~\ref{app:lob_calibration}.  Specifically, starting from the simulated cluster catalog described above, we bin the simulated clusters in richness and redshift bins in a way that is identical to that done in the real data.  We then calculate the mean halo mass of the resulting galaxy clusters, and scatter it according to the observational noise in our cluster mass calibration.  This ``noise-scattering'' properly accounts for correlated uncertainties due to systematics.  The end result is a simulated data vector of cluster abundances and mean cluster masses that can be used as an input to our cosmology pipeline.  
Figure~\ref{fig:mock_res} shows the posteriors from our pipeline when applied to this simulated data set.  The input cosmology is shown as the intersection of the horizontal and vertical lines in each plane, which describe the input parameters of the simulation.  The good agreement between these input parameters and our analysis posteriors demonstrate that our pipeline is working as intended.

\bibliography{database}

\end{document}